\documentclass{aa}
\usepackage{xcolor}
\usepackage[breaklinks=true, colorlinks, citecolor={blue}, urlcolor={blue}]{hyperref}
\usepackage{afterpage}
\usepackage[varg]{txfonts}
\usepackage{supertabular}
\usepackage[normalem]{ulem}
\usepackage{csquotes}
\usepackage{units}
\usepackage{mathtools}
\usepackage{float}
\usepackage{natbib,graphicx}
\usepackage{txfonts}
\usepackage{amsmath}
\usepackage{footmisc}
\usepackage{soul}
\usepackage{lineno}
\usepackage{multirow}
\usepackage{multicol}
\usepackage{amsfonts}
\usepackage{eucal}
\usepackage{natbib,twoopt}
\bibpunct{(}{)}{;}{a}{}{,} %% natbib format for A&A and ApJ
\makeatletter
\newcommandtwoopt{\citeads}[3][][]{\href{http://adsabs.harvard.edu/abs/#3}%
{\def\hyper@linkstart##1##2{}%
\let\hyper@linkend\@empty\citealp[#1][#2]{#3}}}
\newcommandtwoopt{\citepads}[3][][]{\href{http://adsabs.harvard.edu/abs/#3}%
{\def\hyper@linkstart##1##2{}%
\let\hyper@linkend\@empty\citep[#1][#2]{#3}}}
\newcommandtwoopt{\citetads}[3][][]{\href{http://adsabs.harvard.edu/abs/#3}%
{\def\hyper@linkstart##1##2{}%
\let\hyper@linkend\@empty\citet[#1][#2]{#3}}}
\newcommandtwoopt{\citeyearads}[3][][]%
{\href{http://adsabs.harvard.edu/abs/#3}
{\def\hyper@linkstart##1##2{}%
\let\hyper@linkend\@empty\citeyear[#1][#2]{#3}}}

\makeatother
\bibpunct{(}{)}{;}{a}{}{,}

\bibpunct{(}{)}{;}{a}{}{,}

\newcommand{\order}[1]{} %to impose the correct order in reference with same first author and same year

\def\gsim{ \lower .75ex \hbox{$\sim$} \llap{\raise .27ex \hbox{$>$}} }

\def\lsim{ \lower .75ex\hbox{$\sim$} \llap{\raise .27ex \hbox{$<$}} }

\begin{document}
\title{Pre-merger alert to detect the very-high-energy prompt emission from binary neutron-star mergers: Einstein Telescope and Cherenkov Telescope Array synergy}

\titlerunning {Detection perspective of early VHE counterparts of GW events in ET era}
\author{Biswajit Banerjee\inst{\ref{inst1},\ref{inst2}, \ref{inst3}}\thanks{corresponding authors: \\ B. Banerjee, G. Oganesyan, and M. Branchesi; \\ biswajit.banerjee@gssi.it, gor.oganesyan@gssi.it, \\marica.branchesi@gssi.it}
\and Gor Oganesyan\inst{\ref{inst1},\ref{inst2}}$^*$
\and Marica Branchesi\inst{\ref{inst1},\ref{inst2},\ref{inst3}}$^*$
\and Ulyana Dupletsa\inst{\ref{inst1},\ref{inst2}}
\and Felix Aharonian\inst{\ref{inst6}, \ref{inst7}}
\and Francesco Brighenti\inst{\ref{inst1}}
\and Boris Goncharov\inst{\ref{inst1},\ref{inst2}}
\and Jan Harms\inst{\ref{inst1},\ref{inst2}}
\and Michela Mapelli\inst{\ref{inst4}, \ref{inst5}}
\and Samuele Ronchini\inst{\ref{inst1},\ref{inst2}}
\and Filippo Santoliquido\inst{\ref{inst4}, \ref{inst5}}
}
\institute{Gran Sasso Science Institute, Viale F. Crispi 7, I-67100, L'Aquila (AQ), Italy
\label{inst1}
\and
INFN - Laboratori Nazionali del Gran Sasso, I-67100, L’Aquila (AQ), Italy
\label{inst2}
\and 
INAF - Osservatorio Astronomico d’Abruzzo, Via M. Maggini snc, I-64100 Teramo, Italy
\label{inst3}
\and 
Dublin Institute for Advanced Studies, 31 Fitzwilliam Place, D04 C932, Dublin 2, Ireland
\label{inst6}
\and 
Max-Planck-Institut fur Kernphysik, PO Box 103980, D-69029 Heidelberg, Germany
\label{inst7}
\and 
Dipartimento di Fisica e Astronomia ’G. Galilei’, Università degli studi di Padova, vicolo dell’Osservatorio 3, I-35122, Padova, Italia
\label{inst4}
\and 
INFN, Sezione di Padova, via Marzolo 8, I-35131, Padova, Italia
\label{inst5}
}
\abstract{
The current generation of very-high-energy $gamma-$ray (VHE; E $>$ 30\,GeV) detectors (MAGIC and H.E.S.S.) have recently demonstrated the ability to detect the afterglow emission of GRBs. However, the GRB prompt emission, typically observed in the 10 keV-10 MeV band, has so far remained undetected at higher energies. 
Here, we investigate the perspectives of multi-messenger observations to detect the prompt emission of short GRBs in VHE. Considering binary neutron star mergers as progenitors of short GRBs, we evaluate the joint detection efficiency of the Cherenkov Telescope Array (CTA) observing in synergy with the third generation of gravitational wave detectors, such as the Einstein Telescope (ET) and Cosmic Explorer (CE). In particular, we evaluate the expected capabilities to detect and localize gravitational wave events in the inspiral phase and to provide an early warning alert able to drive the VHE search. We compute the amount of possible joint detections by considering several observational strategies and  demonstrate that the sensitivities of CTA make the detection of the VHE emission possible even if it is several orders fainter than the one observed at 10 keV–10 MeV. We discuss the results in terms of possible scenarios of the production of VHE photons from binary neutron star mergers by considering the GRB prompt and afterglow emissions.}


\maketitle
%making single column:
%\onecolumn 
%---------------------
\section{Introduction}\label{sec:intro}
Gamma-ray bursts (GRBs) are extremely luminous events ($\rm\sim\,10^{52}$~erg/s) occurring at cosmological distances. The emission in the prompt phase ranges from milliseconds to several minutes in the keV-MeV range \citep{Kouveliotou1993}. The following multi-wavelength afterglow emission lasts from minutes to months and is observed from radio to {very-high-energy gamma-rays (VHE; E $>$ 30\,GeV)}. While the afterglow emission is interpreted as deceleration of the GRB jet in the circumburst medium \citep{Paczynski1993,Meszaros1997,Sari1998}, the prompt emission is thought to originate from the internal dissipation of the ultra-relativistic jet either via shocks \citep{Narayan1992,Rees1994} or magnetic re-connection \citep{Drenkhahn2002,Lyutikov2003,Zhang2011}. Given the unknown origin of the observed GRB spectra, it is not yet clear if GRB jets are primarily baryonic or magnetic in nature \citep[see][for a review]{Piran2004,Kumar2015,ZhangBook}. Some authors support the scenario for which the GRB jets are dominated by kinetic energy \citep{Paczynski1986,Goodman1986,Shemi1990,Meszaros1992,Rees1992,Levinson1993}, and thus the prompt emission is produced either below the photosphere via radiation-dominated shocks \citep{Eichler2000,Ghirlanda2003,Ryde2005,Peer2006,Giannios2012,Beloborodov2013} or above via optically thin shocks driven by the magnetic turbulence \citep{Narayan1992,Rees1994}. Others suggest that GRB jets are dominated by the Poynting flux \citep{Usov1992,Thompson1994,Meszaros1997} and thus dissipated via magnetic re-connection \citep{Drenkhahn2002,Lyutikov2003,Zhang2011}. In most of the scenarios, excluding the sub-photospheric models, the synchrotron radiation from the non-thermal population of electrons (or protons \citealt{Ghisellini2020} but also see \citealt{Florou2021}) is considered to make most of the observed 10 keV - 10 MeV emission \citep{Rees1994,Sari1996,Tavani1996}. However, the physical parameter space of the source is still unclear \citep{Lloyd2000,Derishev2001,Bosnjak2009,Nakar2009} given the synchrotron line of death problem \citep{Preece1998,Ghisellini2000}. The reshaped spectra of the Synchrotron Self Compton (SSC) component \citep{Papathanassiou1996,Sari1997,Pilla1998,Ando2008,Bosnjak2009} by the pairs \citep{Guetta2003,Peer2004,Razzaque2004} and/or the high energy components produced by the photo-pion interactions \citep{Asano2007,Gupta2007,Asano2009} are 
expected to give signatures in the VHE domain during the prompt phase. The intensity of this emission depends on the strength of the magnetic field, the size of the emission region, the bulk Lorentz factor, and the acceleration process (protons vs electrons energy gain). Therefore, detection or even upper limits on the VHE emission during the prompt emission phase are critical to establishing the nature of GRB jets. \newline
The recent detection of the VHE emission from the GRB afterglows by Major Atmospheric Gamma Imaging Cherenkov Telescope system \citep[MAGIC;][]{2016APh....72...76A, 2016APh....72...61A}  and High Energy Stereoscopic System (H.E.S.S\footnote{https://www.mpi-hd.mpg.de/hfm/HESS/pages/about/telescopes/}) opened up new possibilities of observing these energetic transients.
Thanks to the improvement in the sensitivities and  
smaller
response time of the current generation of Imaging Atmospheric Cherenkov Telescopes (IACTs), we are now able to detect the GRB afterglow emission  in the TeV band (E$>$ 1 TeV) by the MAGIC and H.E.S.S. telescopes, respectively, as shown for the GRB 190114C \citep{2019Natur.575..455M} and GRB 180720B \cite{2019Natur.575..464A}. The detection of GRB 190829A \citep{2021Sci...372.1081H} by the H.E.S.S. Collaboration at energies above 100 GeV shows a similar decay profile for the X-rays and VHE components supporting the same emission nature.  
There were also attempts to detect the VHE emission from the nearby short GRB 160821B using MAGIC  \citep[i.e.,][]{2021ApJ...908...90A}. However, the detection significance is below 4$\sigma$ despite of the shortest slew time of 24 sec achieved by the MAGIC telescope with respect to any other ground-based TeV instruments to date. 
The prompt/early VHE emission from short GRBs has been recently
searched by analyzing the High Altitude Water Cherenkov telescope (HAWC) observations. However, looking at the data within 20\,s from the burst of 47 short GRBs detected by \textit{Fermi, Swift} and {\it Konus} satellites and laying in HAWC field of view, no detection was found in the energy range of 80-800\,GeV \citep{2015ICRC...34..715L}. \newline
It is noteworthy that for the GRB 221009A (the highest fluence GRB ever detected) 
the \textit{Fermi}  Large Area Telescope (LAT) detected a high-energy counterpart starting after about 200 s from the \textit{Fermi}/GBM trigger time, and LHAASO reported the detection of {several thousand VHE photons up to 10 TeV and beyond} within 2000\,s from the trigger-time \citep{LHAASO}. The LHAASO experiment shows a significant improvement over the present generation (e.g., HAWC) of water Cherenkov detectors with the help of two primary components: the water Cherenkov detector (WCDA), operating in the energy range of 0.3-10\,TeV, and the KM2A array, sensitive to energies above 10\,TeV. 
LHAASO \citep[][and references therein]{2019arXiv190502773C}, which covers almost more than 18\% of the sky with almost full duty cycle, is a promising facility to detect the emission from short GRB in survey mode if the VHE emission peaks above 1\,TeV.  \newline
The detection of the short and faint gamma-ray burst GRB 170817A \citep{2017ApJ...848L..13A, 2017ApJ...848L..14G, 2017ApJ...848L..15S} associated with the first gravitational-wave signal GW170817 observed by the Advanced LIGO~\citep{Aasi2015} and Virgo~\citep{Acernese2015} detectors from a binary neutron star merger 
 \citep{2017PhRvL.119p1101A} marked the beginning of a new era of multi-messenger astronomy including GWs \citep{2017ApJ...848L..12A}. The  multi-wavelength observations from the first seconds to several months after the merger have shed light on the origin of short GRBs as products of binary neutron star (BNS) mergers and on the properties of relativistic jets in GRBs \citep{2017ApJ...848L..12A,2017Sci...358.1579H,2017Natur.551...71T, Lyman2018, Alexander2018,Mooley2018, Ghirlanda2019}.\newline
Despite the search by MAGIC, H.E.S.S., and HAWC starting a few hours to several days after the BNS merger, no VHE counterpart was detected for GW170817 \citep{MAGIC:2021rwu,2017ApJ...850L..22A,2020ApJ...894L..16A,Galvan-Gamez:2019xpm}. Also, other gravitational-wave signals have been followed up by VHE instruments without a successful detection so far \citep{2019ICRC...36..743M,Seglar-Arroyo2019ICRC}. This is mainly due to the difficulties of searching over the large sky localization of the gravitational-wave signals, the slow response time (which is a combination of alert time, observatory slew time, and time required to scan the GW sky-localization), and the limited volume of the Universe observed by the GW instrument.
The present generation IACTs are, in principle, capable of following up the alerts from GW events \citep{2019ICRC...36..743M,Seglar-Arroyo2019ICRC}. However, the sky localization of the GW events are around or more than an order of magnitude larger \citep{LRR2020} than their field of view (FoV); the FoV of MAGIC \citep{2016APh....72...76A} and H.E.S.S. \footnote{\url{https://www.mpi-hd.mpg.de/hfm/HESS/pages/about/telescopes/}} are around 3.5$^\circ$ and 5$^\circ$, respectively. The VHE is a beamed emission, and only an observer aligned with the jet is expected to observe it. Within the volume of the Universe currently observed by LIGO, Virgo, and KAGRA, the probability to detect face-on mergers (systems with the orbital plane perpendicular or partially perpendicular to the line of sight) is quite low \citep[see e.g.][]{Colombo2022,Patricelli2022,Perna2022}.\newline
The next generation of VHE observatories will make it possible to access a larger Universe. They represent a significant and valuable advancement in the search for GW counterparts thanks to the better sensitivities, capabilities to monitor large sky-regions, and the rapid response and slew time. The Cherenkov telescope array (CTA\footnote{\url{https://www.cta-observatory.org/}}) will be capable of observing GRB candidates with unprecedented sensitivity \citep{2019scta.book.....C}. The Northern site of the {Cherenkov telescope array (CTA),} CTA-N, will consist of large-size telescopes (LST) and Medium size telescopes (MST). The Southern site will be equipped with MST and small-size telescopes (SST) 
{with the possibility that a couple of} 
LSTs will be added. The LST and MST array will be able to cover FoV up to 50 deg$^2$. In addition, implementation of the divergent pointing \citep{2015ICRC...34..725G, 2019ICRC...36..664D, 2022Galax..10...66M} {\bf which consists in each telescope
pointing to a position in the sky that is slightly offset, in the outward direction, from
the center of the field of view, can lead to an ever larger FoV of at least 100 deg$^2$.}
{The SST is capable of covering a FoV larger than 50 deg$^2$, but it is sensitive to a lower energy range which starts at 1 TeV and begins to perform better only above 5 TeV.} 

Other works investigate the perspectives to detect the VHE afterglow expected from short GRBs and gravitational-wave signals associated with BNS mergers detected by the current GW detectors \citep{Patricelli2021,Patricelli2018,Seglar-Arroyo2019,2019MNRAS.490.3476B,2021hgwa.bookE...5S}. In this paper, we evaluate the perspectives to detect the prompt VHE counterpart proposing optimal observational strategies to detect this emission with the next-generation observatories. In particular, we evaluate the joint detection capabilities of CTA working in synergy with the next generation of gravitational-wave detectors, such as Einstein Telescope \citep[ET-][]{Punturo:2010zz,Maggiore:2019uih} and Cosmic Explorer \citep[CE-][]{Evans:2016mbw,Reitze:2019iox,CEHS}. It has been recently discussed and demonstrated that it is possible to detect BNS during the inspiral phase before the merger and send early warning alerts~\cite[see e.g.][]{2012ApJ...748..136C,2020ApJ...905L..25S, 2021ApJ...910L..21M}. The next-generation GW detectors will greatly improve the sensitivity at lower frequencies, making it possible also to have good sky localizations minutes before the merger \citep{nitz2021bnsloc,Chan2018}. This translates into providing alerts to the VHE observatory, with an estimate of the localization of the source, in time to slew the VHE instrument to the source and enable a unique opportunity to infer the physics of prompt emission of GRBs. To assess the prospects for joint detection by exploiting the use of early warning alerts of GW events detected before the BNS mergers, we develop an end-to-end simulation that, starting with an astrophysically-motivated population of BNS mergers and modeling their GW emission, evaluates the detection and sky-localization capabilities at different pre-merger times for ET working as a single observatory or in a network of observatories including the current and the next generation of GW detectors. We then estimate the number of possible joint GW/VHE detections using CTA.
Since the facilities such as LHAASO, and SGWO \citep{2020MNRAS.497.3142L} are not pointing instruments (a constant fraction of the sky is always visible), 
{pre-merger alerts do not potentially make improvements} for the observation of the VHE counterpart of the GW events.\newline
The paper is organized as follows. Section \S 2 describes the formalism and methodology for estimating the detection rate of GW/VHE. It starts with a description of the method to evaluate the detection rate and sky-localization of pre-merger gravitational wave signals from the population of BNSs observed by ET as a single observatory or by ET included in several GW detector networks. We then describe the capabilities of the CTA array to detect the VHE emission of short GRBs and the observational strategies to detect the prompt and early emission of BNS mergers. 
Section \S 3 describes the results for different observational strategies. Section \S 4 discusses the plausible emission models capable of producing the VHE signal from short GRBs. Finally, in Section \S 5, we present our conclusions.\newline
% ================================================================================
% ==================================  METHOD  ==================================== 
\section{Methodology to estimate the joint GW/VHE detections}
\subsection{Population of binary neutron stars}\label{pop}
\begin{figure}[ht!]
\centering
            \includegraphics[width=9cm, height=9cm]{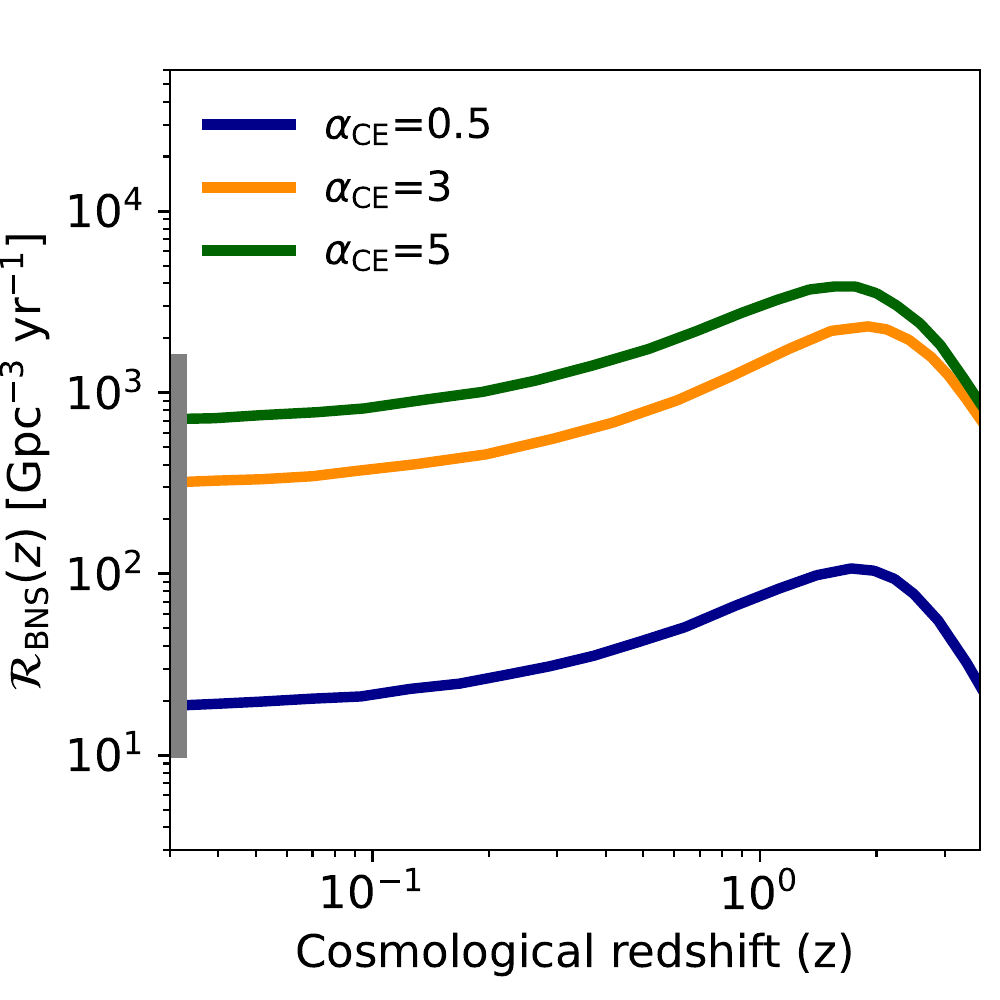}
            \caption{Evolution of the merger rate density in the comoving frame as a function of redshift for the BNS populations used in the present work. Our fiducial population is obtained with a common envelope efficiency $\alpha_{CE}$=3 and is represented by the orange solid line. The pessimistic and optimistic populations are obtained with $\alpha_{CE}$=0.5 and 5 and are shown by blue and green solid lines, respectively. The grey area shows the $90\%$ credible interval of the local merger rate density, as inferred from the first three observing runs of LIGO, Virgo, and KAGRA \citep{GWTC3pop}.}
            \label{fig:POP}
        \end{figure} 
We generate a population of merging BNSs considering systems formed from isolated binary star evolution via a common envelope as described in \cite{Santoliquido2021}. The cosmic merger rate density is built using the semi-analytic code {\sc cosmo$\mathcal{R}$ate} \footnote{\url{https://gitlab.com/Filippo.santoliquido/cosmo_rate_public}} \citep{Santoliquido2020} which combines catalogs of isolated compact binaries obtained using the population-synthesis code MOBSE\footnote{\url{https://gitlab.com/micmap/mobse_open}}
\citep{Mapelli2017,Giacobbo2018,GiacobboMapelli2018} with data-driven models of star formation rate (SFR) density and metallicity evolution. Here, we adopt the SFR and average metallicity evolution of the Universe from \cite{MadauFragos2017}, and a metallicity spread $\sigma_{\rm Z}=0.3$. We describe electron-capture supernovae as in \cite{Giacobbo2019} and assume the delayed supernova model \citep{Fryer2012} to decide whether a core-collapse supernova produces a black hole or a neutron star. When a neutron star forms from either a core-collapse or an electron-capture supernova, we randomly draw its mass according to a uniform distribution between 1 and 2.5 M$_\odot$. This mass distribution is consistent with the gravitational-wave observations showing broad and flat mass distribution for NSs in binaries \citep{GWTC3pop}. Our catalog of synthetic BNS mergers is based on a fiducial scenario that adopts a common envelope ejection efficiency parameter, $\alpha_{CE}$, equal to 3. We model natal kicks as \begin{equation}\label{eq:kick}
v_{\rm kick}=f_{\rm H05}\,{}\frac{m_{\rm ej}}{\langle{}m_{\rm ej}\rangle{}}\,{}\frac{\langle{}m_{\rm NS}\rangle{}}{m_{\rm rem}},
\end{equation} where $f_{\rm H05}$ is a random number extracted from a Maxwellian distribution with one-dimensional root-mean-square $\sigma_{\rm kick} = 265$ km~s$^{-1}$, $m_{\rm rem}$ is the mass of the compact remnant (neutron star or black hole), $m_{\rm ej}$ is the mass of the ejecta, while $\langle {m}_{\mathrm{NS}}\rangle $ is the average 
{neutron star (NS)}
mass, and $\langle {m}_{\mathrm{ej}}\rangle $ is the average mass of the ejecta associated with the formation of a NS of mass $\langle {m}_{\mathrm{NS}}\rangle $ from single stellar evolution. This kick model, introduced by \cite{gm2020}, is able to match both the proper motions of young pulsars in the Milky Way \citep{hobbs2005} and the BNS merger rate density estimated from the LIGO--Virgo collaboration \citep{GWTC3pop}. The local astrophysical merger rate of our fiducial scenario is 365 $\rm Gpc^{-3} yr^{-1}$ which is consistent with the astrophysical rates inferred from studying the population of compact binary mergers detected during the first, second, and third run of observations of LIGO and Virgo and corresponding to different mass distribution models \citep{GWTC3pop}. The union of 90\% credible intervals for the different models in \cite{GWTC3pop} gives a BNS merger rate between 10 and 1700 $\rm Gpc^{-3} yr^{-1}$. As shown in \cite{Santoliquido2021} the common envelope efficiency determines one of the main uncertainties for the number of BNS mergers per year, with larger values of $\alpha_{CE}$ translating into higher merger BNS efficiency. In order to evaluate the impact of the uncertainties of the BNS merger rate normalization on our results, we build other two catalogs of synthetic BNS mergers assuming $\alpha_{CE}$ equal to 0.5 and 5. Throughout the paper, we call the BNS catalog obtained with $\alpha_{CE}$=0.5 and $\alpha_{CE}$=5 the pessimistic and optimistic scenario catalog, respectively. As shown in Fig. \ref{fig:POP} the local merger rates of these populations are still consistent with the range constrained by the LIGO and Virgo observations. 

We consider non-spinning systems, as the NS spin is expected to be small in compact binaries that will merge within a Hubble time, as observed through the electromagnetic channel \citep{Burgay2003}. We generate an isotropic distribution in the sky and a random inclination of the orbital plane with respect to the line of sight. 
\subsection{Gravitational-wave signal detection and parameter estimation}
\label{GWsimu}
The next-generation detectors aim at making the low-frequency band below 10 Hz accessible. ET will be built underground and it is expected to cover frequencies down to 2\,Hz. Extending observations at low frequencies enables to increase the  in-band duration of BNS signals, offering two key advantages for the successful detection of prompt/early electromagnetic counterparts: 1) to accumulate enough signal-to-noise ratio (SNR) before the merger to make pre-merger detection and early warning possible; 2) to significantly improve the sky-localization accuracy by using the imprint on the signal amplitude of the time variation of the detector's response due to the daily rotation of Earth.

In order to evaluate the pre-merger detection and sky-localization capability for ET as a single observatory or included in a network of detectors, we use the Fisher-matrix approach implemented in {\it GWFish} \citep{Dupletsa2023}\footnote{The code is publicly available at this repository \url{https://github.com/janosch314/GWFish}}. The code estimates the uncertainties on the measured source parameters from simulated gravitational wave observations, taking into account the effects of the time-dependent detector response and the Earth’s rotation on long-duration signals, such as the ones from BNS coalescences. 

We build a simulation injecting a GW signal for each BNS merger of the population described in Sect.~\ref{pop} up to  redshift z=1.5. This redshift is conservatively larger than the maximum distance up to which CTA will be able to detect the VHE prompt emission expected for short GRBs. We inject $2.7\times10^{5}$ BNS signals, corresponding to the number of BNS mergers up to redshift z=1.5 that we can observe at Earth in one year with a perfect detector according to our fiducial scenario. For the pessimistic and optimistic scenarios, we inject $2.0\times10^{4}$ and $4.0\times10^{5}$, respectively. The inspiral GW signal for each merging BNS is constructed using a post-Newtonian formalism, in particular the TaylorF2 waveforms \citep{Buo2009}. The SNR is computed by {\it GWFish} during the inspiral, applying a high-frequency cutoff at 4 times the frequency of the innermost stable circular orbit \citep[see][for details]{Dupletsa2023}. A network SNR larger than 8 is used to select each BNS detection.  

\begin{table*}[t]
\centering
\renewcommand{\arraystretch}{1.5}
\begin{tabular}{| l | c | c | c |c |} \hline
Telescope & Components  & Energy band [TeV] 	& FoV [deg$^2$] & t$^{120^\circ}_{\rm slew}$ [s] \\ \hline
MAGIC	& 2	(North)                       & 0.03-$\sim$10 	    & $\sim$7 	         & $\sim$20  \\ \hline
CTA-LST  & 4  (North) +  2$^*$ (South) 	   &  0.02- $\sim$5  & 13$^{**}$	         & 20  \\  \hline
CTA-MST  & 9  (North) + 14 (South)      &  0.15-5   	        & 44$^{**}$	         &  90  \\  \hline
CTA-SST  & 37 (South)		           &  5-300 	 	        & $>$50          &  60 \\ \hline
\end{tabular}
\caption{Detector specification of CTA (alpha-configuration) as compared to the current generation IACT, i.e., MAGIC. The first column indicates the telescope name, and the second, third, fourth, and fifth columns correspond 
to the expected number of telescopes, the covered energy band, the field of view, and the slewing time, t$_\text{slew}$, to re-point the telescopes. 
$^*$Initially not included in the CTA-array and recently funded d by PNRR program by the Italian government.
$^{**}$ As described in the text, a FoV of 10 deg$^2$ and 30 deg$^2$ are used for the LST and MST in the present analysis.}
\label{table:CTA}
\end{table*}

We consider five GW detector configurations: ET as a single triangular-shape 10 km arm-length detector located in Sardinia (one of the possible European site candidates to host ET), ET plus a network of five second-generation detectors, ET plus two Voyager detectors (located in the current USA LIGO sites), ET plus CE (L-shaped 40 km arm length detector located in the USA), and ET plus two CE (one located in the USA and one in Australia). For the second-generation network, we consider Advanced LIGO, Virgo, and KAGRA with the optimal sensitivity (phase plus) expected for the fifth run of observations as in \cite{LRR2020}, and the same version of the LIGO detector in India. For ET we use the ET-D sensitivity curve \citep{Hild2011etd}. For Voyager and CE(40 km), the sensitivity given in \cite{Adhikari2020} and \cite{Evans2021}, respectively. For each GW detector as well as for each of the three combinations of high and low-frequency interferometers of ET, {we assume a duty cycle of 85\% \citep{2023arXiv230315923B} }. 

We evaluate the sky-localization and other parameters of the detected sources 15, 5, and 1 minutes before the merger and at the merger time for the different detector configurations. These pre-merger times are appropriate both to select events with a suitable pre-merger sky-localization to be observed by the CTA, and to have adequate time for the CTA to respond to the trigger, to point and observe the sky-localization to detect the prompt/early VHE emission (see Sect.~\ref{sec:CTAsection}). We then focus on face-on events (the orbital plane perpendicular, assumed to be aligned with the jet, within 10 degrees with respect to the line of sight).  These are the events for which we  expect to detect the VHE counterpart.   

{Our analysis is based on the assumption of a Gaussian noise background in the GW detectors. 
In a more realistic scenario, other backgrounds need to be accounted for; they are: 1) a stochastic GW background of unresolved compact binaries that might affect ET analyses below 20Hz, 2) an overlap between individual resolvable signals and its potential impact on signal analyses, 3) instrumental noise transients (glitches). The nonstationary stochastic GW background might reduce the signal-to-noise ratio of GW observations at times (in average, it is weaker than instrument noise). However, the triangular configuration of ET makes it possible to assess the impact of the GW background and mitigate it \citep{GoncharovNitzHarms}.  A recent study suggests that the overlap between individual resolvable signals will not have an important effect on signal analyses \citep{Samajdar}. Instrument glitches can in principle affect signal analyses, but effective glitch mitigation methods are under development, and we can assume that optimized signal plus noise Bayesian analyses will be available when ET starts operation \citep[e.g.,][]{PhysRevD.103.044013}.}

\subsection{CTA array specification and observational strategies}
\label{sec:CTAsection}
{CTA is expected to increase our capabilities to perform a follow-up and detect transient events \citep{2019scta.book.....C} thanks to the unprecedented sensitivity, field of view, and rapid slew to any given direction.} The complete CTA will consist of Large-Sized Telescopes \citep[LST;][]{2021arXiv210903515L}, Medium-Sized Telescopes \citep[MST;][]{2017AIPC.1792h0002P} and Small-Sized Telescopes \citep[SST;][]{2015arXiv150806472M}. During the first construction phase, the approved configuration is called alpha-configuration \footnote{https://www.cta-observatory.org/science/ctao-performance/}. This configuration will consist of 14 Medium-Sized Telescopes \citep[MSTs][]{2017AIPC.1792h0002P} and 37 Small-Sized Telescopes \citep[SSTs;][]{2015arXiv150806472M} in the southern site at the Paranal Observatory (Chile). The Northern site at the Roque de los Muchachos Observatory (Spain) is expected to host Large-Sized Telescopes \citep[LSTs;][]{2021arXiv210903515L} and 9 MSTs. {The specification of the different size telescopes within the alpha-configuration are given in Table \ref{table:CTA}. LSTs, MSTs, and SSTs are designed for covering different science cases. The array of SSTs has the largest sky coverage ($>$50 deg$^{2}$), whereas the array of 4 LSTs has the smallest sky coverage ($\sim$13 deg$^{2}$). While SST effectively covers events with energies from 5 to 300 TeV, the LST and MST target lower energy events from 20 and 150 GeV, respectively. Partial CTA operation has recently started with one LST in the Northern site \citep{2021arXiv210903515L}, covering the energy band of 10~GeV to 10~TeV.}

{Building the optimal CTA follow-up of GW signals from BNS mergers requires taking into account duration, luminosity, energy band of the expected VHE counterpart combined with slewing time (t$_{\rm slew}$), the field of view (FoV) and sensitivity of CTA.
Since the expected energy band for the VHE counterparts of GW events is sub-TeV, we consider the use of LSTs and MSTs in the present work, {excluding} SSTs from the analysis. Although the SST array is expected to have larger sky coverage, it does not cover the energy band below 1 TeV. 
CTA is expected to operate in a hybrid mode with LST and MST (individual sub-arrays) observing together or separately \citep{2011ExA....32..193A, 2019scta.book.....C}. In this work, we consider separately the individual components of CTA (LSTs and MSTs) in order to increase the effectiveness of operation taking into account the different slewing times. 
We assume a FoV of 10 deg$^2$ and slewing time (t$_{\rm slew}$) of 20\,s for the LST sub-array. 
Although the Southern LST is not yet guaranteed, we consider LSTs located both in the Northern and Southern Hemispheres. The MST sub-array (one located in the Northern and one in the Southern Hemispheres) is assumed to have a FoV of 30 deg$^2$ and slewing time (t$_{\rm slew}$) of 90\,s. 
Our assumption of a smaller FoV, compared to the design one, for both LST and MST sub-arrays accounts for the reduction in the angular resolution and energy reconstruction capability for the off-axis events \citep{2019scta.book.....C}. We consider a duty cycle of 15\%. We also assume a 50\% reduction on the sky visibility taking into account that the sub-arrays are hardly capable of observing the sky above the zenith angle of 60$^\circ$.}

\begin{figure*}[h]
\centering
            \includegraphics[width=0.80 \linewidth, height=8cm]{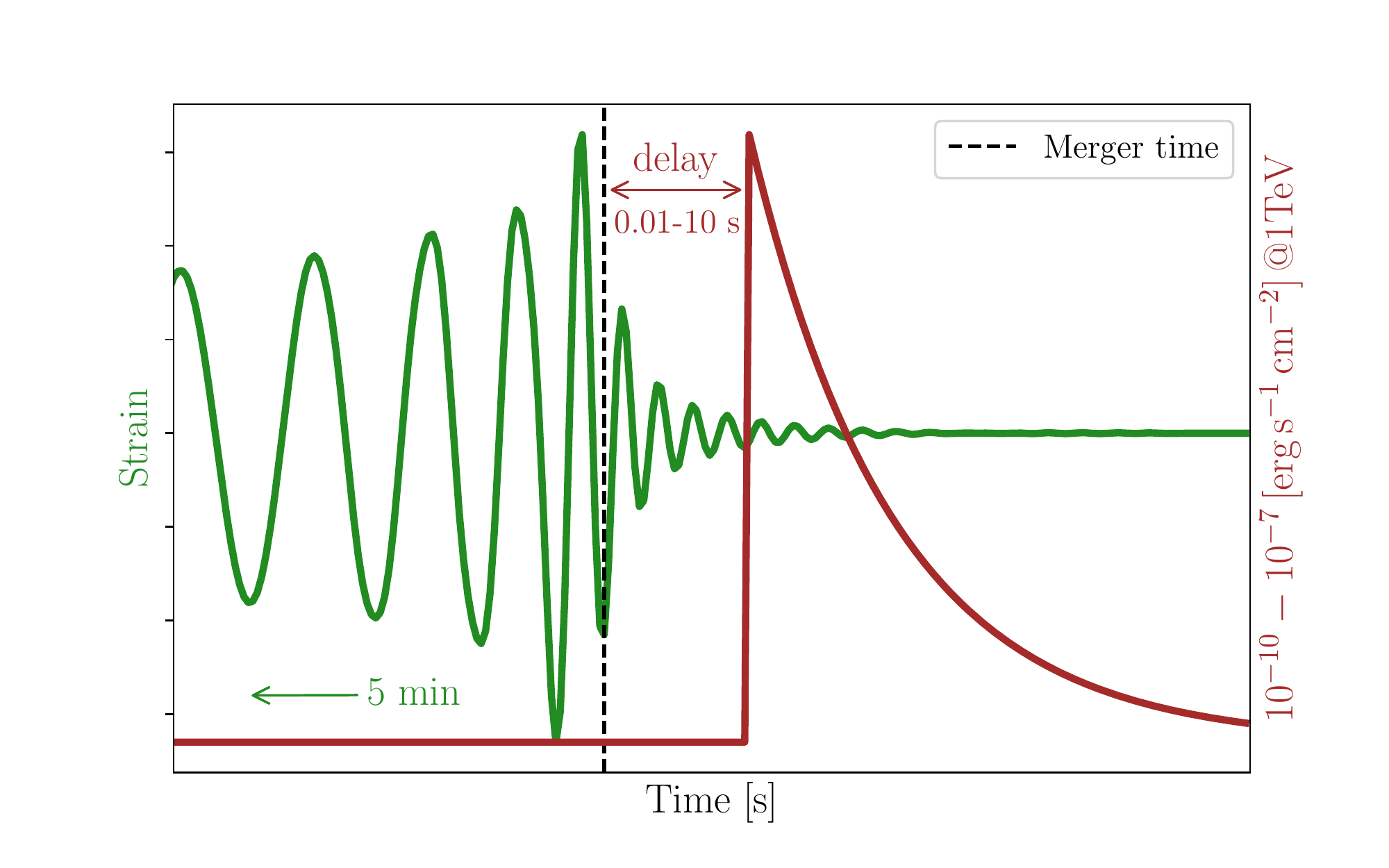}
            \caption{General observation scheme to detect the VHE prompt emission phase of the BNS merger. The low-frequency observations made possible by the next generation of GW detectors will enable to detect an inspiraling BNS system and localizing it before the merger. Pre-merger alert for the event is sent and the VHE detectors can rapidly point to the target during the merger. {The delay between the GW and VHE emission is assumed to be within our exposure time of 20 seconds}.
            }
            \label{fig:Sketch}
        \end{figure*}

We then explore three observational strategies {to follow up events triggered by the GW network}:
\begin{itemize}
    \item {\it direct pointing} of events with sky-localization smaller than the FoV to detect the prompt emission; 
    \item {\it one-shot observation strategy}, which consists of following up triggers using a single observation randomly located in the sky-localization uncertainty of the GW signal to detect the prompt emission (here, we also consider the possibility to use divergent pointing, see Sect.~\ref{divpoint};  
    \item {\it mosaic strategy}, which tiles the sky-localization being more effective to detect the afterglow emission.  
\end{itemize}

For each event, we consider the total time to be spent consisting of a) time to respond to the trigger (t$_{\rm alert}$), b) slewing time (t$_{\rm slew}$), and c) exposure time (t$_{\rm exp}$). 
{In order to detect the prompt emission, we consider a single exposure of 20\,s starting around the merger time. This exposure is larger than the delay of a few seconds expected between the GW and the prompt gamma-ray emission of short GRBs \citep{2019FrPhy..1464402Z}, but enables  to detect possible VHE emission with a larger delay without preventing to detect a signal with smaller latency. Indeed, during the post-processing of the observed data, the signal can be extracted by analyzing shorter exposure (for example 2\,s).}
This exposure enables us to sample isotropic energy of $10^{50}$ erg in the 0.2-1\,TeV up to redshift of 1 (see Section~\ref{sect:VHEmethod} and Fig.~\ref{fig:FigLumDl}) and to follow-up several GW triggers.  In order to reach the source location before the merger takes place and capture the prompt emission, we consider pre-merger alerts. The response time and slewing time can be in principle reduced to 1 minute (t$_{\rm alert}$+t$_{\rm slew}$) for the LST sub-array, thanks to its rapid slewing time of 20\,s. Thus, in a very optimistic scenario, the LST sub-array can follow up  (even) 1-minute pre-merger alerts and reach the source location at the merger time. Instead, due to a longer slewing time for MST of about 90\,s, longer pre-merger time is required for detecting the prompt emission. A minimum pre-merger alert time of 5 minutes is considered for MST in our study. {More details on the observational strategies and time required to follow up GW events are given in Sect.~\ref{CTAdetect}.} The sketch of the proposed observational scheme is shown in Fig.~\ref{fig:Sketch}.

\begin{figure*}
\centering
            \includegraphics[width=.6\linewidth]{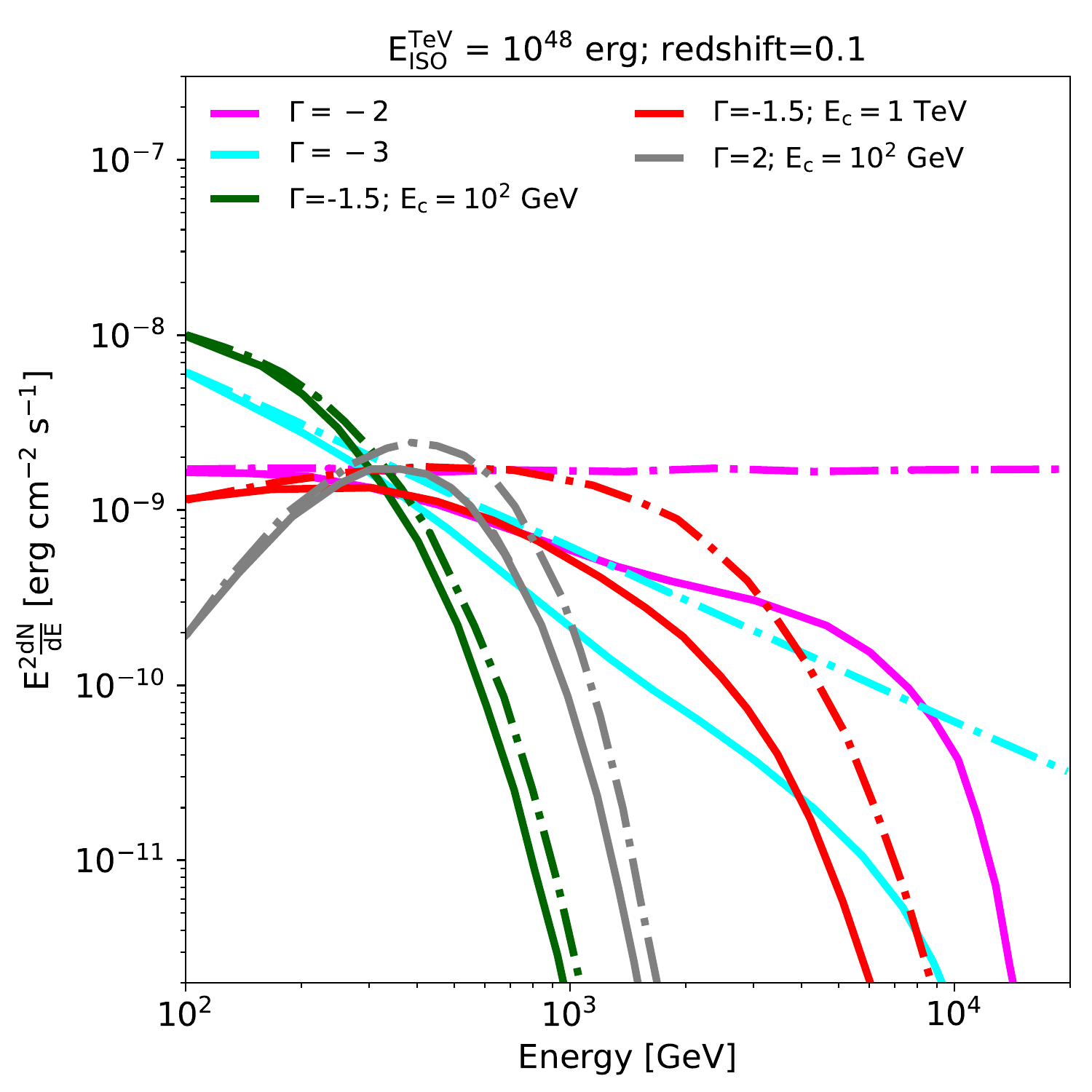}
            \caption{The intrinsic (assumed) and observed spectra  for VHE transient events with specific spectral indices ($\Gamma$) and cutoff (E$_c$) for the isotropic energy  E$^{\rm TeV}_{\rm ISO}$=10$^{48}$ erg in the 0.2-1 TeV band and redshift of 0.1. The observed spectra are corrected for the EBL attenuation following the prescription of \cite{2011MNRAS.410.2556D}. The dot-dashed and solid lines represent the intrinsic and the EBL-attenuated observed spectra, respectively. {The shape of the observed spectra with respect to the intrinsic ones differ more for harder (extending up  to higher energy) spectra than the softer ones due to the larger EBL absorption.}}
            \label{fig:FigSEDintobs}
        \end{figure*}       

\subsection{Detection of GRBs in VHE band by IACTs}\label{sect:VHEmethod}

\begin{figure}[ht!]
\centering
            \includegraphics[width=9cm, height=9cm]{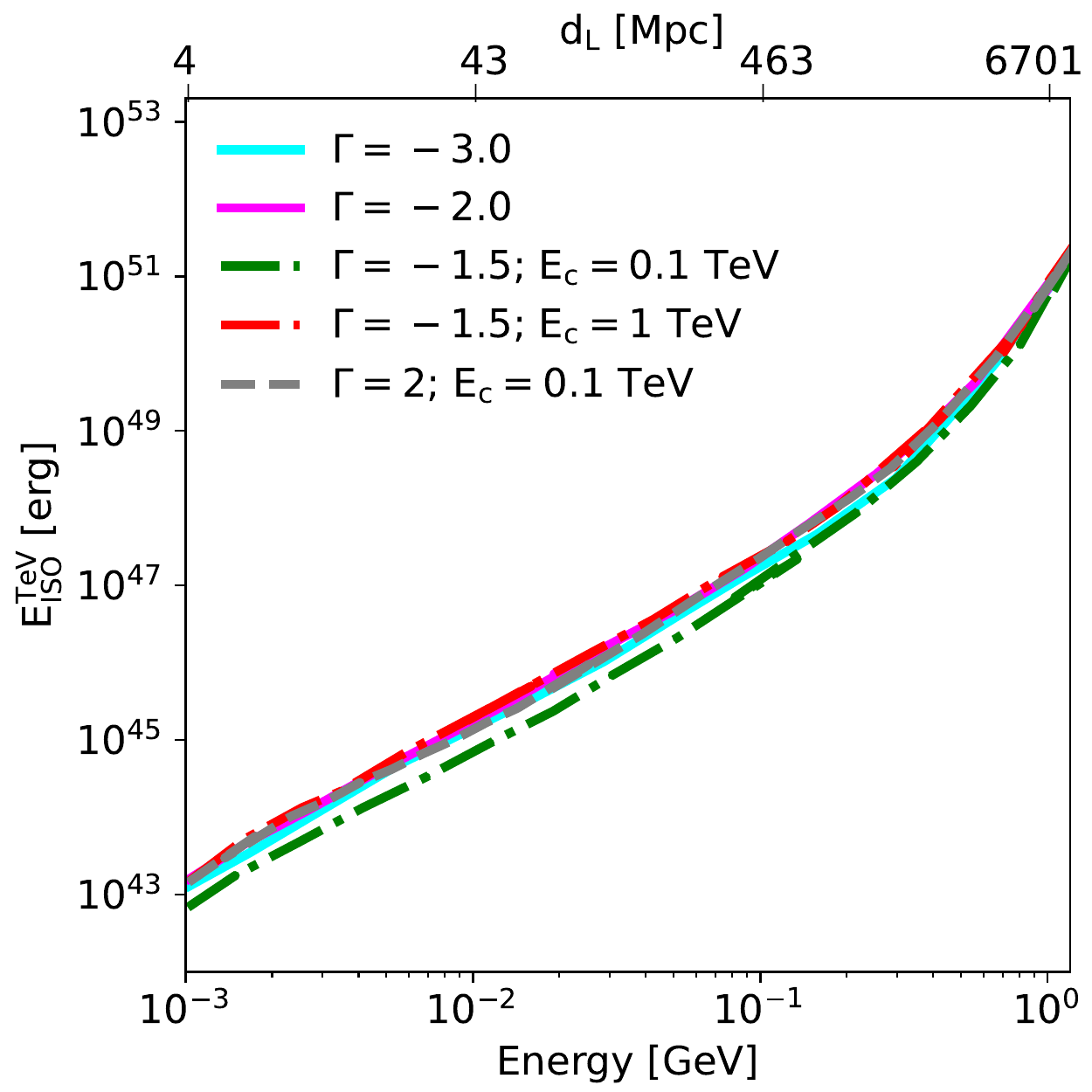}
            \caption{The lower limit on isotropic energy (E$_{\rm ISO}$) in the range 0.2-1\,TeV as a function of redshift for CTA-N (alpha-configuration including 4 LSTs and 9 MSTs). We consider a short constant VHE emission of 10 seconds, and we use a 
            {threshold}
            on the significance ($\sigma$) of 5$\sigma$ and on excess gamma-ray events (N$_{ex}$) larger than 10 to define a detection. A conservative energy threshold of E$>$200\,GeV (hence a slightly higher limit on  E$^{\rm TeV}_{\rm ISO}$) is preferred as the observation condition such as weather-condition, presence of the moon, zenith angle dependence impact the energy threshold. Observed spectra depending on the E$^{\rm TeV}_{\rm ISO}$, $\Gamma$, E$_c$ and redshift are shown in Fig. \ref{fig:FigSEDintobs}.}
            \label{fig:FigLumDl}
        \end{figure}

To evaluate the isotropic energy from short GRBs sampled by CTA as a function of the redshift, we assume an intrinsic spectrum of the VHE emission of the form 

\begin{center}
\begin{equation}
\rm     \dfrac{dN}{dE} \propto \left( \dfrac{E}{E_0} \right)^{\Gamma} \times exp(-E_c),
     \end{equation}
\end{center}
 where $\Gamma$ is the spectral index, {E$_{\rm 0}$ is the energy scale}, and  E$_{\rm c}$ represents the intrinsic cut-off energy.

The observed spectrum is evaluated as follows
\begin{center}
\begin{equation}
\rm     \dfrac{dN}{dE}=N_0 \left( \dfrac{E}{E_0} \right)^{\Gamma} \times exp(-E_c),
     \end{equation}
\end{center}
where the normalization N$_0$ is given by:

\begin{equation}
\centering
\rm    {\rm N_{0}(E)}=  \dfrac{E_{\rm ISO}\times (1+z)}{ 4 \pi d^2_L {\int^{E_2/(1+z)} _{E_1/(1+z)}} ~dE ~E ~exp[-\tau (E,z)]  \left(\dfrac{E}{E_0}\right)^{\Gamma}  exp(-E_c) },
\end{equation}
 where $\rm exp[-\tau (E,z)]$ is the extragalactic background light (EBL) correction factor \citep{2011MNRAS.410.2556D}, $E_{\rm ISO}$ is the isotropic energy in the VHE gamma-ray band (0.2 - 1 TeV) for a duration of the burst of 10 seconds, and z the redshift of the source.

We simulate a number of GRB spectra varying the isotropic energy (E$_{\rm ISO}$) in the range [10$^{42}$ - 10$^{53}$ erg] and redshift in the range [0.001 - 1.3] for three cut-off energy E$_{\rm c}$ 100\,GeV, 1\,TeV and 10\,TeV. The cut-off is assumed for the indices $-$1.5 and 2.0. We also consider two cases, $\Gamma$= 2.0 \cite[similar to the VHE afterglow detected by MAGIC for GRB 190114C ][]{2019Natur.575..455M} and 3.0 without assuming any intrinsic cut-off. Fig.~\ref{fig:FigSEDintobs} shows the intrinsic and observed spectrum for several spectral indices and cut-off energies for a source at z=0.1. 

Using the observed spectrum, we estimate the number of excess events N$_{\rm ex}$ and the significance of detection ($\sigma$) from the MAGIC performance paper \cite{2016APh....72...76A}. The significance of detection ($\sigma$) is obtained using the Li \& Ma method \citep{1983ApJ...272..317L} for all the grid points [E$_{\rm ISO}$, z]. We consider the simulated GRB as detected when  $\sigma>$5 and N$_{ex}>$10.
In order to obtain the detection limit of CTA, we scale the number of excess and background events using the Crab-signal rate observed by MAGIC by the ratio of collection area for specific energy bins. The collection area for MAGIC and CTA are obtained from \cite{2016APh....72...76A} and CTA-webpage\footnote{\url{https://www.cta-observatory.org/science/ctao-performance/}}, respectively. The background events for CTA as a function of energy are obtained also from CTA-webpage$^{\textcolor{red}{4}}$, and later is converted into rate [events/ min] by multiplying by the point spread function (PSF) of CTA as a function of energy. Fig.~\ref{fig:FigLumDl} shows 
the detection limit on the isotropic energy (E$_{\rm ISO}^{\rm TeV}$) for a VHE emission of 10\,s in the 0.1 - 1 TeV as a function of redshift considering a  detection threshold of $\sigma>$5 and N$_{ex}>$10  for several spectral indices and cut-off energies. The detection limit is obtained for the sensitivity of CTA-N (alpha-configuration including 4 LSTs operating with 9 MSTs). 

The operation of LST as an independent array might increase the detection limit by a factor of 2-3 \citep{2013APh....43..171B} in the energy band of 0.2-1\,TeV. MST is more sensitive than LST in the 0.2 - 1\,TeV band and the detection limit does not change significantly with respect to the entire CTA array. As a comparison, we highlight that the afterglow emissions detected by MAGIC for GRB 190114C and by H.E.S.S. for GRB 180720B correspond to isotropic energies of $2\times 10^{52}$ erg and $2\times 10^{54}$ erg, respectively \citep{2019Natur.575..455M,2019Natur.575..464A}, which are largely above our detection limit.

{The detection of gamma-rays of energy above $\sim$20 GeV is based on the indirect technique of detecting atmospheric Cherenkov light produced by the VHE photons coming from the astrophysical TeV emitters (point sources or extended sources). The extensive air showers (EAS) produced by the hadrons act as a background and might mimic a transient signal. However, there are solid analyses to reject this background which are already implemented in the data analysis technique of current generation detectors (such as MAGIC, H.E.S.S). In the present analysis, we take into account the background and rely on the random-forest method used by the MAGIC telescope system \citep{2016APh....72...76A}. On the basis of our current VHE observations, the VHE gamma-ray sky is not polluted with the presence of several sources given that they are not in the vicinity of any known extended sources. The astrophysical contaminants, which can potentially mimic the VHE counterpart of the GW signals are expected to be removed easily.}

\section{Results}
\subsection{Pre-merger detections and sky-localization}

The results of the simulations evaluating the detection rate and pre-merger sky-localization are summarized in Table~\ref{table:skylocall} \& \ref{table:skyloc}, where we show ET as a single observatory, and ET included in different networks: ET plus the second-generation detectors LIGO-Livingston, LIGO-Hanford, LIGO-India, Virgo, and KAGRA (ET+LVKI+), ET plus two Voyager in the USA (ET+2VOY), ET plus CE in the USA (ET+CE), ET plus two CE, one in USA and one in Australia (ET+2CE). The number of detections per year for a specific threshold of sky-localization ($\Omega$ ($90\%$ c.l.) = 0.1, 1, 10, 30, 100, and 1000 deg$^2$) is given for three different pre-merger times (15, 5, and 1 minute(s) before the merger) and at the merger time. The quoted numbers refer to the fiducial population ($\alpha_{\rm CE}$=3). The pessimistic ($\alpha_{\rm CE}$=0.5) and optimistic ($\alpha_{\rm CE}$=5) scenarios are given in square brackets. Table \ref{table:skyloc} shows the detections per year of simulated BNS mergers with a viewing angle {($\theta_{\rm v}$; the angle between the line of sight and the perpendicular to the orbital plane of the BNS system)} smaller than 10$^{\circ}$. Since the VHE emission is expected to be beamed along the jet, these events are the ones for which we expect a VHE counterpart to be detectable.

\begin{figure*}
\centering
            \includegraphics[width=0.24 \linewidth, height=5cm]{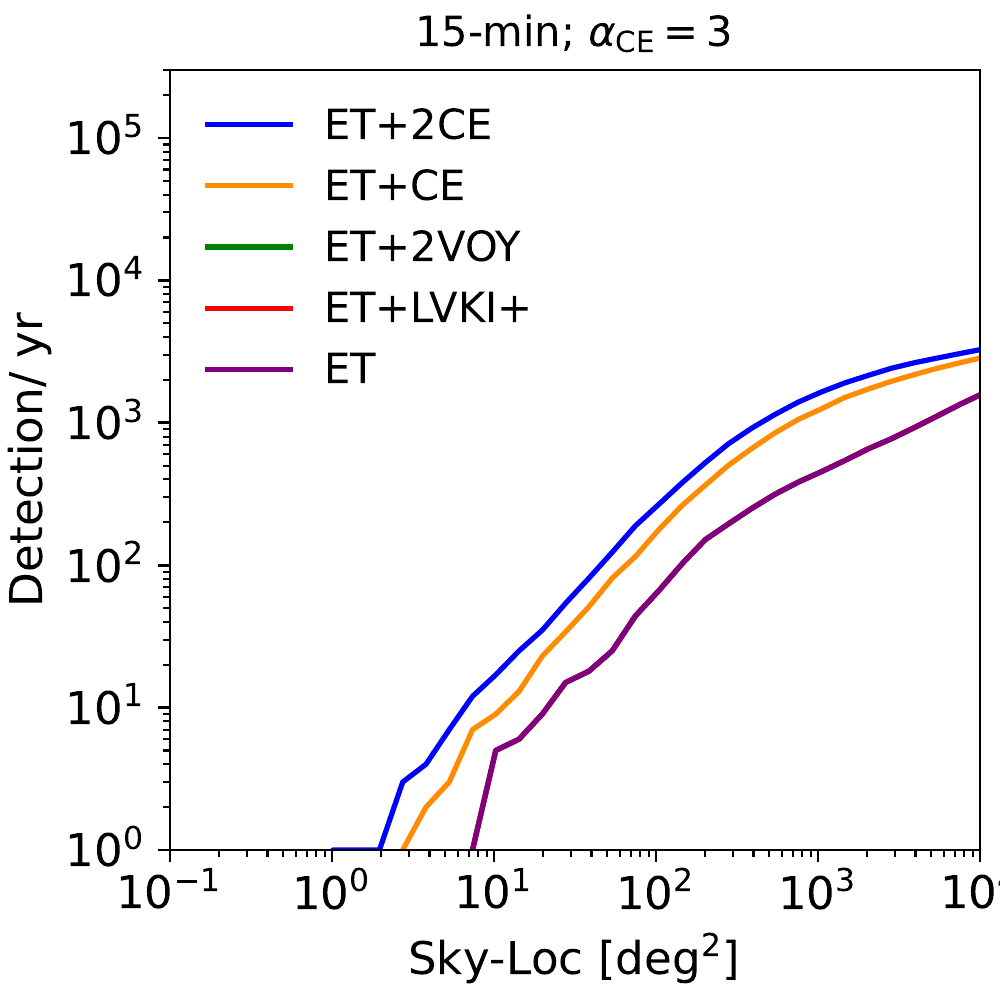}
            \includegraphics[width=0.24 \linewidth, height=5cm]{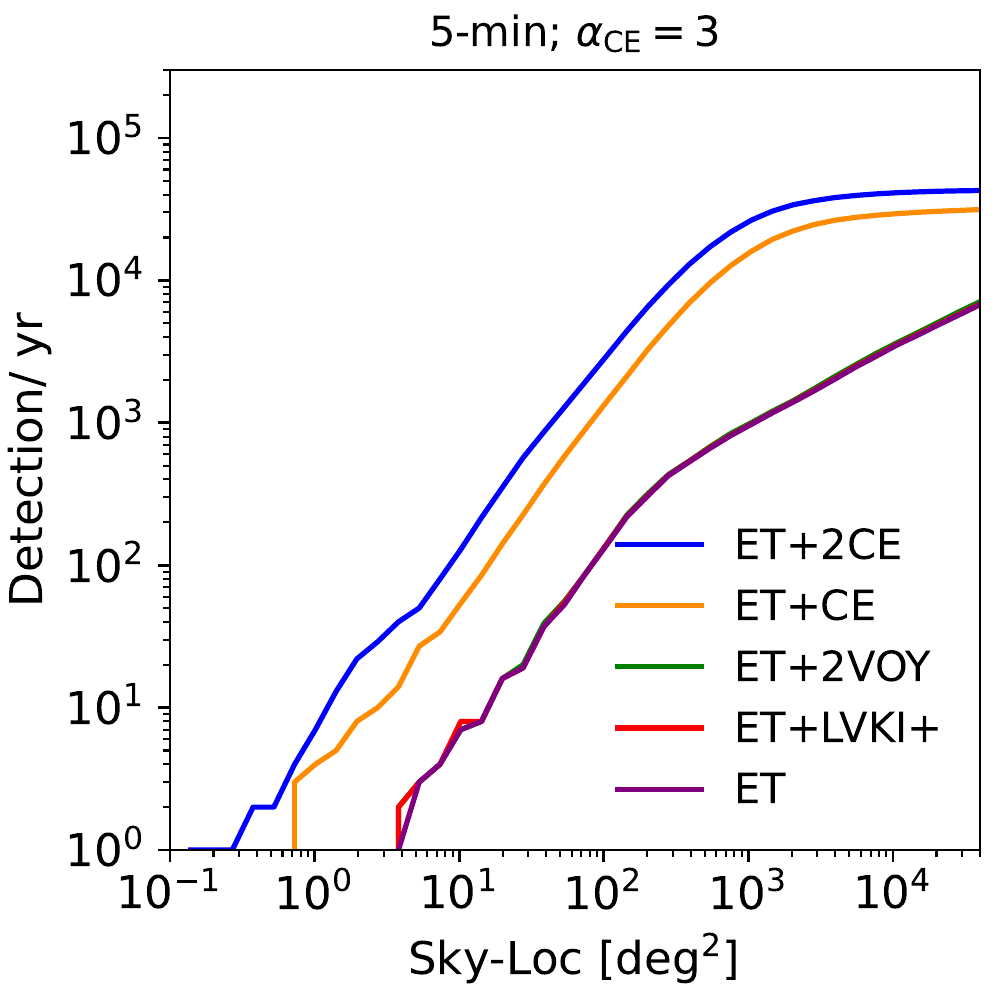}
            \includegraphics[width=0.24 \linewidth, height=5cm]{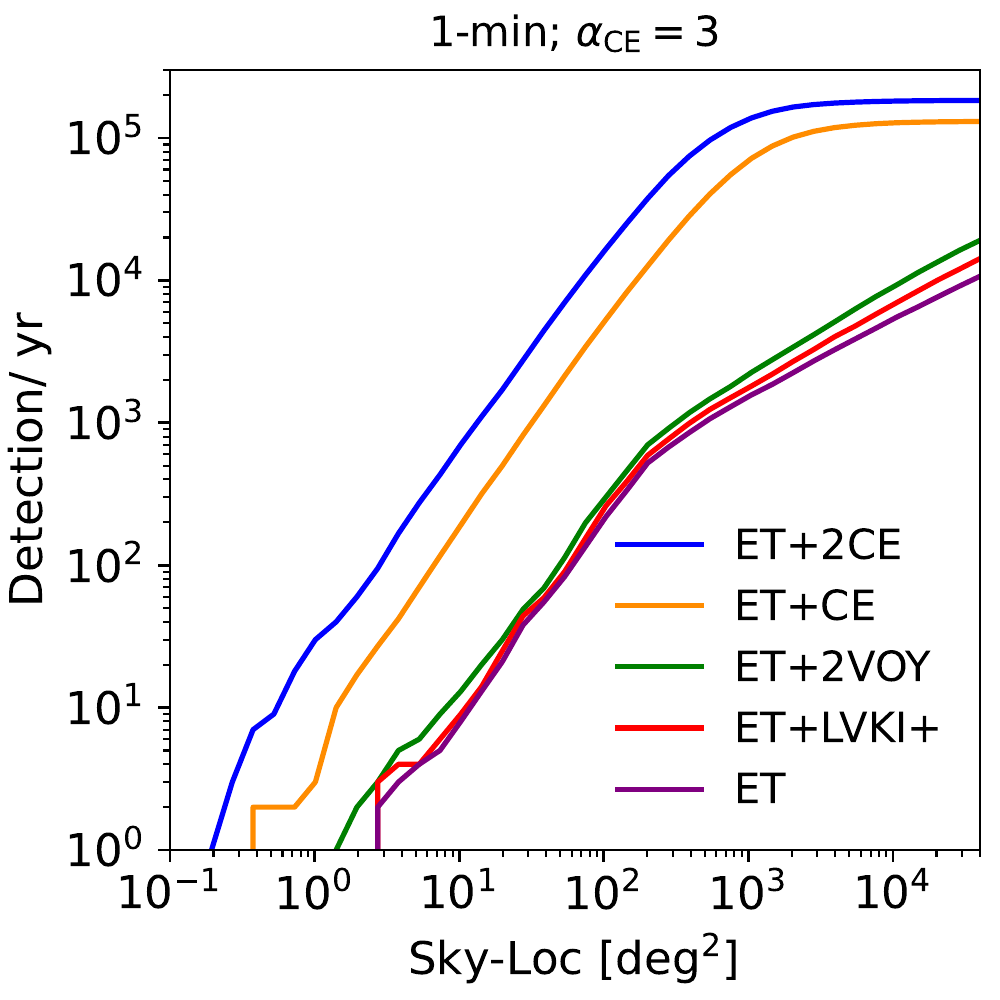}
            \includegraphics[width=0.24 \linewidth, height=5cm]{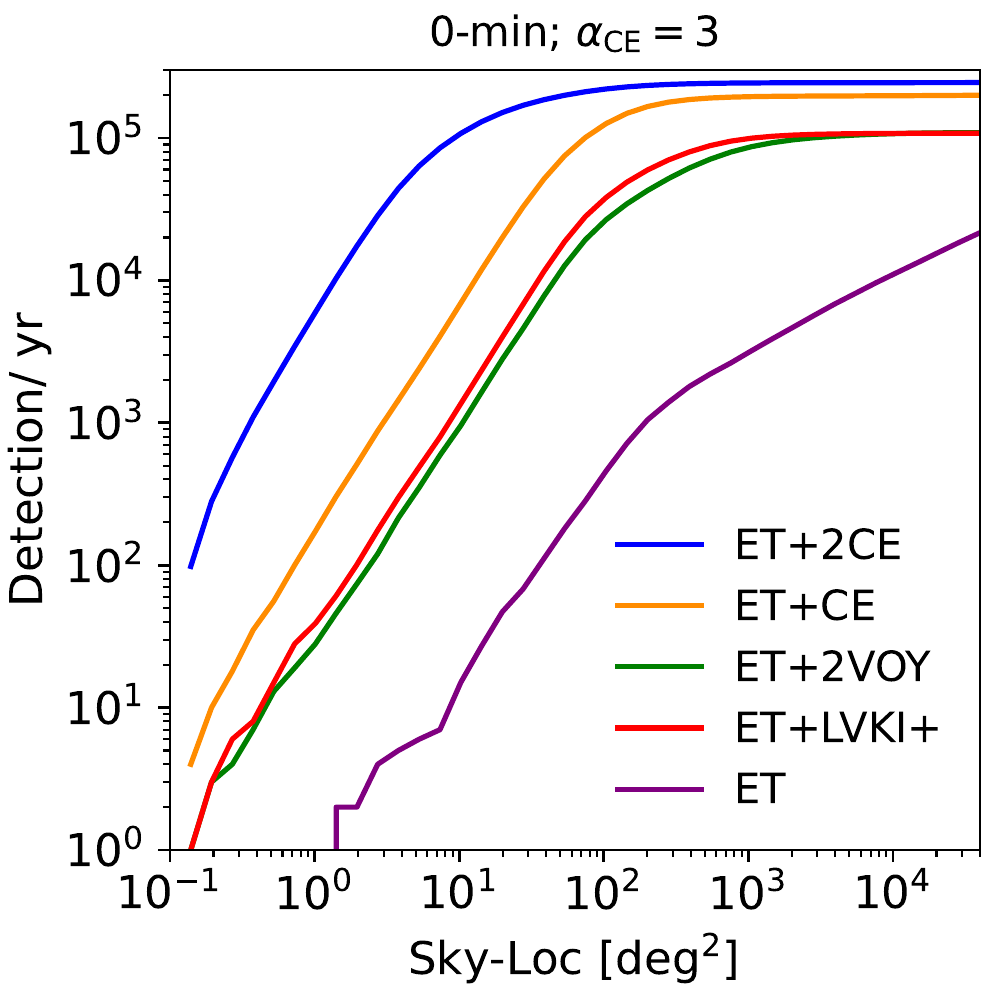}
            
            \includegraphics[width=0.24 \linewidth, height=5cm]{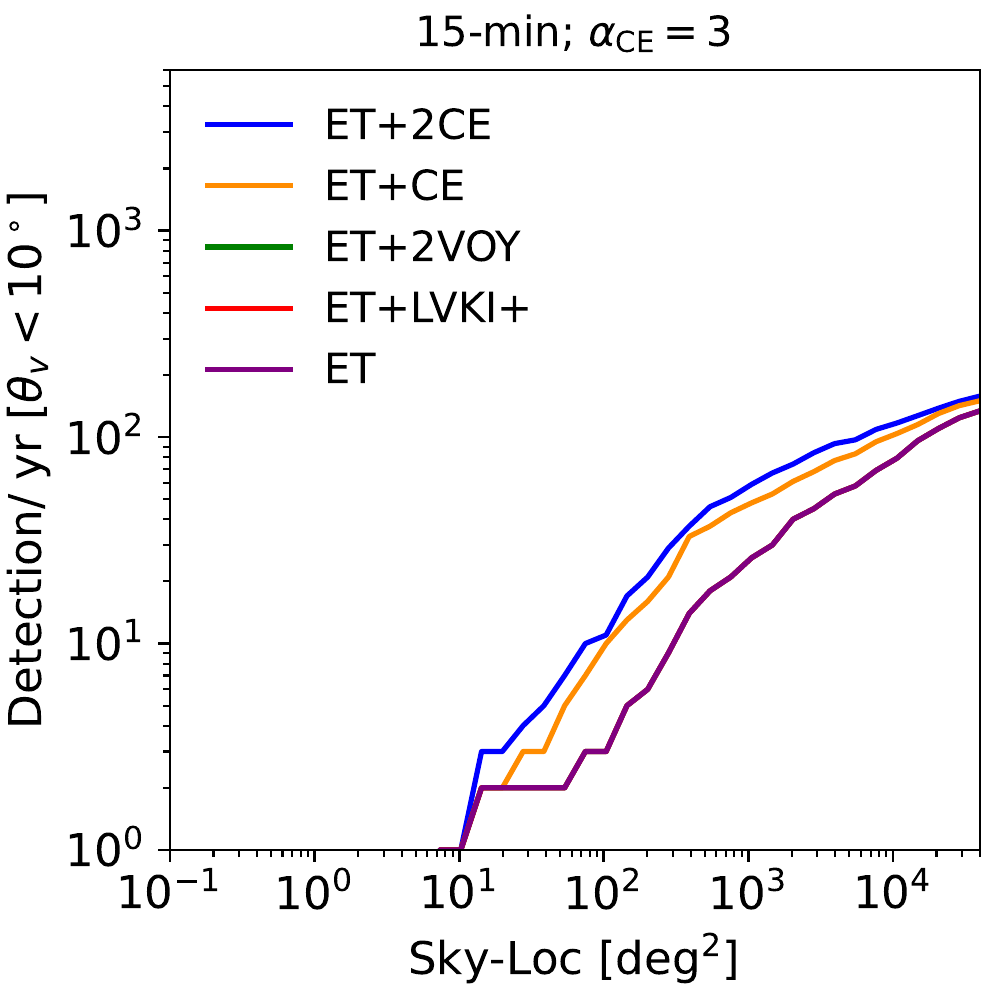}
            \includegraphics[width=0.24 \linewidth, height=5cm]{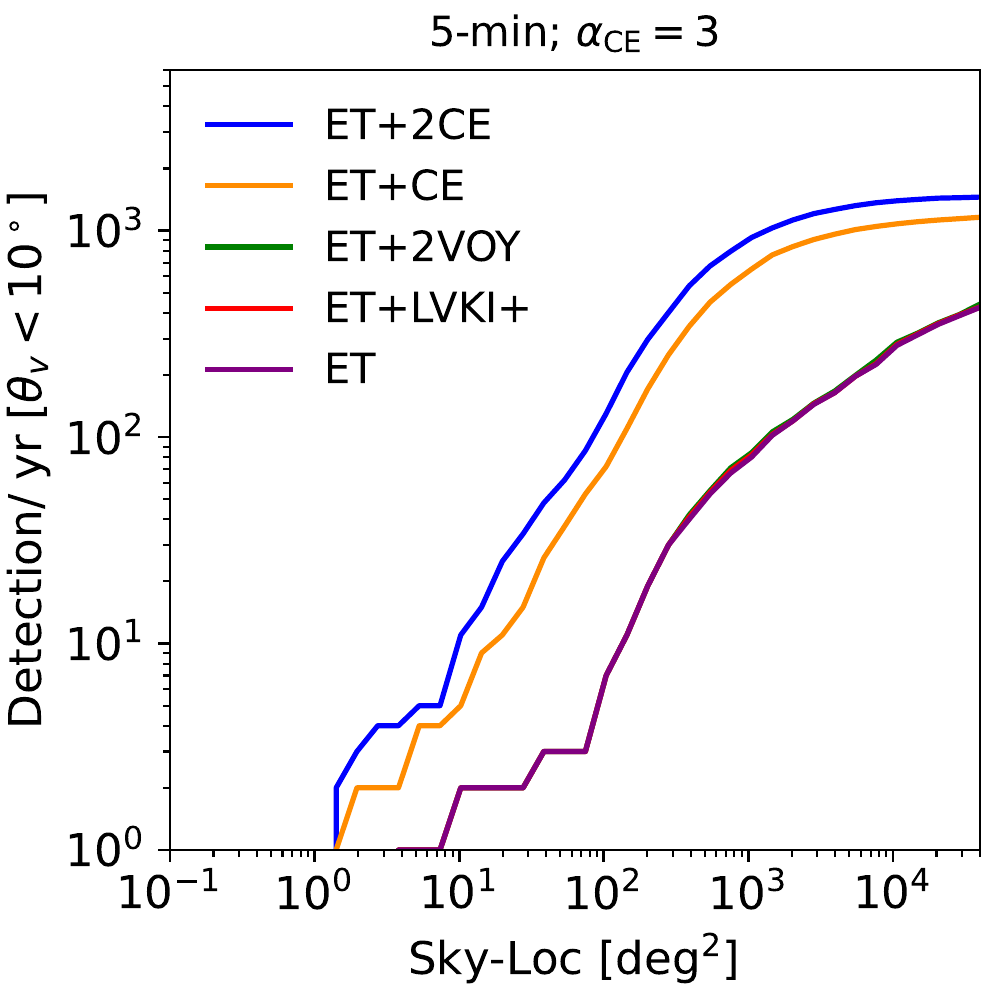}
            \includegraphics[width=0.24 \linewidth, height=5cm]{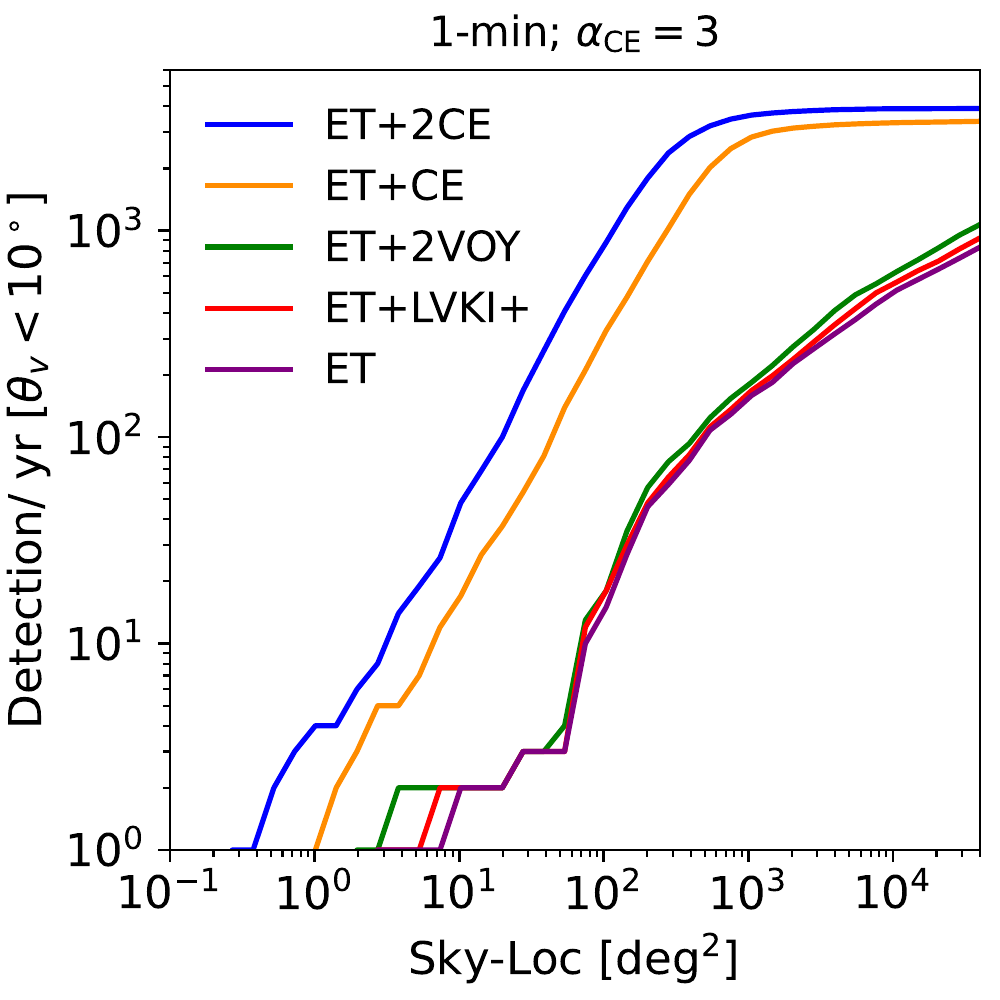}
            \includegraphics[width=0.24 \linewidth, height=5cm]{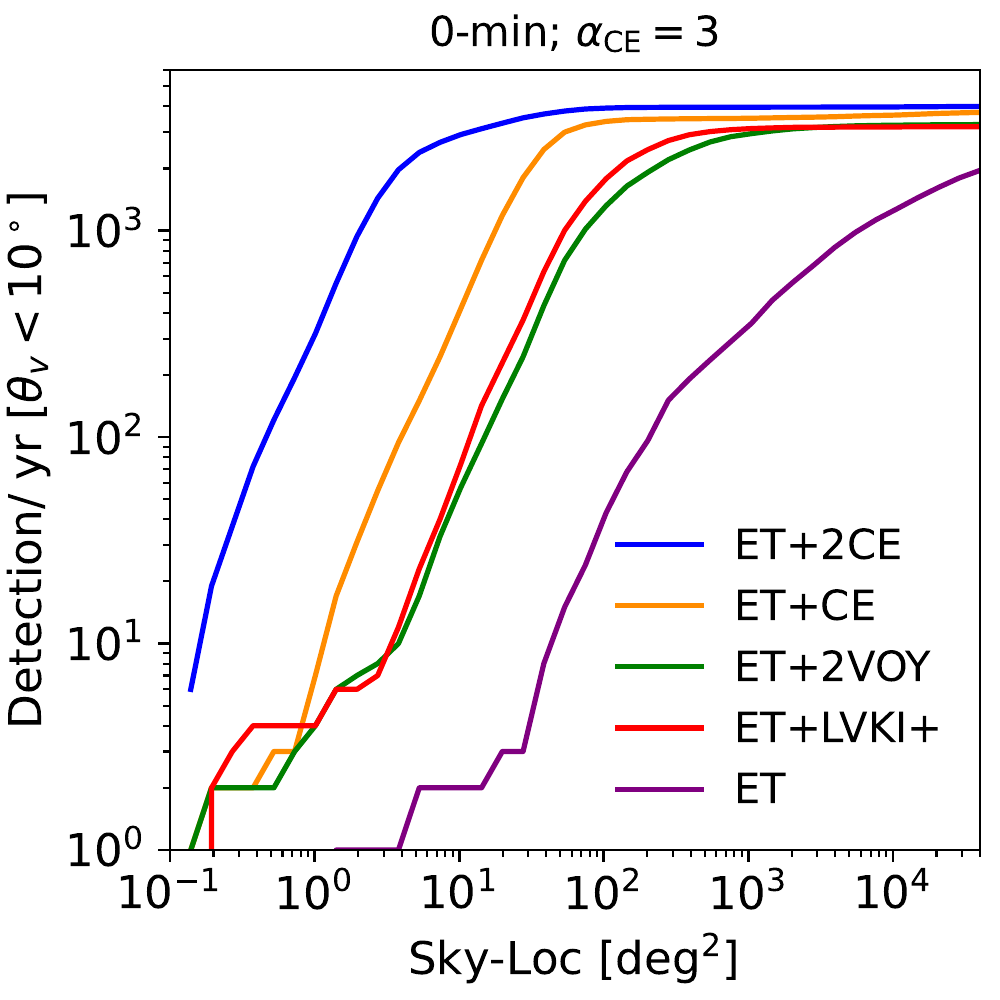}
            
            \caption{The cumulative number of detections (SNR$>$8) per year for different networks of GW detectors considering 15, 5, and 1 minutes before the merger and at the merger time. The top panels show the detections considering BNS systems with all orientations. The bottom panels show the detections of BNS systems with a  viewing angle smaller than 10$^{\circ}$ (on-axis events), a fraction of which are expected to produce detectable VHE emissions. For each of the simulations, the injected BNSs are within redshift z=1.5. The quoted detection numbers refer to the fiducial population and are obtained assuming a duty cycle of 0.85 as described in the text. For the 15 and 5 minutes pre-merger scenarios, the ET+LVKI+ and ET+2VOY do not show any significant difference with respect to ET as a single observatory (the red and green lines lie under the purple line). }
            \label{fig:BNSrateSN3}
        \end{figure*}

\begin{table*}
\centering
\renewcommand{\arraystretch}{1.5}
\begin{tabular}{| c | l | r | r | r | r |}
\hline
\multirow{2}{*}{Detector} & {\hspace{0.3cm}$\Omega$}      & \multicolumn{4}{c|}{All orientation BNSs} \\ \cline{3-6}
                          & [deg$^{2}$]    & 15 min                  &    5 min                   & 1 min           &     0 min             \\ \hline    
\multirow{3}{*}{ET}       &	 10 &	 4 [0, 4]               &	 5 [0, 9]                & 8 [0, 11]       &	 14 [1, 27]\\ \cline{2-6}
                         &	 30 &	 16 [0, 22]             &    25 [2, 40]              & 42 [3, 72]      &     81 [6, 157]\\ \cline{2-6}
			             &	 100        &	 63 [4, 117]               &	 130 [8, 255]             & 208 [16, 435]    &	 436 [33, 919]\\ \cline{2-6}
			             &	 1000       &	 445 [26, 1024]             &	 948 [61, 2225]           & 1511 [89, 3429]  &	 3130 [194, 7021]\\\hline
\multirow{4}{*}{ET+LVIK+}&	1   &	  n.d.                 &	 n.d.                 & n.d.         &	 38 [3, 91]\\\cline{2-6}
			             &	 10         &	 4 [0, 4]                 &	 6 [0, 9]                 & 9 [0, 13]        &	 1296 [72, 3094]\\\cline{2-6}
			             &	 30         &	 16 [0, 22]    & 25 [2, 40] & 47 [3, 89] & 7790 [418, 17106] \\ \cline{2-6} 
			             &	 100&	 63 [4, 117]&	 131 [8, 256]&	 244 [16, 503]&	 37046 [2034, 78383]\\\cline{2-6}
			             &	 1000& 445 [26, 1024]&	 956 [61, 2237]&	 1764 [107, 4047]&	 99040 [5312, 203579]\\\hline
\multirow{4}{*}{ET+2VOY}&	 1&	  n.d.&	 n.d.&	 n.d.&	 30 [3, 112]\\\cline{2-6}
			             &	10&	 4 [0, 4]&	 6 [0, 9]&	 11 [2, 21]&	 927 [105, 4200]\\\cline{2-6}
			             &	30&	 16 [0, 22] & 26 [2, 41] & 55 [3, 125] & 5202 [603, 23345]\\\cline{2-6} 
			             &	100&	 63 [4, 117]&	 132 [9, 267]&	 292 [22, 751]&	 25775 [2161, 84612]\\\cline{2-6}
		               &  1000&  445 [26, 1025]&	 984 [63, 2339]&	 2189 [163, 6222]&	 85799 [5342, 205575]\\\hline
\multirow{4}{*}{ET+CE}&	 1&	  n.d.&	 4 [0, 3]&	 3 [0, 11]&	 177 [9, 400]\\\cline{2-6}
			&	 10&	 12 [0, 13]&	 51 [2, 112]&	 185 [10, 430]&	 6656 [366, 14836]\\\cline{2-6}
			&	 30&	 37 [1, 66] & 253 [15, 587] & 915 [47, 2107] & 36782 [2022, 78357]\\\cline{2-6}
			&	 100&	 168 [11, 369]&	 1325 [73, 3034]&	 5075 [263, 11255]&	 123303 [6422, 250439]\\\cline{2-6}
			&	 1000& 1229 [69, 2862]&	 15497 [896, 34487]&	 69423 [3703, 144222]&	 194834 [10065, 388038]\\\hline
\multirow{5}{*}{ET+2CE}	&	 0.1&	  n.d.&	 n. d.  &	 n. d. &	 158 [9, 354]\\ \cline{2-6}
			&	 1&	1  [0, 3]&	 7 [0, 8]&	 30 [0, 58]&	 5999 [348, 13383]\\\cline{2-6}
			&	 10&	16 [0, 22]&	 125 [7, 320]&	 675 [41, 1570]&	 105931 [5628, 215840]\\\cline{2-6}
			&	 30&	58 [2, 119] & 624 [39, 1446] & 3070 [164, 7023] & 173679 [9097, 348009] \\\cline{2-6}
			&	 100&	247 [19, 550]&	 2784 [150, 6498]&	 15910 [867, 34921]&	 219966 [11419, 438414]\\\cline{2-6}
			&	 1000& 1640 [91, 3831]&	 25848 [1494, 57007]&	 135482 [7130, 276082]&	 243459 [12537, 483247]\\\hline
\end{tabular}
\caption{The number of detected (network SNR $>$ 8) BNS mergers per year within z=1.5 for different GW detector configurations (ET as a single detector, ET plus second generation detectors including phase plus LIGO-L, LIGO-H, LIGO-I, Virgo, and KAGRA, ET plus two Voyager located in USA, ET plus CE(40 km) located in USA, ET plus 2 CE(40 km) located one in USA and one in Australia). Three pre-merger scenarios (15, 5, and 1 minute(s) before the merger) and the scenario at the time of the merger are shown in different columns.  For each detector configuration, the rows give the number of detections with sky-localization ($90\%$c.l) within 10, 30, 100, and 1000 deg$^2$. When the number of detections is not negligible, we add also rows for 1 and 0.1 deg$^2$. We show the numbers of detected BNS mergers for our fiducial BNS population ($\alpha_{CE}$=3) and in the square bracket the numbers for the pessimistic ($\alpha_{CE}$=0.5) and optimistic BNS population  ($\alpha_{CE}$=5). The numbers of GW detections per year are obtained assuming a duty cycle of 0.85 as described in the text. {We use "n.d." to indicate no detection.}} \label{table:skylocall}
\end{table*}

\begin{table*}
\centering
\renewcommand{\arraystretch}{1.5}
\begin{tabular}{| c | l | r | r | r  | r |}
\hline
\multirow{2}{*}{Detector} & {\hspace{0.3cm}$\Omega$}      & \multicolumn{4}{c|}{On-axis BNS [$\theta_{v} < 10^{\circ}$]} \\ \cline{3-6}
        &   [deg$^{2}$]    & 15 min                  & 5 min                  & 1 min      & 0 min         \\ \hline    
\multirow{3}{*}{ET}	  
	&10	  &    1 [0, 1]     & 1 [0, 1] & 2 [0, 2] & 2 [0, 4] \\  \cline{2-6}
	&30	  &    2 [0, 3]     & 2 [0, 4] & 3 [0, 9] & 3 [0 15] \\  \cline{2-6}
	&100	  &    3 [0, 9] & 6 [0, 23] & 13 [0, 42] & 40 [5, 99] \\  \cline{2-6}
	&1000	  &    26 [0, 64] & 77 [5, 181] & 154 [11, 340] & 346 [24, 807] \\  \hline

\multirow{4}{*}{ET+LVKI+}
	& 1	  &    n.d.     &  n.d.   &  n.d.   & 4 [0, 4] \\  \cline{2-6}
	& 10	  &    1 [0, 1] & 2 [0, 1] & 2 [0, 2] & 71 [7, 169] \\  \cline{2-6}
	& 30	  &    2 [0, 3] & 2 [0, 4] & 3 [0, 11]& 421 [26, 995] \\  \cline{2-6}
	& 100	  &    3 [0, 9] & 6 [0, 23] & 15 [0, 46] & 1745 [97, 3776] \\  \cline{2-6}
	& 1000  &    26 [0, 64] & 78 [5, 181] & 163 [12, 359] & 3119 [182, 6487] \\  \hline

\multirow{4}{*}{ET+2VOY}
	& 1	  &    n.d.     &  n.d.   &  n.d.   & 4 [0, 8] \\  \cline{2-6}
	& 10	  &    1 [0, 1]     & 2 [0, 1] & 2 [0, 3] & 54 [10, 226] \\  \cline{2-6}
	& 30	  &    2 [0, 3] & 2 [0, 4] & 3 [0, 12] & 280 [39, 1287] \\  \cline{2-6}
	& 100	  & 3 [0, 9] & 7 [0, 25] & 18 [1, 49] & 1290 [111, 3813] \\  \cline{2-6}
	& 1000  & 26 [0, 64] & 81 [5, 181] & 181 [14, 462] & 2939 [188, 6624] \\  \hline

\multirow{4}{*}{ET+CE} 

	& 1 	  &    n.d.     &  n.d.   & 1 [0, 0] & 8 [2, 26] \\  \cline{2-6}
	& 10	  &    2 [0, 2]     & 5 [0, 5] & 17 [0, 27] & 397 [29, 913] \\  \cline{2-6}
	& 30	  &    3 [0, 4]  & 16 [0, 33] & 57 [3, 139] & 1964 [103, 4134] \\  \cline{2-6}
	& 100	  &    8 [0, 17] & 71 [2, 165] & 314 [15, 613] & 3376 [195, 6715] \\  \cline{2-6}
	& 1000  &   48 [2, 105] & 632 [39, 1470] & 2800 [171, 5749] & 3504 [204, 6974] \\  \hline
\multirow{5}{*}{ET+2CE} & 0.1 &    n.d.     &  n.d.   &  n.d.   & 8 [1, 19] \\  \cline{2-6}
	& 1 	  &    n.d.     &  n.d.   & 4 [0, 2] & 321 [21, 762] \\  \cline{2-6}
	& 10 	  & 2 [0, 1] & 11 [0, 18] & 47 [4, 99] & 2909 [172, 5797] \\  \cline{2-6}
	& 30 	  & 5 [0, 6] & 37 [0, 70] & 184 [11, 394] & 3558 [202, 7096] \\  \cline{2-6}
	& 100	  & 11 [0, 24] & 128 [10, 298] & 846 [49, 1838] & 3929 [227, 7841] \\  \cline{2-6}
	& 1000  & 58 [3, 128] & 904 [62, 2091] & 3608 [215, 7245] & 3971 [229, 7919] \\  \hline
\end{tabular}
\caption{Same as Table \ref{table:skylocall} showing the events with viewing angle smaller than 10$^{\circ}$.}
\label{table:skyloc}
\end{table*}  

\begin{figure*}
\centering
            \includegraphics[width=0.32 \linewidth, height=5.8cm]{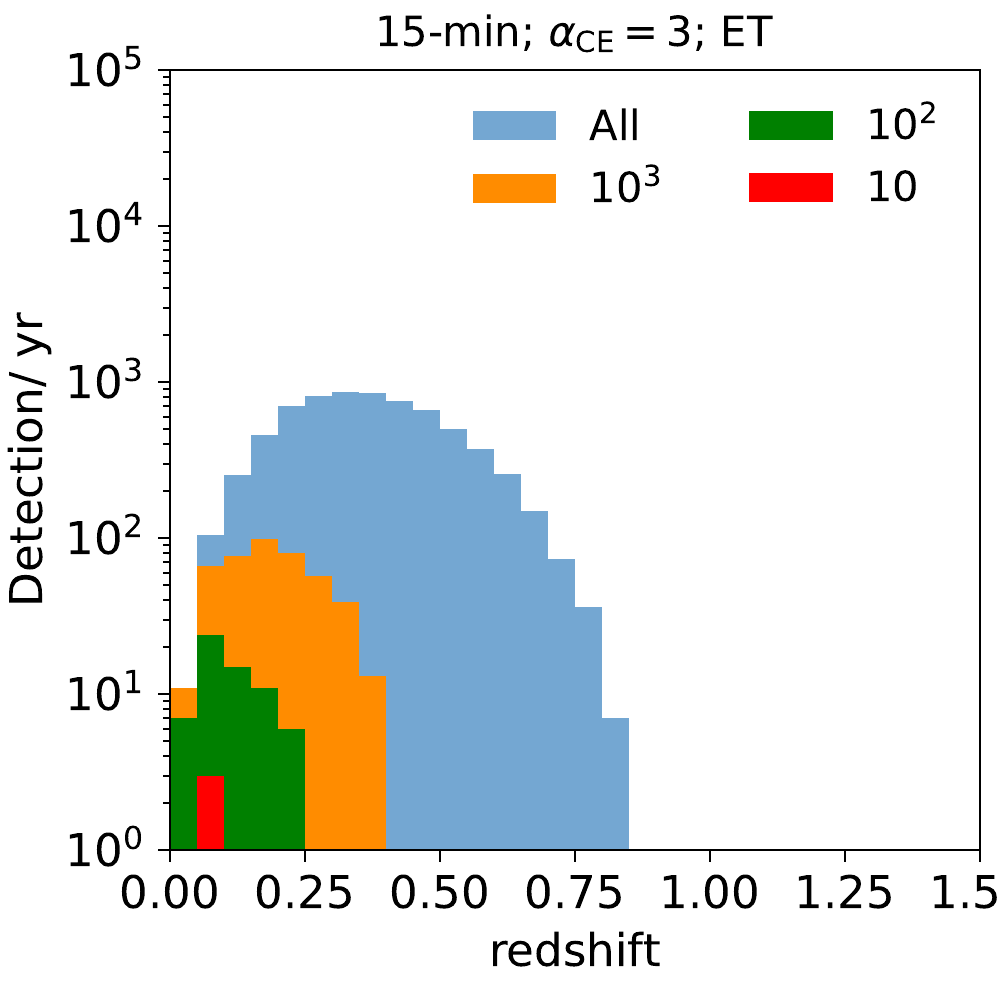}
            \includegraphics[width=0.32 \linewidth, height=5.8cm]{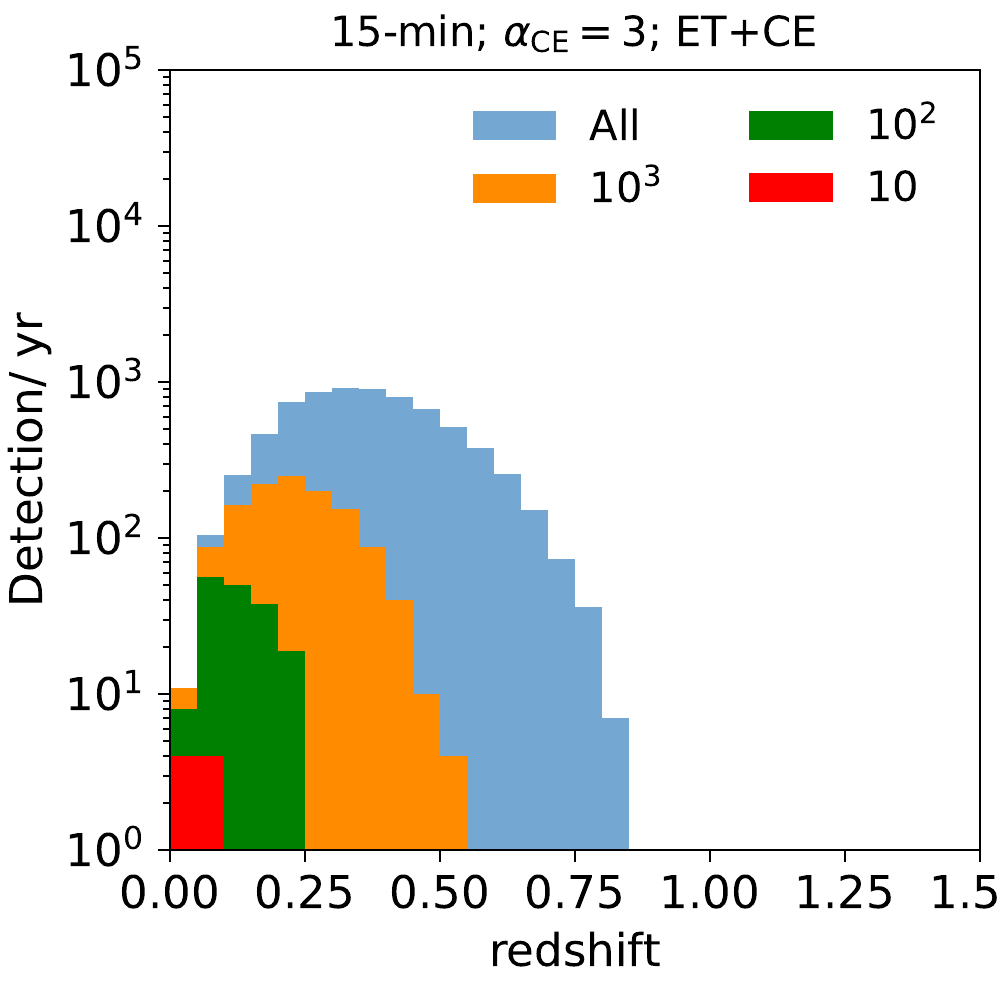}
            \includegraphics[width=0.32 \linewidth, height=5.8cm]{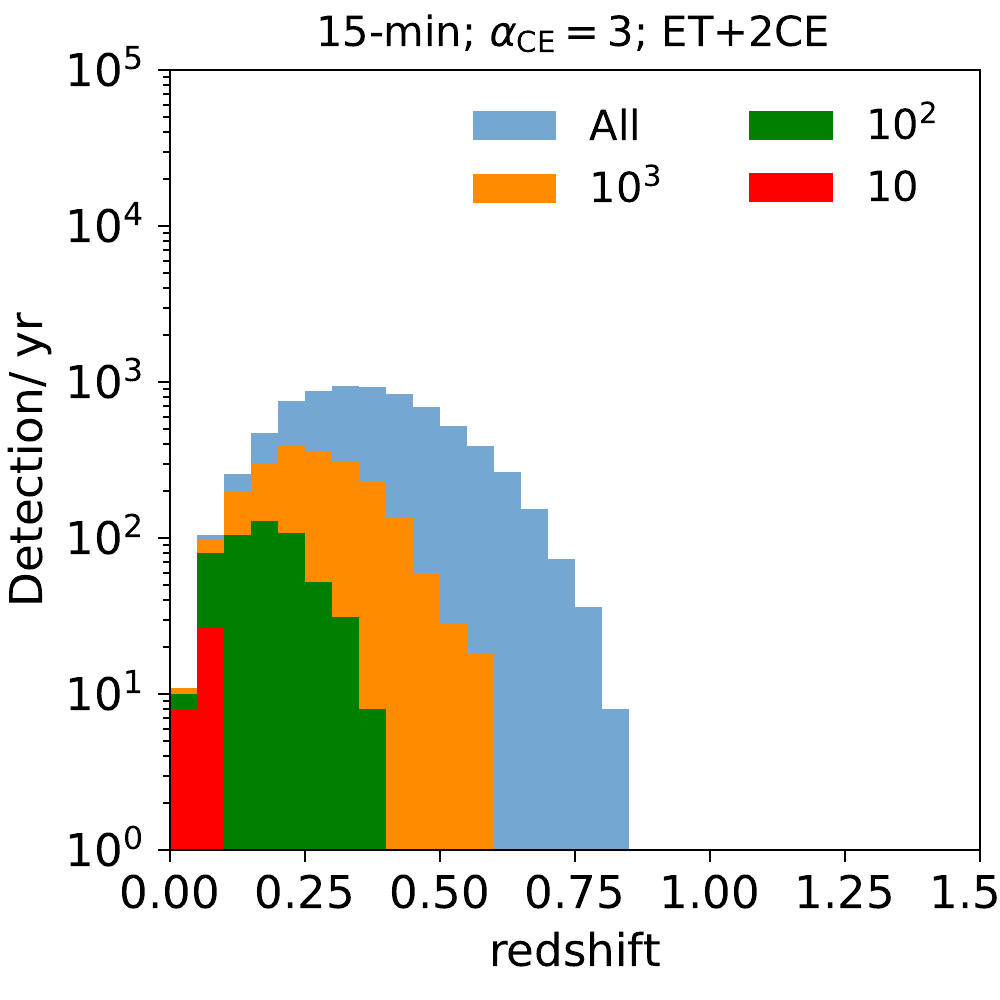}
            \includegraphics[width=0.32 \linewidth, height=5.8cm]{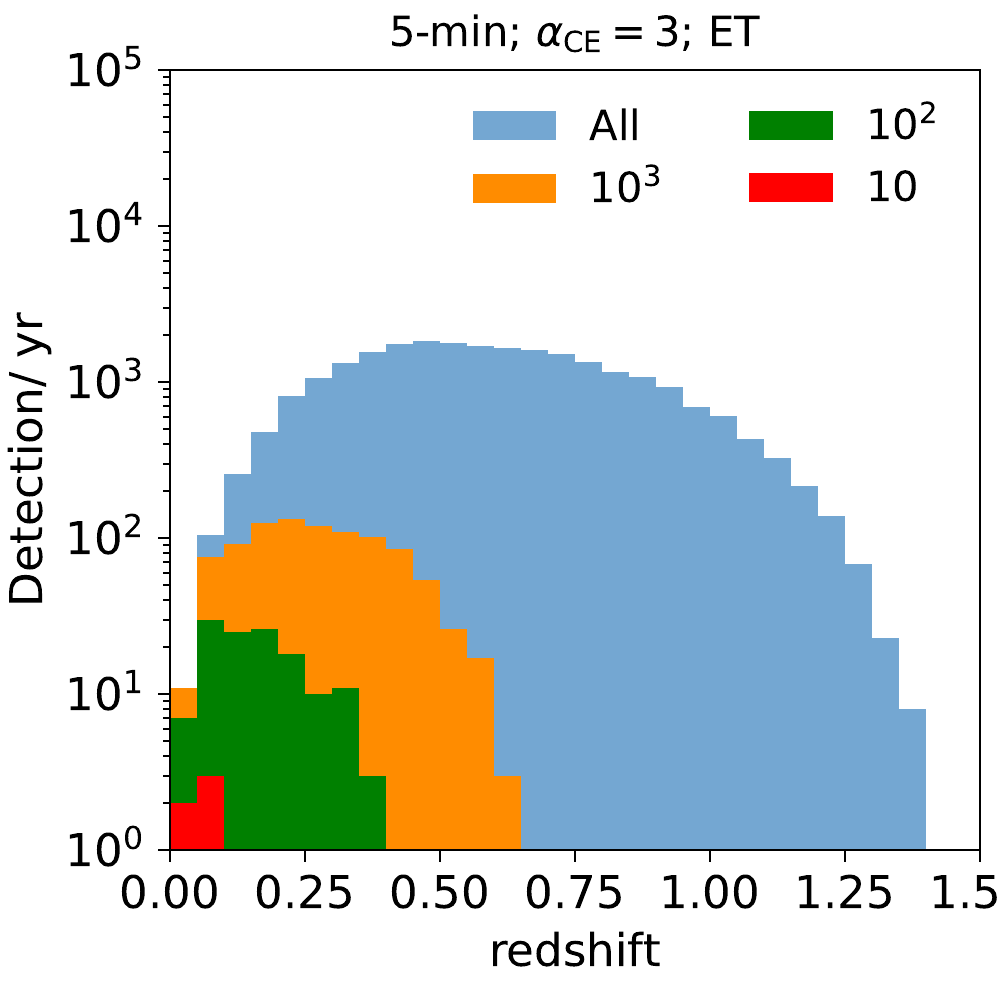}
            \includegraphics[width=0.32 \linewidth, height=5.8cm]{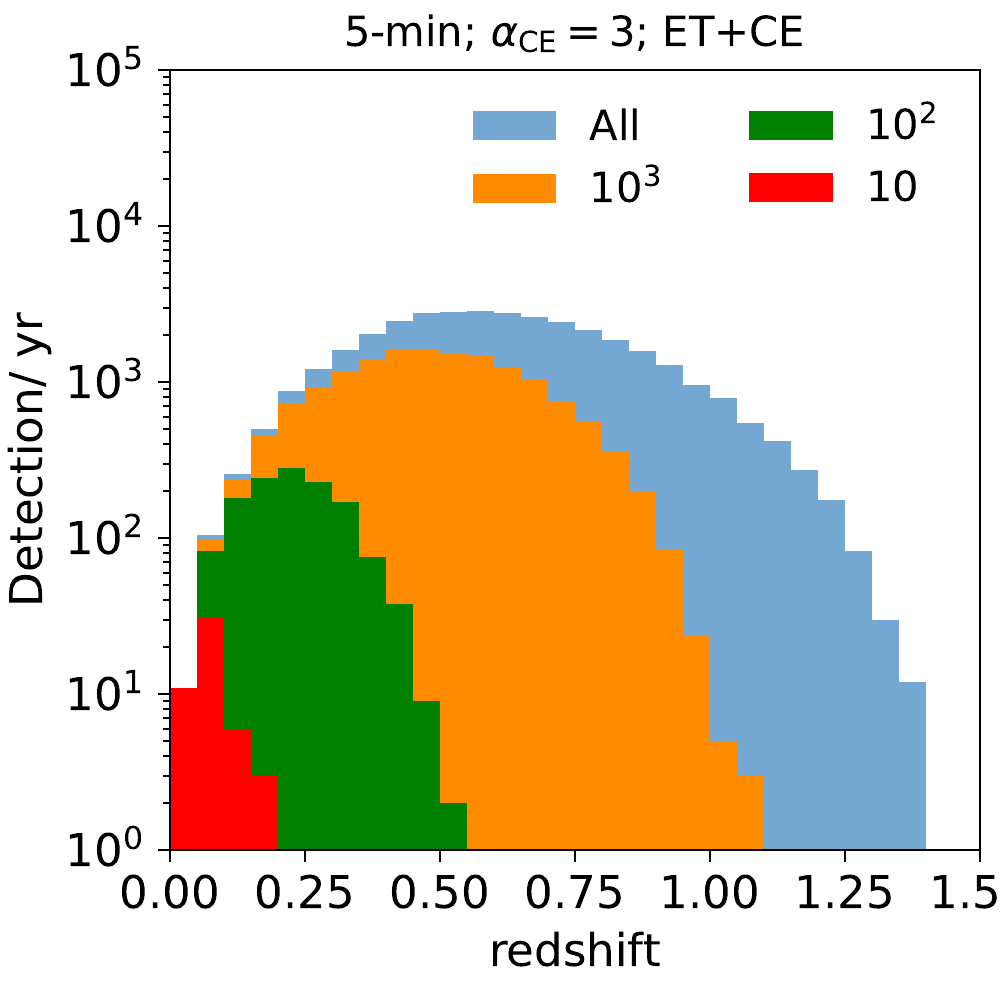}
            \includegraphics[width=0.32 \linewidth, height=5.8cm]{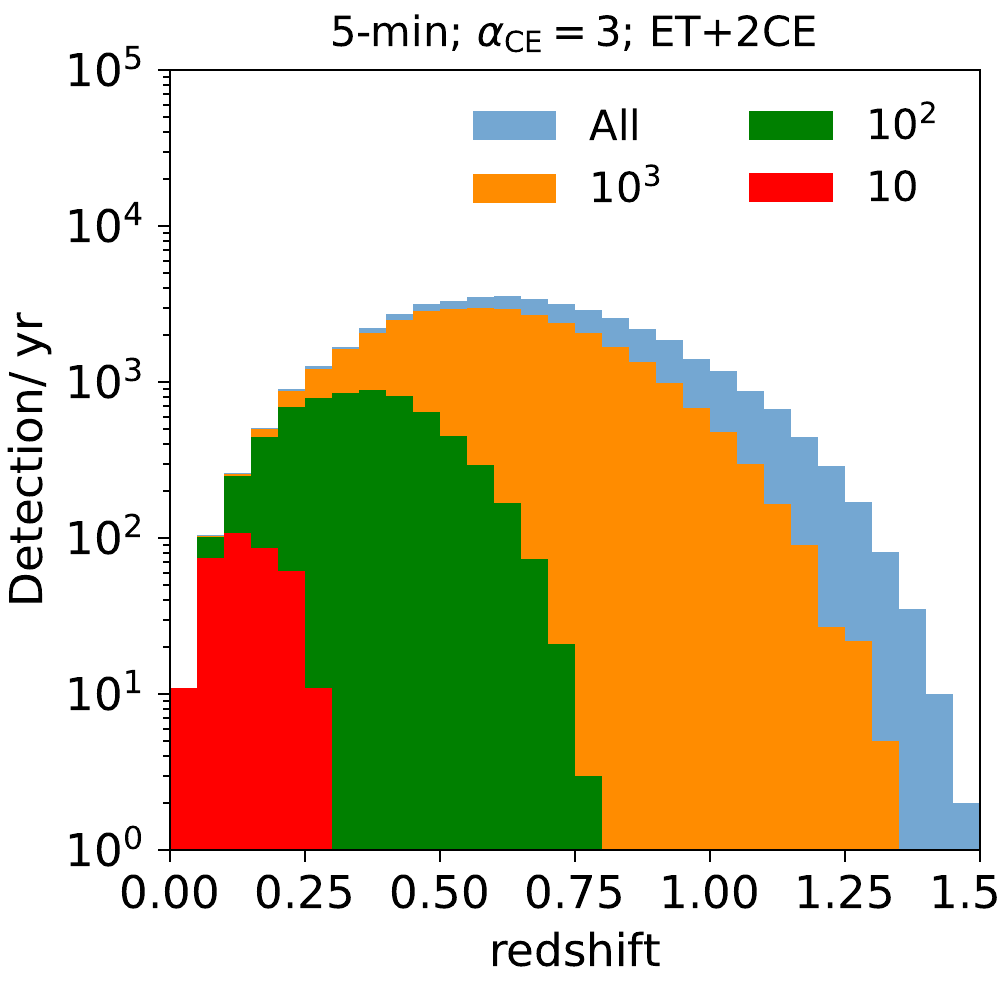}
            \includegraphics[width=0.32 \linewidth, height=5.8cm]{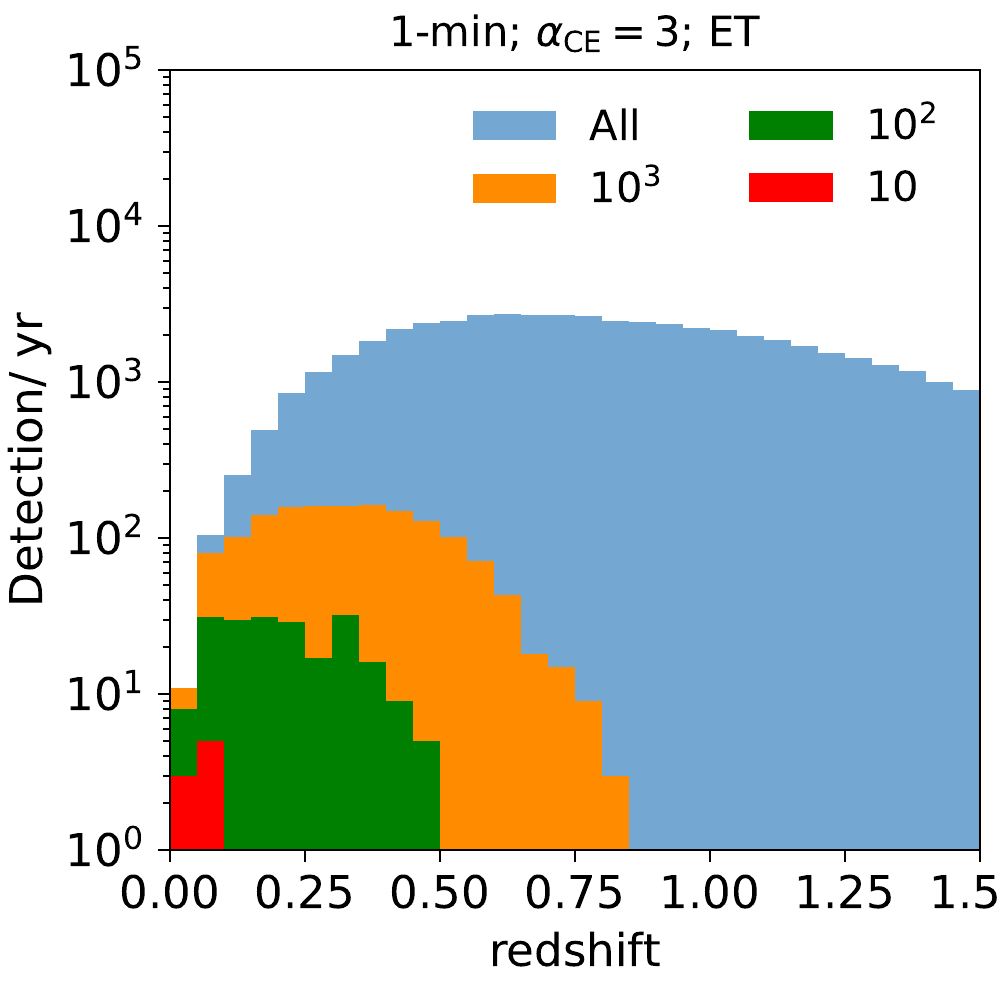}
            \includegraphics[width=0.32 \linewidth, height=5.8cm]{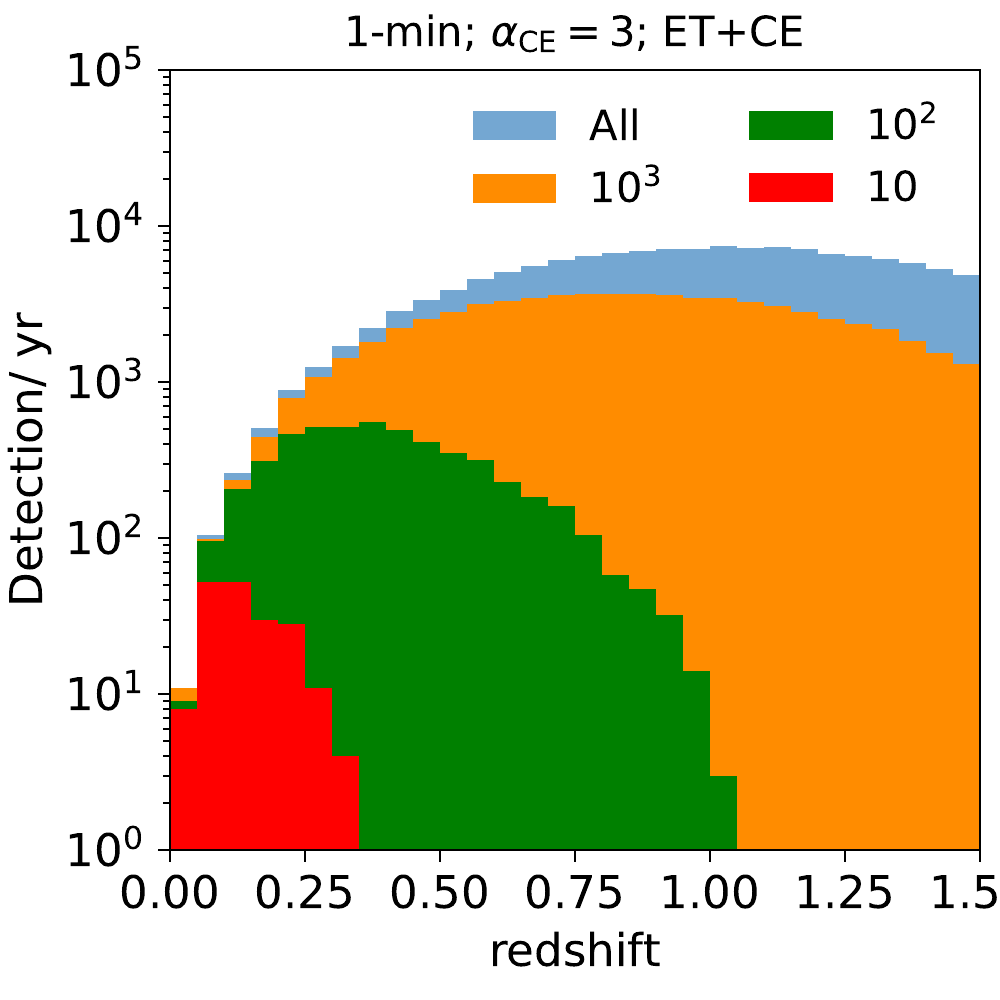}
            \includegraphics[width=0.32 \linewidth, height=5.8cm]{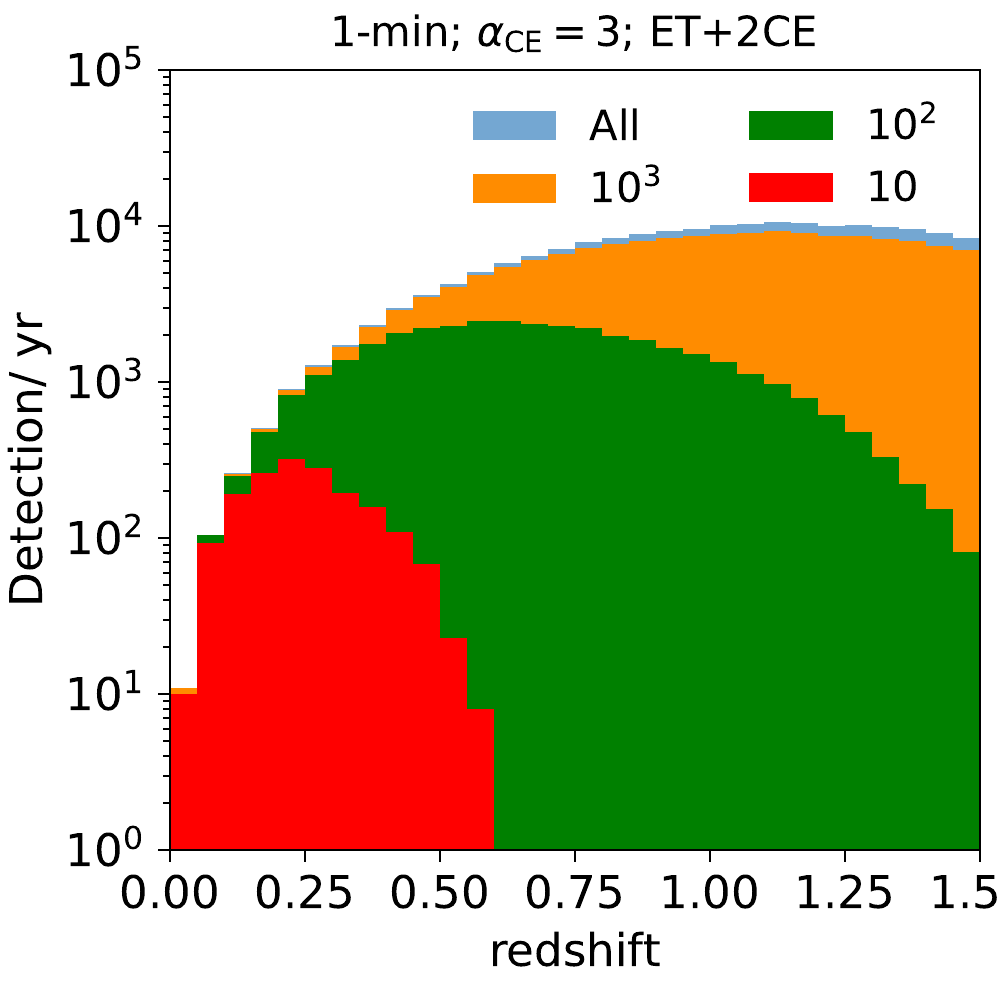}
            \includegraphics[width=0.32 \linewidth, height=5.8cm]{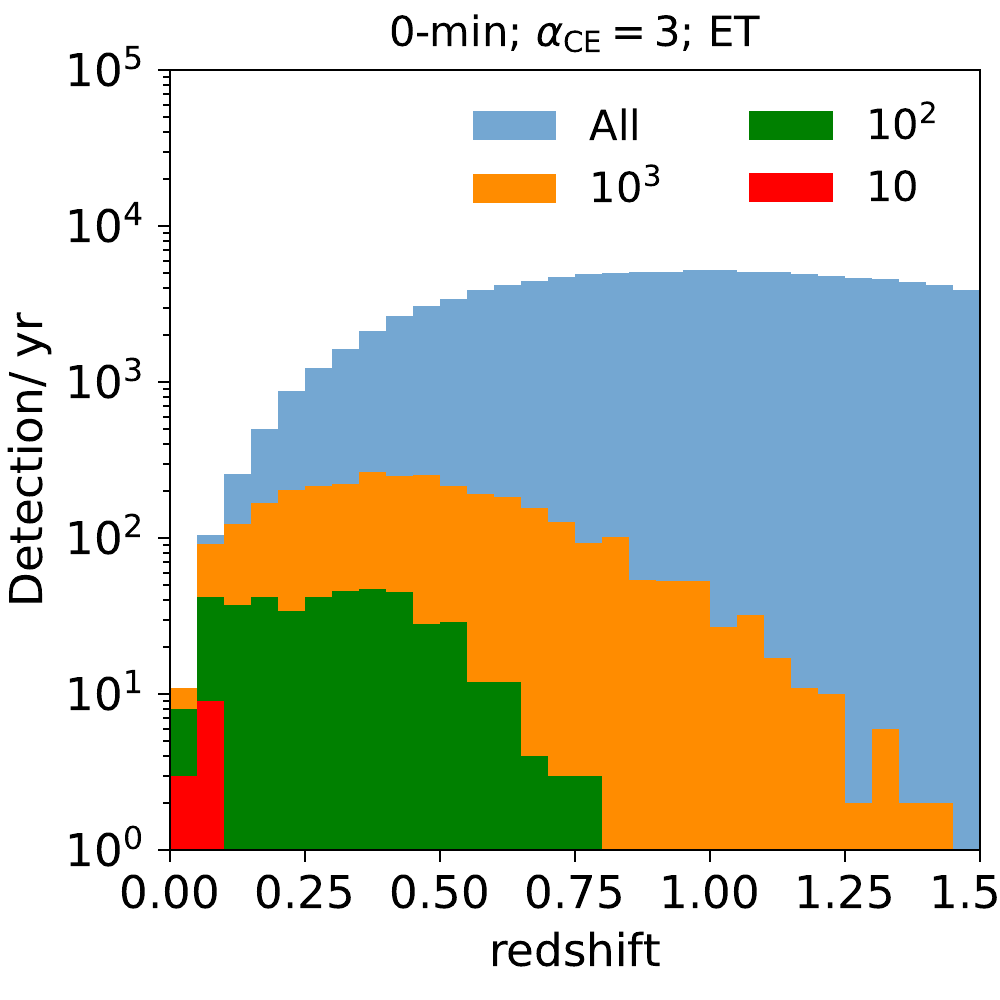}
            \includegraphics[width=0.32 \linewidth, height=5.8cm]{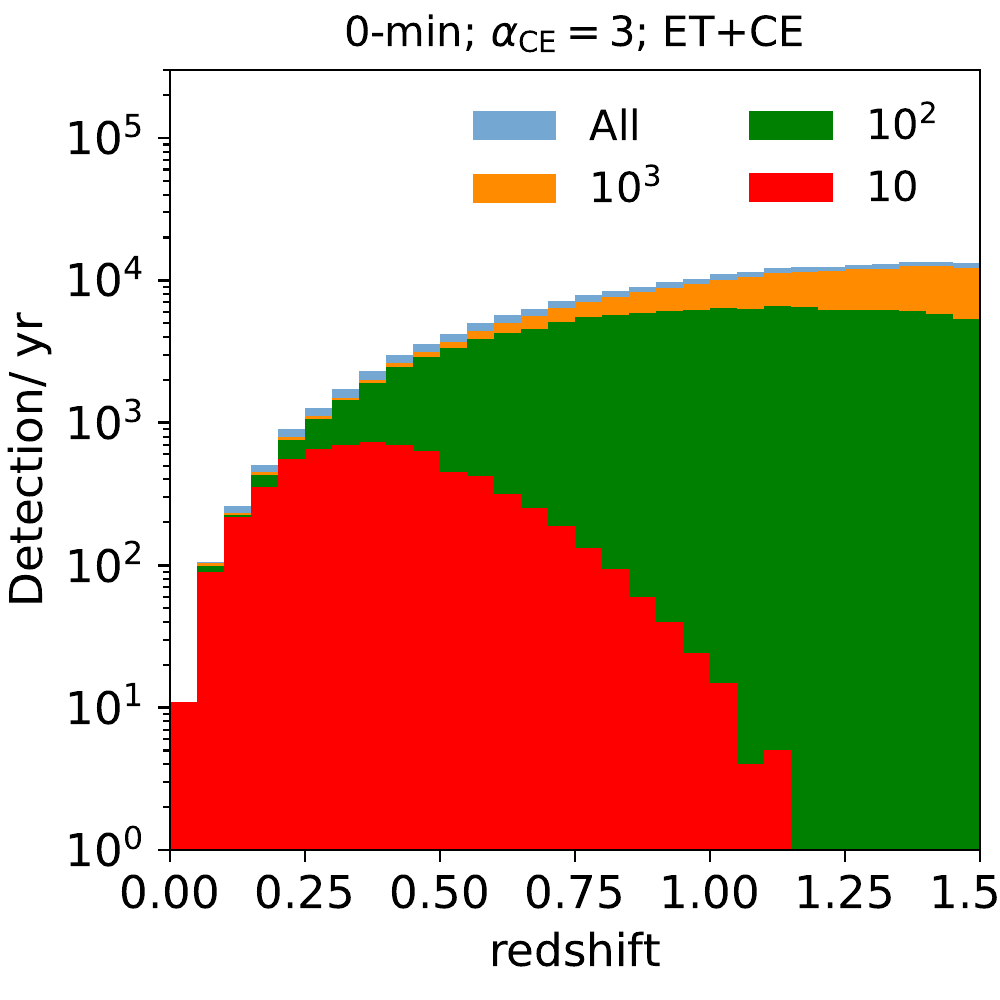}
            \includegraphics[width=0.32 \linewidth, height=5.8cm]{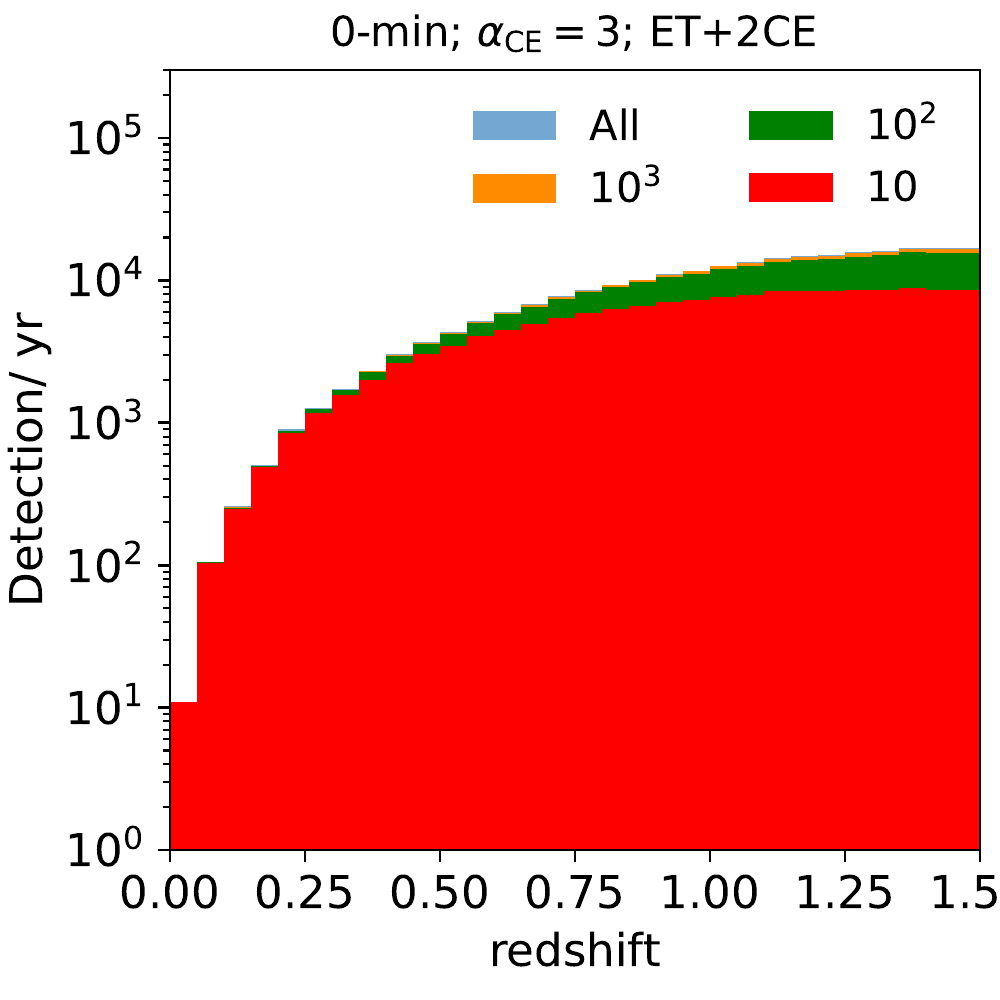}
            \caption{Redshift distribution of the sky-localization uncertainty (given as 90$\%$ credible region) for three detector configurations: ET, ET+CE and ET+2CE. The absolute numbers refer to the fiducial BNS population sample and detections within a redshift of 1.5 per year of observation assuming a duty cycle of 0.85 as described in the text. The panels show the detections and the corresponding sky-localizations as a function of the redshift 15, 5, and 1 minute(s) before the merger and at the merger time. The blue histogram indicated with "all" shows all the detected sources.}
            \label{fig:SLvsz}
        \end{figure*}

The cumulative distribution of detections per year up to redshift equal to 1.5 as a function of the sky-localization is shown in Fig.~\ref{fig:BNSrateSN3} for the different detector configurations. For 15 and 5 minutes pre-merger scenarios, the cumulative distributions for ET as a single observatory, ET+LVKI+, and ET+2VOY are the same, indicating that ET is the main observatory that localizes in the network. The presence of second-generation detectors or the two Voyagers improves the  sky-localization one minute before the merger, and the improvement becomes largely significant at the merger time. The presence of CE in the network significantly improves the sky-localization capabilities pre-merger of ET also 15 minutes before the merger. {As shown in Fig.~\ref{fig:BNSrateSN3}, when ET is included in a network of next-generation GW detectors, the cumulative number of detections tends to flatten for sky-localizations larger than 1000 deg$^2$. This is due to the fact that the network localizes most of the events better than this value (see Fig.~\ref{fig:SLvsz}).} 

The sky-localization capabilities of ET, ET+CE, and ET+2CE, at five and at one minutes before the merger and at the time of the merger  are shown in Fig.~\ref{fig:SLvsz} as a function of redshift for the fiducial population of the BNS ($\alpha_{\rm CE}$=3). The number of well localized ($\Omega < 100$ deg$^2$) are not negligible (order of hundred) already 5 minutes before the merger and up to z=0.4 for ET as a single observatory. This number increase to thousands of detections up to z=0.5 for ET+CE and ET+2CE. One minute before the merger several thousands of detection have sky-localizations $\Omega < 100$ deg$^2$ for ET+CE (ET+2CE) up to z=1.0 (1.3), and hundreds have sky-localizations $\Omega < 10$ deg$^2$ up to z around 0.4 for ET+CE and ET+2CE.

Our pre-merger sky-localization results are in agreement with \citealt{nitz2021bnsloc} and \citealt{li2021skyloc}. For example, despite the different BNS population and detection criteria (SNR$>$12 and 100\% duty cycle), \citealt{li2021skyloc} find 7 and 210 events per year for sky-localization of $<$1 and 10 deg$^2$ detected by ET+2CE 5 minutes before the merger. These numbers match with this study: $\mathcal{O}$(10) and $\mathcal{O}$(100). Also in the case of ET alone, 5 minutes before the merger \citealt{nitz2021bnsloc} find 6 and 94 detection per year with sky-localization smaller than 10 and 100 deg$^2$, which is again in agreement with our numbers $\mathcal{O}$(10) and $\mathcal{O}$(100), respectively. \citealt{li2021skyloc} found several hundreds of detection with sky-localization smaller than 30 deg$^2$ before the merger for ET+2CE as in the present work. The number of events detected and localized at the time of the merger are in agreement with the extensive work of \citealt{Ronchini2022} and \citealt{Iacovelli2022} (taking into account that the present work analyzed events up to z=1.5).

{Our follow-up observational strategies for the next generation GW detectors are based on the presence of low-latency pipelines and infrastructures able to detect GW event candidates and send  public alerts in almost real-time as currently done by the LIGO, Virgo and KAGRA collaborations \citep{LRR2020,2019ApJ...875..161A}. We assume an alert time (t$_{\rm alert}$) of 30\,s covering the time to detect an event, to transmit and receive the alert. The current low-latency detection pipelines are able to detect an event within 10 seconds \citep{2022PhRvD.105b4023C}. They perform a matched-filter search for binary merger signals using a bank of gravitational-wave template waveforms and give in low-latency a first estimate of the source parameters (including sky-localization and viewing angle). Currently, the latency to send an alert is dominated by the semi-automated detector characterization and data quality checks which bring the detection alert latency to a few minutes, but efforts are ongoing to reduce this time to an order of seconds. We consider a t$_{\rm alert}$ = 30 seconds appropriate for the next generation detector to include possible delay in the transmission/receipt.  In the case of data quality check delays similar to the current ones, the only observational strategy whose results could be negatively impacted is the one of LST following 1-minute pre-merger alert (the strategy called LST-c in Table 5) which we consider as the most risky strategy in our work.}

\subsection{CTA observational strategies and detectability}
\label{CTAdetect}
The detection of VHE emission from BNS merger is currently challenging because of (a) the large sky-localization of GW signals relative to the FoV of IACTs, (b) the long delay 
between the merger time and the GW alert time and response time of IACTs, and (c) the small volume of the Universe up to which GW detectors are able to observe BNS merger, which makes the probability of detecting on-axis events from which VHE is expected small. The study presented in the paper shows that the era of ET and CE can mark a paradigm shift mostly because of the ability to provide pre-merger alerts with a good sky-localization even 15 minutes before the merger. In addition, the effectiveness of the VHE counterpart search will increase due to the improved sensitivity of the next generation GW detectors, which increase the number of on-axis event detections, and the large FoV, unprecedented sensitivity, and short slewing time of CTA.

We assume that the prompt VHE emission originating from the processes described in \S\ref{sec:theory} is short-lived and detectable in an observational window of 20\,s around the merger time. We focus on the pre-merger alert scenarios of 15, 5, and 1 minute(s) before the merger. We consider an alert time (t$_{\rm alert}$) of 30\,s corresponding to the communication of the alert among the GW detector network and CTA, the CTA-LST (-MST) slew time (t$_{\rm slew}$) of 20\,s (90\,s), and an exposure time (t$_{\rm exp}$) of 20\,s. 
We also add t$_{\rm add}$ 10\,s which includes a possible repositioning and the uncertainty on the estimation of the merger time. The total CTA time for one single observation is t$_{\rm obs}$=t$_{\rm slew}$+t$_{\rm add}$+t$_{\rm exp}$=50\,s and 120\,s for LST and MST, respectively.

For the one-minute pre-merger alerts, we only consider the LST array, since it has a faster slew-time of less than 20\,s. However, the chances of detection are reduced by the smaller sky-coverage of LST which has a FoV of around 10 deg$^2$. The slew-time of 90\,s for MST makes it impossible to follow the one-minute pre-merger alerts.  In contrast, the MST array is appropriate for following up the 5 and 15-minute pre-merger alerts. Although the number of GW detection to be followed up is smaller than in the case of one-minute pre-merger alerts, the FoV of 30\,deg$^2$ is an advantage.

In the following, we examine the results for four observational strategies: the direct pointing of well-localized events, the one-shot observation strategy, the divergent pointing, and the mosaic strategy. 

\subsubsection{Direct pointing of well-localized events}
We explore the direct pointing strategy by selecting events with sky-localization smaller than 10 and 30 deg$^2$ taking into account the adopted FoV for the LST and MST arrays, respectively.  

From Table~\ref{table:skylocall}, the number of events with sky-localization smaller than 10 deg$^2$ to be followed up by the LST array is around ten per year one minute before the merger for ET, ET+LVKI+, and ET+2VOY considering the fiducial population. This number increases to a few hundred (several hundred) for ET+CE (ET+2CE). Among these events, as shown in Table~\ref{table:skyloc}, the number of events on-axis (namely, the events with a viewing angle smaller $10^{\circ}$ from which we expect to observe the VHE), is negligible for ET, ET+LVKI+, and ET+2VOY, and it becomes of the order of a few (several) tens for ET+CE (ET+2CE). 

The number of on-axis events with sky-localization smaller than 30 deg$^2$ five minutes before the merger is negligible for ET, ET+LVKI+, and ET+2VOY. This number becomes a few tens for ET+CE and ET+2CE (see Table~\ref{table:skyloc}). These events can be detected in VHE by using the MST array following up a few hundred (several hundred) of 5 minutes of pre-merger alerts (see Table~\ref{table:skylocall}).  

To estimate the actual number of joint detections, it is necessary to take into account the CTA duty cycle of $15\%$ and the CTA visibility. Considering that CTA telescopes are able to observe sources with an elevation larger than 30 deg, this reduces the visible sky by a factor of 2. Another factor to consider is the fraction of BNSs that produce a jet.  Although this fraction is still largely uncertain, studies combining electromagnetic observations of short GRBs and BNS merger rates from GW observations indicate that a $20-50\%$ fraction of BNS mergers produce a jet \citep[e.g.][]{Ronchini2022, Colombo2022,2022arXiv220201656S}. Considering all these factors, the direct pointing strategy is expected to give a few joint detections per year only when ET will operate in a network of three next-generation detectors. No joint detections are expected for the pessimistic BNS population scenario. 

\subsubsection{One-shot observation strategy}
In order to increase the number of events to be followed, we propose a strategy that for each detected event uses a one-shot observation covering an area corresponding to the FoV of CTA-LST and -MST (10 and 30 deg$^2$). 

The percentage of CTA total observational time in one year that would be spent following all the individual events by a one-shot observation randomly positioned in the sky-localization area is shown by the central plots (second column) of Fig.~\textcolor{red}{7} for MST and Fig.~\ref{fig:FigLSTPrediction} for LST. This percentage is evaluated as the fraction of CTA observational time necessary to follow up all the events with sky-localization smaller than the one indicated in the x-axis: 

\begin{equation}
\rm CTA time (\%) = \frac{N(< \Omega) \times t_{obs} \times CTA_{vis}}{CTA_{TOT} }  
\end{equation}
where $\rm N(< \Omega)$ is the number of events with sky-localization smaller than $\Omega$, $\rm t_{obs}$ is the observational CTA time spent for each event, and $\rm CTA_{TOT}$ the total observational time of CTA in one year including the duty cycle of 15\%. The visibility of CTA, $\rm CTA_{vis}=0.5$, takes into account that CTA telescopes are able to observe sources with an elevation larger than 30 deg. 
The $\rm t_{obs}$ is set to 120 seconds for MST and 50 seconds for LST. The observational strategy consists of receiving the pre-merger alert and beginning to slew the telescopes just before the estimated merger time (that we assume to be given as information in the alert). For the MST, we consider only 15 and 5 minutes pre-merger alerts, after receiving the alert, the MSTs start the slew about 100\,s before the merger time. For the LST (thanks to the very rapid slew-time of 20\,s), we also consider the 1-minute pre-merger alerts. In this case, after receiving the alert, the LST slewing starts 30\,s before the merger time given in the alert. 

For the 15 and 5 minutes alerts, we also consider a double-step strategy; for all the events localized with a sky-localization smaller than 1000 deg$^2$ we initially point the center of the sky-localization uncertainty and then we re-point the antennas based on the updated sky-localization obtained 1 minute before the merger which is expected to be significantly smaller. Since the Fisher matrix approach does not give the real shape of the sky-localization uncertainty and the distribution probability within it, we approximate the sky-localization with a circular error-region and a uniform distribution probability to contain the GW source within it.
Given the angular radius of the LST FoV of about 2$^\circ$, the maximum required repositioning for an error-region of 1000 deg$^2$ (angular radius of 18$^\circ$) is $\sim$16$^\circ$ which corresponds to a repositioning time of about $\sim$3\,s. The required repositioning time for MST (angular radius of 3$^\circ$) for a movement of 15$^\circ$ is about $\sim$10\,s. We consider the same repositioning time of t$_{\rm rep} =$\,10\,s for both LST and MST. Our follow-up strategy to detect the prompt/early VHE emission relies on the precision of the merger time which is expected to be given in the alert. The uncertainty on the merger time is estimated by the Fisher matrix analysis to be much smaller than 0.1 second for events localized better than 1000 deg$^2$. Table~\ref{table:obsstrategyMST} and Table~\ref{table:obsstrategyLST} summarize the different observational strategies included in the present analysis, namely 1) following up all the events detected 15 minutes before the merger (MST-a, LST-a), 2) following up  all the events detected 5 minutes before the merger (MST-b, LST-b), 3) following up all the events detected 1 minute before the merger (LST-c), and 4) using the improved sky-localization updated 1 minute before the merger (MST-c, MST-d, LST-d, LST-e).

\begin{table*}[t]
\centering
\renewcommand{\arraystretch}{1.5}
\begin{tabular}{| c | p{2.2cm} | p{2.2cm} |  p{2.8cm} | p{2.8cm} |} \hline
Time  before merger  & MST-a & MST-b & MST-c & MST-d \\ \hline
15 minutes   & Event detected &  &  Event detected with sky-loc < $10^3$ deg$^2$	 &   \\ \hline
14.5 minutes &    Alert received  &  & Alert received & \\  \hline
5 minutes  &  & Event detected &  	 &  Event detected with sky-loc < $10^3$ deg$^2$ \\ \hline
4.5 minutes &     & Alert received  & & Alert received \\  \hline
100\,seconds        &  \multicolumn{4}{c|}{Start slewing} \\  \hline
60\,seconds  &    &   & \multicolumn{2}{c|}{Parameters updated} \\  \hline
30\,seconds &     &  & \multicolumn{2}{c|}{Updates received} \\  \hline
\multirow{3}{*}{10\,seconds}&  \multirow{3}{*}{Sky-loc reached}   & \multirow{3}{*}{Sky-loc reached}  & \multicolumn{2}{c|}{Sky-loc reached} \\ 
\cline{4-5} &   &  & \multicolumn{2}{c|}{Repositioning on the updated sky-loc}\\
\cline{4-5} &   & & \multicolumn{2}{c|}{Updated sky-loc reached} \\
\hline
Merger time & \multicolumn{4}{c|}{20 s of exposure} \\  \hline
& \multicolumn{4}{c|}{t$_{\rm alert}$=30\,s} \\
& \multicolumn{4}{l|}{Assumed time:\hspace{2.85cm}t$_{\rm slew}$=90\,s}  \\ 
& \multicolumn{4}{c|}{t$_{\rm rep}$=10\,s} \\
& \multicolumn{4}{c|}{t$_{\rm exp}$=20\,s} \\ \cline{2-5}
& \multicolumn{4}{l|}{Total CTA time required:\hspace{1.5cm}t$_{\rm obs}$=120\,s}\\ \hline
Results & Fig.~\textcolor{red}{7} (top row) & Fig.~\textcolor{red}{7} (bottom row) & Fig.~\ref{fig:NdetReal} (top left plot) & Fig.~\ref{fig:NdetReal}  (bottom left plot)\\
\hline
\end{tabular}
\caption{The observational strategies of following-up pre-merger alert events with CTA-MST. For all the strategies shown in the table (MST-a, MST-b, MST-c, MST-d), MST starts slewing to the sky location at about 100\,s before the merger time following the alerts and parameters received at 14.5 minutes (MST-a and MST-c) or 4.5 minutes (MST-b and MST-d) before the merger. In the MST-c and MST-d scenarios, the updated sky-localization estimated 1 minute before the merger is used to re-positioning the MST array in about 10\,s. In the MST-c and MST-d scenarios, we follow up only sky-localization smaller than $\rm 10^3 deg^2$ estimated 15 and 5 minutes before the merger, respectively. The table gives the values assumed for the time to detect, transmit, and receive the alert (t$_{\rm alert}$), for the MST slewing time t$_{\rm slew}$, for the MST re-positioning time t$_{\rm rep}$, and for the MST exposure time (t$_{\rm exp}$). In all four strategies, the total CTA-MST time necessary for a one-shot observation is 120\,s.}
\label{table:obsstrategyMST}
\end{table*}

\begin{table*}[t]
\centering
\renewcommand{\arraystretch}{1.5}
\begin{tabular}{| c | p{2.2cm} | p{2.2cm} |  p{2.8cm} | p{2.8cm} | p{2.8cm} |} \hline
Time  before merger  & LST-a & LST-b &LST-c& LST-d & LST-e\\ \hline
15 minutes   & Event detected &  &  &Event detected with sky-loc < $10^3$ deg$^2$	 & \\ \hline
14.5 minutes &    Alert received  &  & &Alert received & \\  \hline
5 minutes  &  & Event detected &  	 &  &Event detected with sky-loc < $10^3$ deg$^2$\\ \hline
4.5 minutes &     & Alert received  & & & Alert received \\  \hline
60\,seconds  &    &   & Event detected&\multicolumn{2}{c|}{Parameters updated} \\  \hline
30\,seconds        &  \multicolumn{2}{c|}{Start slewing}&  Alert received +Start slewing & \multicolumn{2}{c|}{Start slewing}\\  \hline
\multirow{3}{*}{10\,seconds}& \multicolumn{3}{c|}{} & \multicolumn{2}{c|}{Sky-loc reached} \\ 
\cline{5-6} &  \multicolumn{3}{c|}{Sky-loc reached} & \multicolumn{2}{c|}{Repositioning on the updated sky-loc}\\
\cline{5-6} &  \multicolumn{3}{c|}{}& \multicolumn{2}{c|}{Updated sky-loc reached}\\ \hline
Merger time & \multicolumn{5}{c|}{20 s of exposure} \\  \hline
&\multicolumn{5}{c|}{t$_{\rm alert}$=30\,s} \\
& \multicolumn{5}{l|}{Assumed time: \hspace{4.4cm}t$_{\rm slew}$=20\,s}  \\ 
& \multicolumn{5}{c|}{t$_{\rm rep}$=10\,s} \\
& \multicolumn{5}{c|}{t$_{\rm exp}$=20\,s} \\ \cline{2-6}
& \multicolumn{5}{l|}{Total CTA time required: \hspace{3.0cm}t$_{\rm obs}$=50\,s}\\
\hline
Results & Fig.~\ref{fig:FigLSTPrediction} (top row) & Fig.~\ref{fig:FigLSTPrediction} (middle row) & Fig. \ref{fig:FigLSTPrediction} (bottom row)& Fig.~\ref{fig:NdetReal} (top right plot) & Fig.~\ref{fig:NdetReal}  (bottom right plot)\\
\hline
\end{tabular}
\caption{The observational strategies of following-up pre-merger alert events with CTA-LST.
Taking into account the faster slewing of LST (t$_{\rm slew}$) with respect to MST, the slewing to reach the sky-localization starts 30\,s before the merger in all the five strategies (LST-a, LST-b, LST-c, LST-d, LST-e).  The alerts received at 14.5 before the merger are followed in the LST-a scenario, the ones received 4.5 minutes in LST-b, and the ones received 30\,s before the merger in LST-c. Considering an alert time (t$_{\rm alert}$) of 30\,s, also the 1-minute alerts can be followed up (LST-c). The {last two columns} show the scenarios LST-d and LST-e where the LST array follow-up sky-localization smaller than $\rm 10^3 deg^2$ obtained at 15 and 5 minutes and then is re-positioned within the updated sky-localization obtained 1 minute before the merger and received 30 seconds before the merger. The re-positioning time (t$_{\rm rep}$) of 10\,s has also been added to the total observation time for LST-a, LST-b, and LST-c to make the follow-up procedure safer. In all the five cases described above, the total CTA-LST time for one observation is 50\,s.}
\label{table:obsstrategyLST}
\end{table*}

The central plots of Fig.~\textcolor{red}{7}
show the percentage of CTA time necessary to follow up all the events detected by ET alone or in a network of detectors with a one-shot observation of MST. They refer to the fiducial BNS population and the 15 and 5 minutes pre-merger alerts (MST-a and MST-b strategies in Table~\ref{table:obsstrategyMST}). Fig. \ref{fig:FigLSTPrediction} shows the same for LST (LST-a and LST-b strategies in Table~\ref{table:obsstrategyLST}). For LST a plot is added (third row) for the 1-minute pre-merger alerts (LST-c strategy in Table~\ref{table:obsstrategyLST}). 

The follow-up with MST of all the events detected 15 minutes before the merger and with sky-localization smaller than $\rm 10^4 deg^2 $ is possible at the cost of around 4\% CTA observational time for ET+2CE and ET+CE, and around 2\% for ET alone. The number of detected events 5 minutes before the merger is larger than the ones detected 15 minutes before the merger, and the amount of time to follow up all of them is around 40\% and 50\% of CTA time for ET+CE and ET+2CE, and 4\% of the CTA time for ET alone.

Using LST, the CTA time budget will be exhausted by following up all the events with a pre-merger alert of 1 minute and with sky-localization smaller than around 1000 deg$^{2}$ for ET+2CE and with sky-localization smaller than about $\rm 10^4 deg^{2}$ for ET+CE. 
Only 10\% (20\%) of CTA-LST time will be consumed following the 5-minute pre-merger events with sky-localization smaller than about $\rm 10^3 (10^4) deg^{2}$ for ET+CE (ET+2CE).
Due to the smaller observational time for each observation of LST with respect to the MST one, for the 15-minute and 5-minute alerts, the observational time reduces by about a factor of 2 with respect to MST. 

We then estimate the expected number of VHE counterparts detectable with the one-shot observation method. {Since the VHE emission is expected only from on-axis events, we identify all the events injected with $\theta_{\rm v}<$10$^{\circ}$ and detected by the GW detectors in our simulation. Due to the fact that the estimate of $\theta_{\rm v}$ from the GW data will not be precise enough to directly select on-axis events (see Fig.~\ref{fig:FigdTvsT} for the distribution of the uncertainty on the viewing angle), our observational strategy consists on following-up all the GW triggers. The expected number of possible VHE detections per year by observing all the GW triggers is evaluated by summing over all the on-axis events ($\theta_{\rm v}<$10$^{\circ}$) with sky-localization smaller than a threshold $(\Omega)$\footnote{The sum is done over events with sky-localization smaller than a threshold $(\Omega)$ in order to obtain the cumulative number of VHE counterparts as a function of this threshold. This threshold can be used in real observations to decide what GW triggers to be followed} and by assigning to each on-axis event a weight based on its sky-localization, FoV/$\Omega_i$ (its probability to be detected with one-shot observation decreases for larger sky-localization)}:

\begin{equation}
\begin{split}
{\rm N_{VHE} = \sum_{i=1}^{N_{\theta_{\rm v} < 10^{\circ}}(< \Omega)} \frac{FoV}{\Omega_i} \times {D.C.}} \rm \times{CTA_{vis}}
\end{split}
\label{VHEN}
\end{equation}
for each event with ${\rm \Omega_i < FoV}$, the fraction $\rm  {FoV}/{\Omega_i}$ is set equal to 1. D.C. is the CTA duty cycle of 15\% and $\rm CTA_{vis}$ the CTA visibility of 50\%.
While the cumulative distribution of on-axis events $\rm N_{\theta_{\rm v} < 10^{\circ}}$ as a function of sky-localization is given by the lower panels of Fig.~\ref{fig:BNSrateSN3}, the cumulative distribution of the expected number of possible VHE counterpart detection N$_{\rm VHE}$ as a function of sky-localization for the one-shot observation strategy is given by the 
plots in the first column of Fig.~\ref{fig:Fig7MSTPrediction} and \ref{fig:FigLSTPrediction} for MST and LST, respectively. 
By looking at these plots together with those of CTA time (second column), it is possible to estimate the expected number of possible VHE detections following all events with a sky-localization below a certain threshold and the corresponding amount of CTA time required.

Following pre-merger alerts of 15 minutes, we expect to detect around 1 VHE possible counterpart per year by CTA-MST operating with the network of ET and CE (ET+CE and ET+2CE). This number increases to around 10 possible VHE counterparts per year for ET+CE (ET+2CE) following pre-merger alerts of 5 minutes with sky-localization smaller than $\rm 10^3 deg^2$ by using 25\% (40\%) of time of CTA-MST operating with ET+CE (ET+2CE).  

Using CTA-LST, we do not expect detection even with ET+2CE following all the GW events with 15 minutes pre-merger alert. However, we expect around 3 (5) possible detections per year triggered by ET+CE (ET+2CE) using 10\% (20\%) of the CTA time budget and following all the events with sky-localization of $\rm 10^3 deg^2$. Around ten possible VHE counterparts are expected by following   the 1-minute pre-merger alerts with sky-localization smaller than about $\rm 200 deg^2$ detected by ET+CE at the expense of 20\% of the CTA time. Twenty possible VHE detections are expected for ET+2CE by following all the events with sky-localization smaller than $\rm 10^2 deg^2$ detected by ET+CE at the expense of about 10\% of the CTA time. Only following pre-merger alerts of 1 minute can give a few detections for ET as a single observatory or operating in the network ET+LVKI+ and LVK+2VOY. All these numbers are obtained using the fiducial population and considering all BNS launching a successful jet. However, as written in the previous section, on the basis of the current studies only a fraction $20-50\%$ is expected to produce a jet. 

This research can benefit from the reduction of CTA time to be devoted to the follow-up of events while maintaining the same VHE detection efficiency. 
For example, {the viewing angle is estimated from the GW signals in low-latency and it can be used to remove all the off-axis systems from which the VHE emission is not expected.} This makes it possible to reduce the number of events to be followed up by CTA, and thus the CTA time to be spent on this search. {Figure~\ref{fig:FigdTvsT} in the Appendix shows the uncertainty on the viewing angle coming from the analysis of the GW observations as a function of injected $\theta_{\rm v}$ of the BNS system. It can be inferred from the figure that the observed uncertainties on the viewing angle are large and in particular are larger for smaller viewing-angle. Therefore, it is not possible to directly select on-axis events ($\theta_{\rm v} < 10^{\circ}$), but based on the smaller errors on larger viewing-angles, it is safer and more effective to exclude from the follow-up off-axis events. We choose an arbitrary threshold on $\theta_{\rm v} = 45^{\circ}$ which enables us to exclude a large number of off-axis events and to limit the number of excluded on-axis events.} The plots in the right column of Fig.s~\textcolor{red}{7} and \ref{fig:FigLSTPrediction}, show the percentage of CTA time to follow all the events with $\theta_{\rm v}<$  45$^{\circ}$. This selection based on the observed $\theta_{\rm v}$ reduces the total follow-up time by about a factor of 3. As also shown in \citet{Ronchini2022}, to optimize the observational strategy and increase the efficiency of the search in the era of 3G detectors, it will be critical to send information on the source parameters beyond distance and sky-localization which are the only source parameters currently sent in low-latency for LIGO, Virgo, and KAGRA event candidates. 

In addition, we also evaluated the possibility to use updated information on the source parameters; following up the pre-merger alerts of 15 minutes or 5 minutes, we use the updated sky-localization available 1 minute before the merger (see MST-c and MST-d cases in Table~\ref{table:obsstrategyMST}, and LST-d and LST-e in Table~\ref{table:obsstrategyLST}). For this scenario, we consider to followed-up only 
{events with} 
sky-localization smaller than 1000 deg$^2$ during the initial alerts. Figure \ref{fig:Update1min} shows the improvement of the sky-localizations over time from sky-localizations at 15- or 5 minutes to 1 minute. The updated sky-localization at 1-minute clusters around 100 deg$^2$ for the cases of ET+CE and ET+2CE for both 5 and 15-minute scenarios. Although this strategy offers a significant improvement, it counts on a very rapid communication/response to the updated alert and a possible rapid repositioning. Any delay in the response or slewing of CTA could be problematic. Figure \ref{fig:NdetReal} shows the results for this observational strategy providing the possible VHE detections
by MST (left column) and LST (right column). \newline
For CTA-MST, the expected number of possible VHE counterparts using updated information on the 1-minute pre-merger alert sky-localization are 2.5-3 (20-40) events when alerted by ET+CE and ET+2CE 15 minutes (5 minutes) before the merger. These numbers compare to around 1 (10) detections if the 1-minute sky-localization update is not used (see plots on the left column of Fig.~\textcolor{red}{7}). 
The use of updated information on sky-localization can significantly increase the efficiency of this search. 
For LST and 5 minutes pre-merger alerts, 10-20 possible VHE counterparts are expected with CTA operating with ET+CE and ET+2CE. These numbers compare with the few detections expected using the one-shot observation over the sky-localization obtained 5 minutes before the merger (see the central plot on the left column of Fig.~\ref{fig:FigLSTPrediction}). These numbers are comparable to the ones of following up 1-minute pre-merger alerts (see the bottom plot on the left column of Fig.~\ref{fig:FigLSTPrediction}), but this strategy of following up the 5-minute pre-merger alerts and the updated 1-minute sky-localization is safer and it requires net less observational time. 

In this analysis, using the results from a Fisher matrix code, we assume a uniform localization probability distribution among the 90\% credible region. However, the full Bayesian parameter estimation (which is in development for 3G detector era and will be used in low-latency as currently done with LIGO, Virgo, and KAGRA) gives the localization probability in each position of the sky. Thus, this search can be refined and made more efficient by starting the observation from the most probable region of the sky-localization and evaluating the actual probability enclosed within the one-shot observation (namely within the CTA FoV). The formalism described in this section can be used also for other EM observatories by changing the FoV, duty cycle, and visibility.   

\subsubsection{Divergent pointing}
\label{divpoint}
Searching for VHE counterparts can significantly benefit from a larger FoV, which can increase the coverage of sky localization of the GW signals. 
One way to increase the FoV of CTA is to use divergent pointing \citep{2015ICRC...34..725G, 2019ICRC...36..664D, 2022Galax..10...66M}; taking advantage of many telescopes that can point slightly offset from each other the FoV
can become larger by a factor of at least 4-5 at the expense of sensitivity and angular resolution. With the help of an offset alignment of 3$^\circ$ (4$^\circ$), a FoV of 150 (250) deg$^2$ can be achieved with 19 MST as described in \citet{2019ICRC...36..664D}. The angular resolution reduces down to around 0.2$^\circ$ from the target MST angular resolution, whereas the sensitivity of the array worsens by about 20-25\% \citep{2015ICRC...34..725G} in the core energy range. Figure~\ref{fig:DivergentPointing} shows the one-shot strategies using a FoV of 100 deg$^2$. Using the divergent pointing can lead to a total possible detection of 60 (4) per year 
at the expense of 10\% (less than 1\%) of MST time following up events with sky-localization up to 1000 deg$^2$ for the pre-merger alert case of 5 minutes (15 minutes) and following-up only events with $\theta_{\rm v}<45^{\circ}$. These numbers compare to 10 (1) obtained with MST FoV of 30 deg$^2$ at the same amount of MST time expense.

\subsubsection{Mosaic strategy} 
MST is best suited for mosaic strategy due to a three-fold larger FoV as compared to LST. The proposal for this scenario is the following: in order to cover a sky area of 100 deg$^2$, we consider three pointing of MST which requires a total of 60\,s. The slew between these three pointing requires $\sim$ 5\,s, considering the slew time of MST to be 90\,s to move to any points in the visible sky. Considering the same strategy of using pre-merger alerts and being on source at the merger time, the mosaic strategy maintains the same detection efficiency for short ($< 20 \rm sec$) prompt/early VHE emission but it requires a 30\% larger amount of CTA time with respect to the one-shot observation. However, this strategy becomes more efficient with respect to the one-shot observation strategy for longer signals such as the afterglow emission. 

With respect to the divergent pointing the mosaic strategy has the advantage of not reducing the sensitivity. However, the divergent pointing has the significant advantage of the larger FoV; to cover the same area the mosaic strategy needs more observational time.
Knowing the emission properties, in particular, the expected flux decaying, it would be possible to precisely compare the mosaic and divergent observational strategies (by assuming a larger exposure for detecting also signals longer than 20 s). The emission properties are largely uncertain to make precise estimates. 

It is worth noting that in case of longer signal, the mosaic strategy can also benefit by the detection of the classical GRB prompt emission in the KeV-MeV by high-energy satellites able to localize the source \citep{Ronchini2022}.

\begin{figure*}[t]
\centering
            \includegraphics[width=0.33 \linewidth, height=7cm]{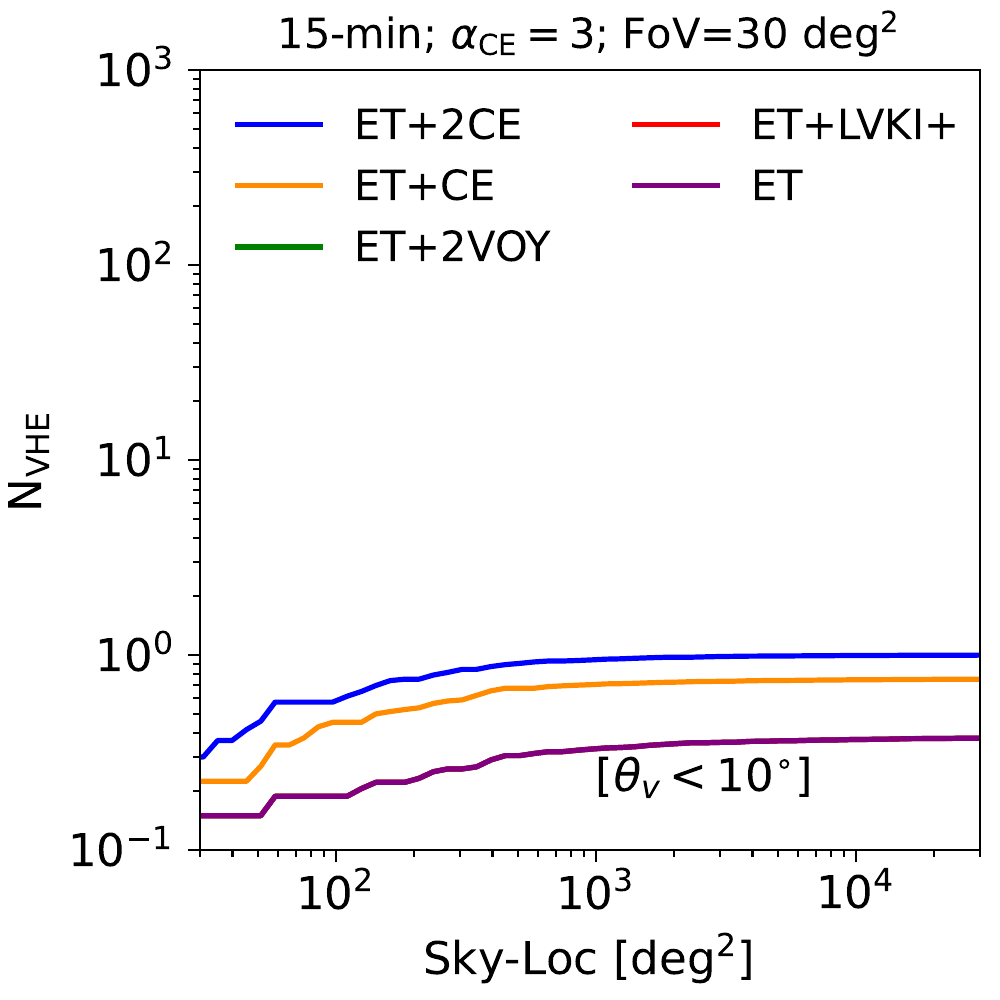}
            \includegraphics[width=0.33 \linewidth, height=7cm]{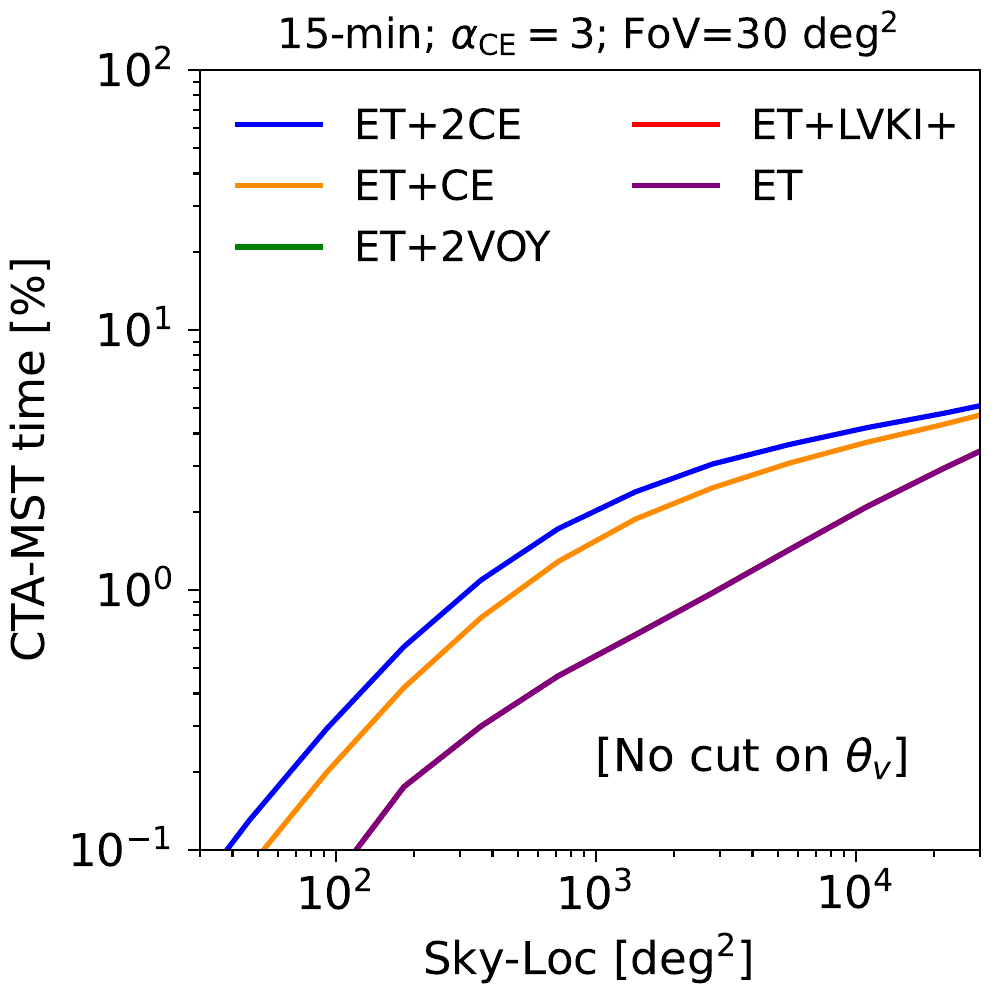}  
            \includegraphics[width=0.33 \linewidth, height=7cm]{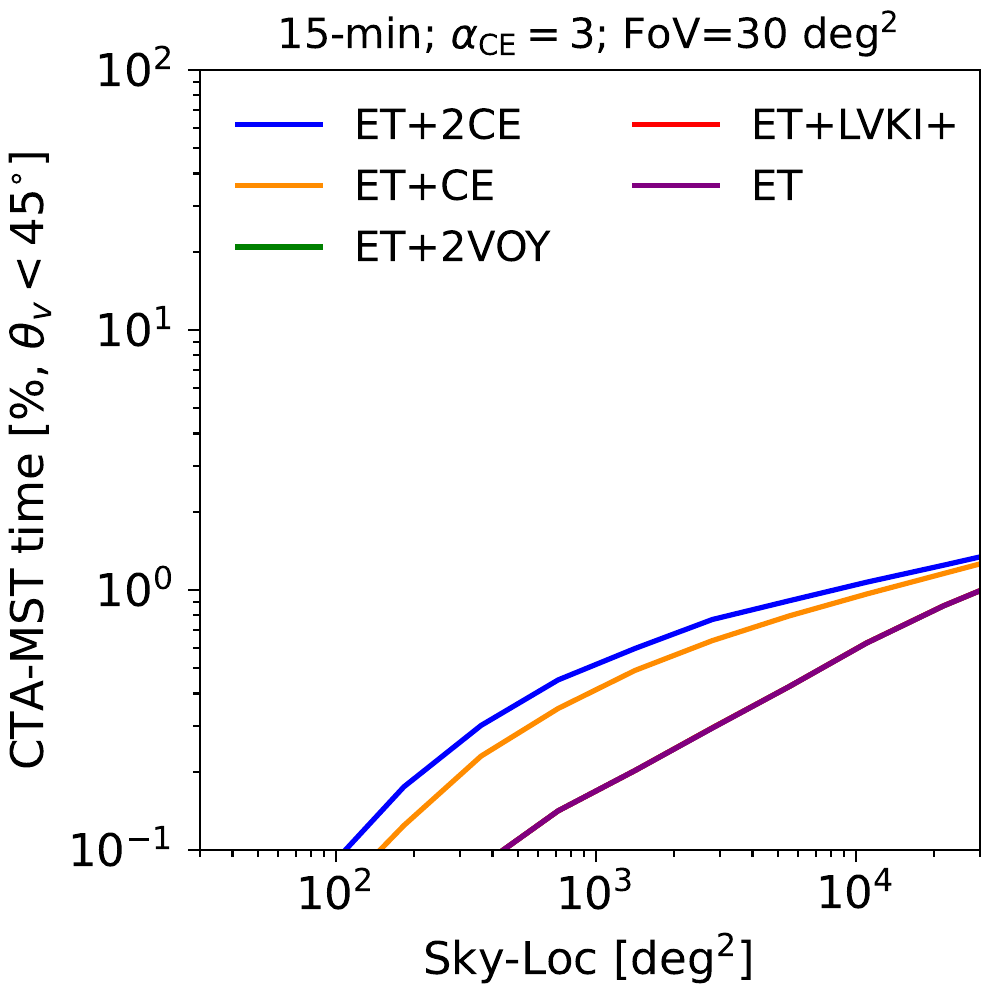}  
            \includegraphics[width=0.33 \linewidth, height=7cm]{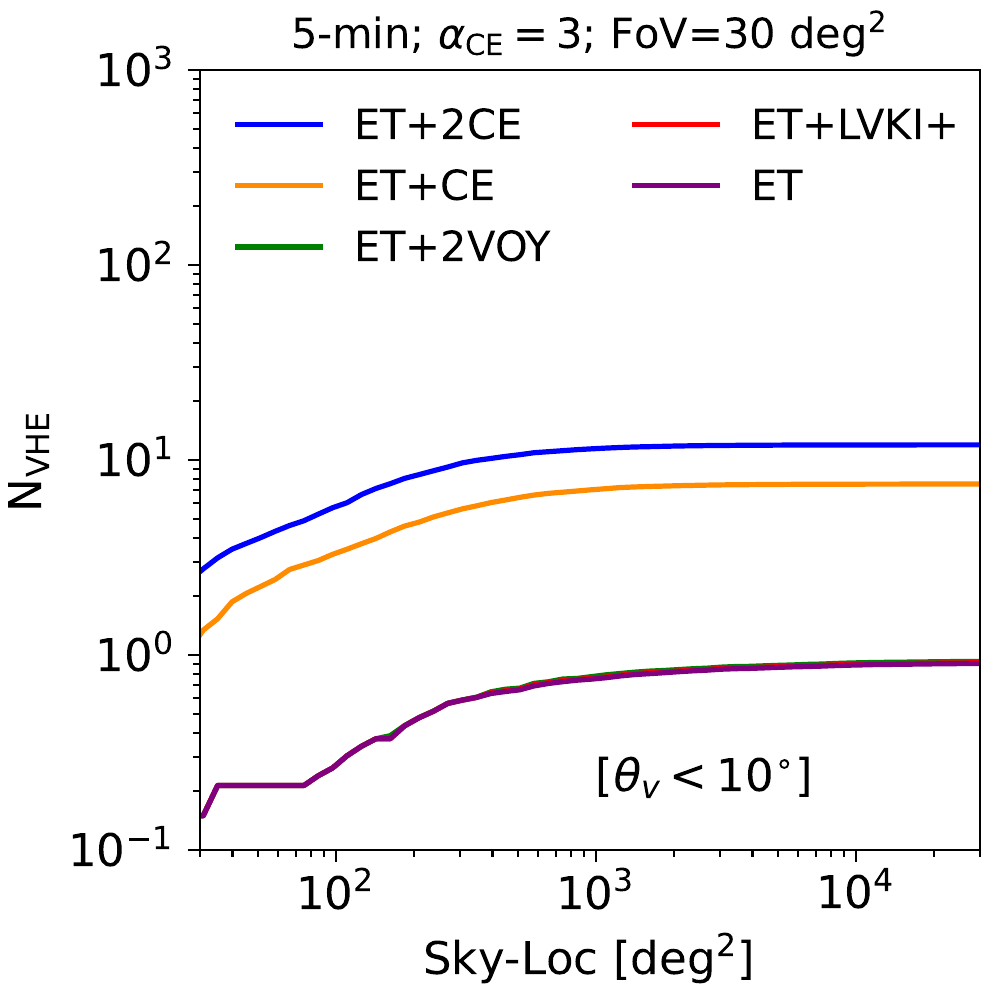}
            \includegraphics[width=0.33 \linewidth, height=7cm]{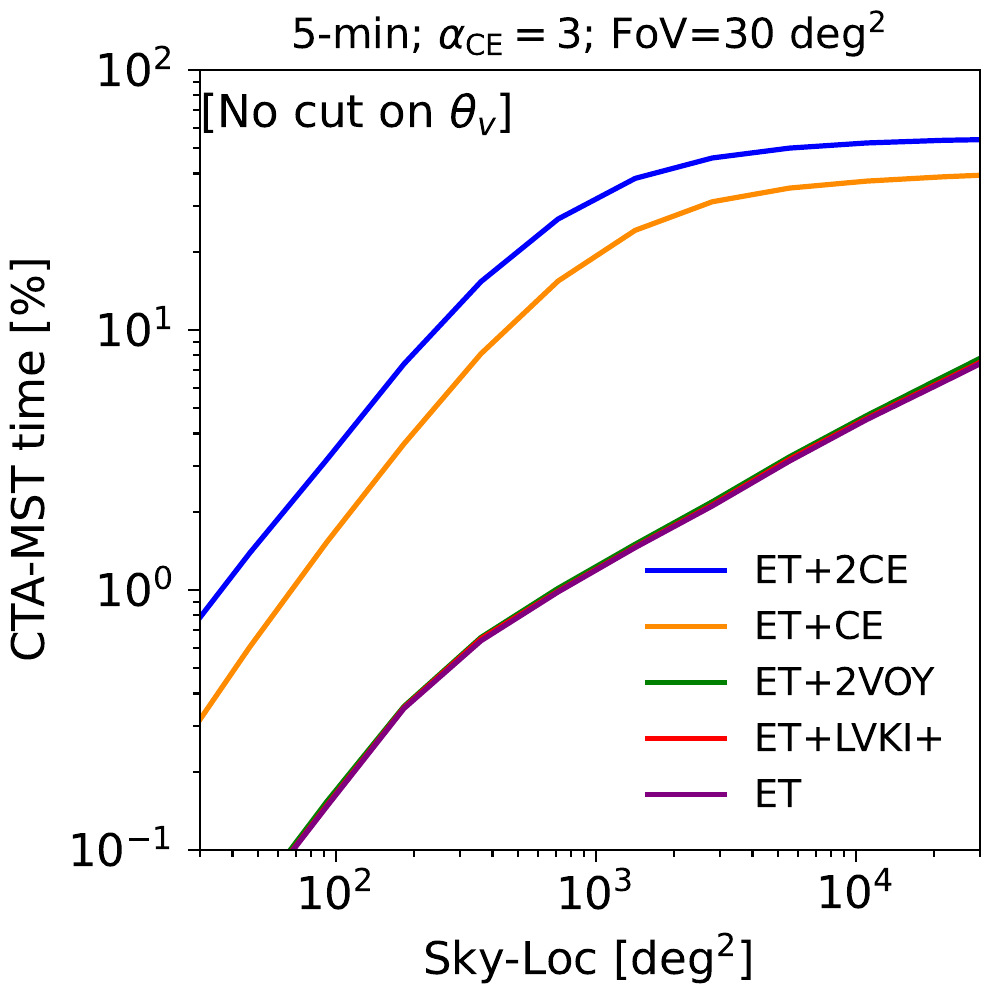}
            \includegraphics[width=0.33 \linewidth, height=7cm]{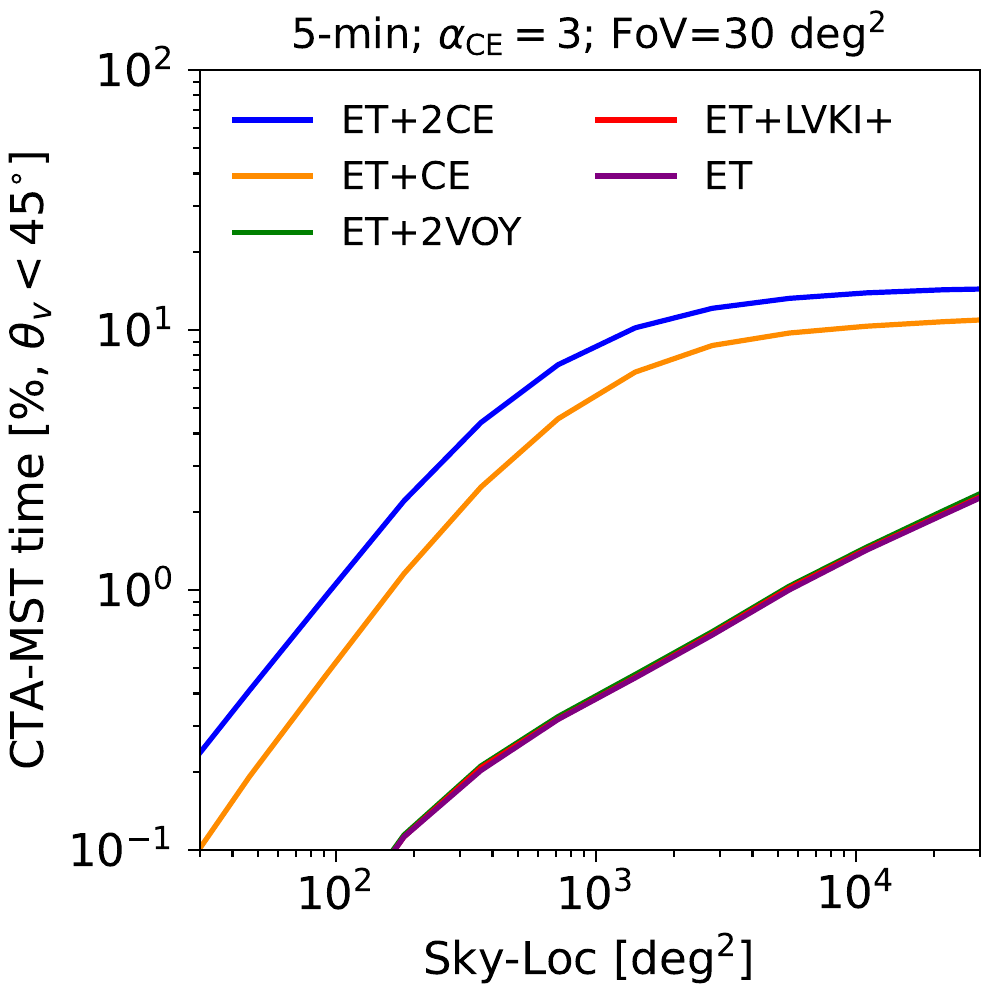}
            \caption{The expected numbers of detection by CTA-MST  using the one-shot observation strategy are shown by the plots in the left column. {The estimates of the number of possible VHE counterparts are based on the on-axis BNS systems (systems injected with $\theta_{\rm v} < 10 deg$ in our simulations) and evaluated as described in the text (see Eq.~\ref{VHEN})}. These estimates assume that all BNS produce a jet. They take into account the sky-localization ($\rm \Omega_i$) of each event, the MST field of view of 30\,deg$^2$, the CTA duty cycle of 15\%, and the CTA visibility limited to a zenith angle larger than 60$^\circ$ (minimum elevation of 30$^\circ$). The fraction of CTA time spent following all the GW alerts with the sky-localization smaller than the one indicated in the x-axis is given by the plots in the central column. For observing each event (independent of the sky-localization), we consider the observational time (t$_{\rm obs}$) given by the sum of slewing time (t$_{\rm slew}$=90\,s), an additional re-positioning time (t$_{\rm rep}$=10\,s), and the exposure time (t$_{\rm exp}$=20\,s). The plots on the right column show the CTA time when only triggers with {an observed $\theta_{\rm v}$<45$^{\circ}$} are followed up resulting in a significant reduction of CTA time.}
            \label{fig:Fig7MSTPrediction}
        \end{figure*} 
\begin{figure*}
\centering

            \includegraphics[width=0.33 \linewidth, height=7cm]{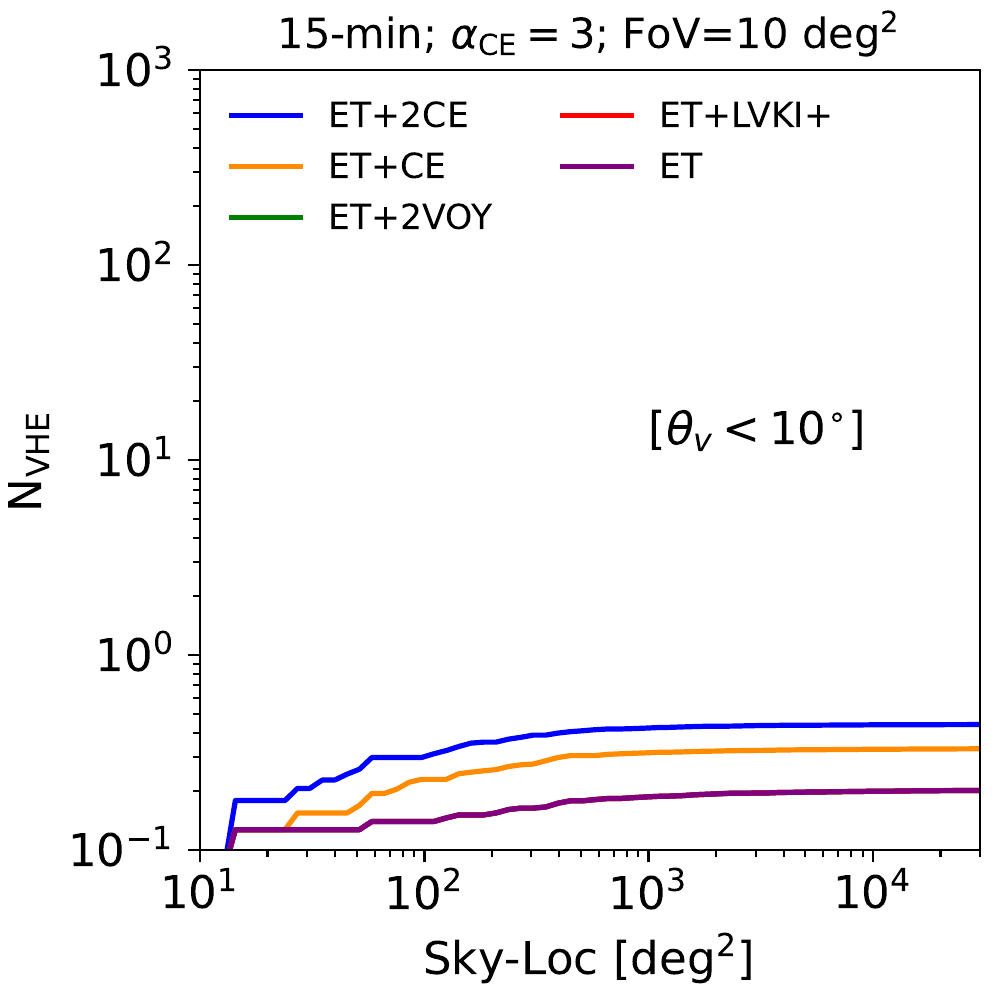}
            \includegraphics[width=0.33 \linewidth, height=7cm]{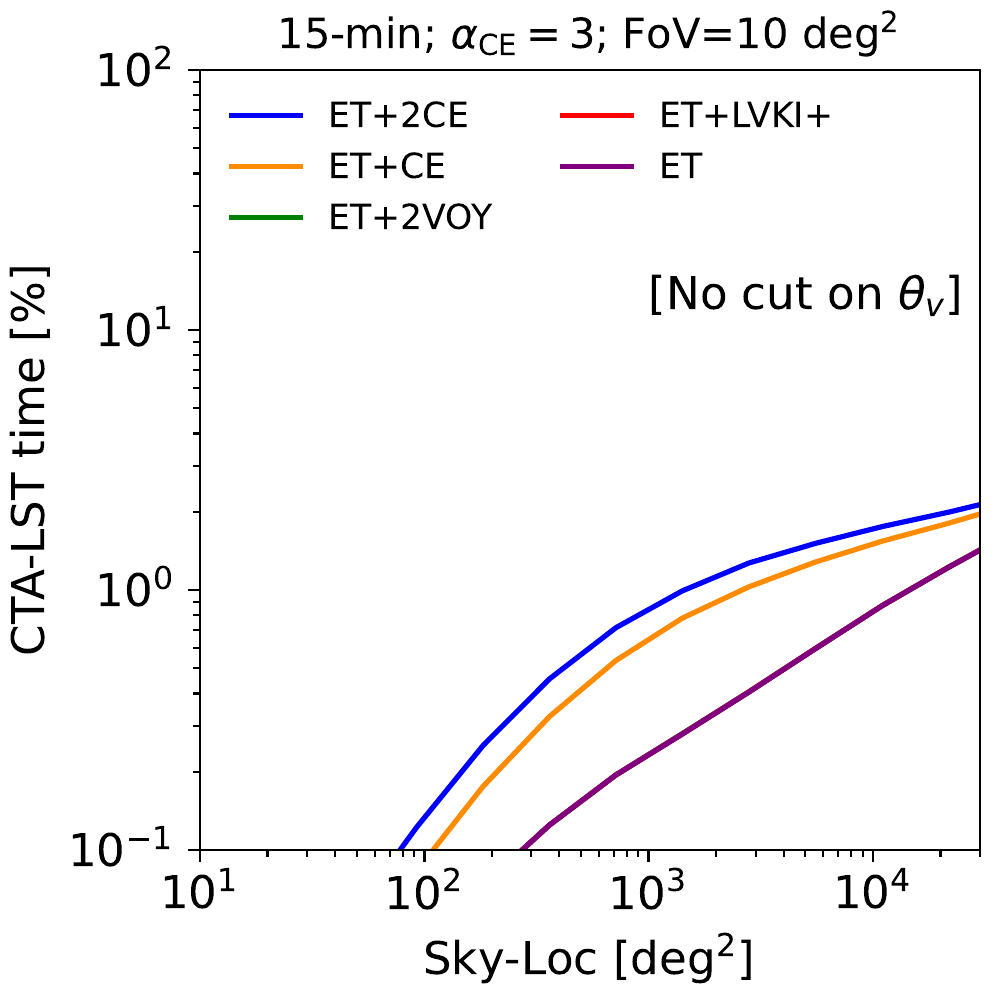}
            \includegraphics[width=0.33 \linewidth, height=7cm]{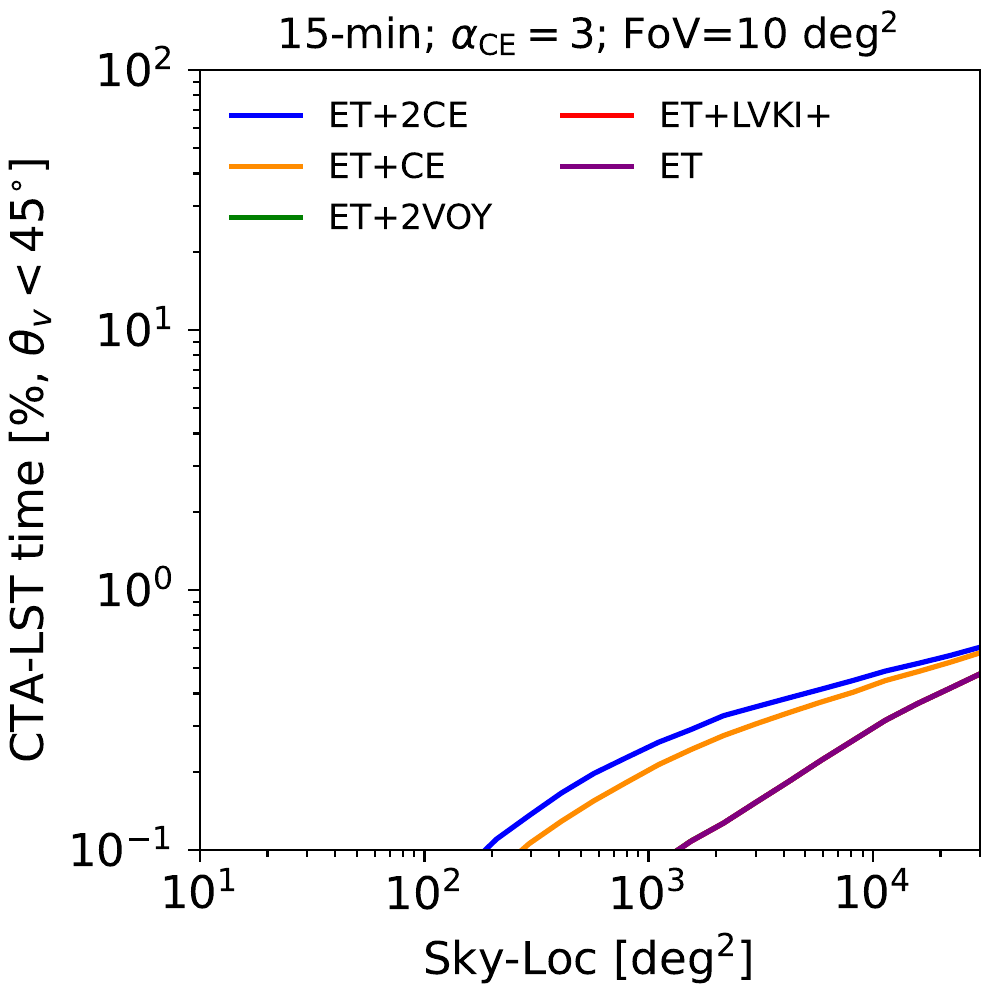}
            
            \includegraphics[width=0.33 \linewidth, height=7cm]{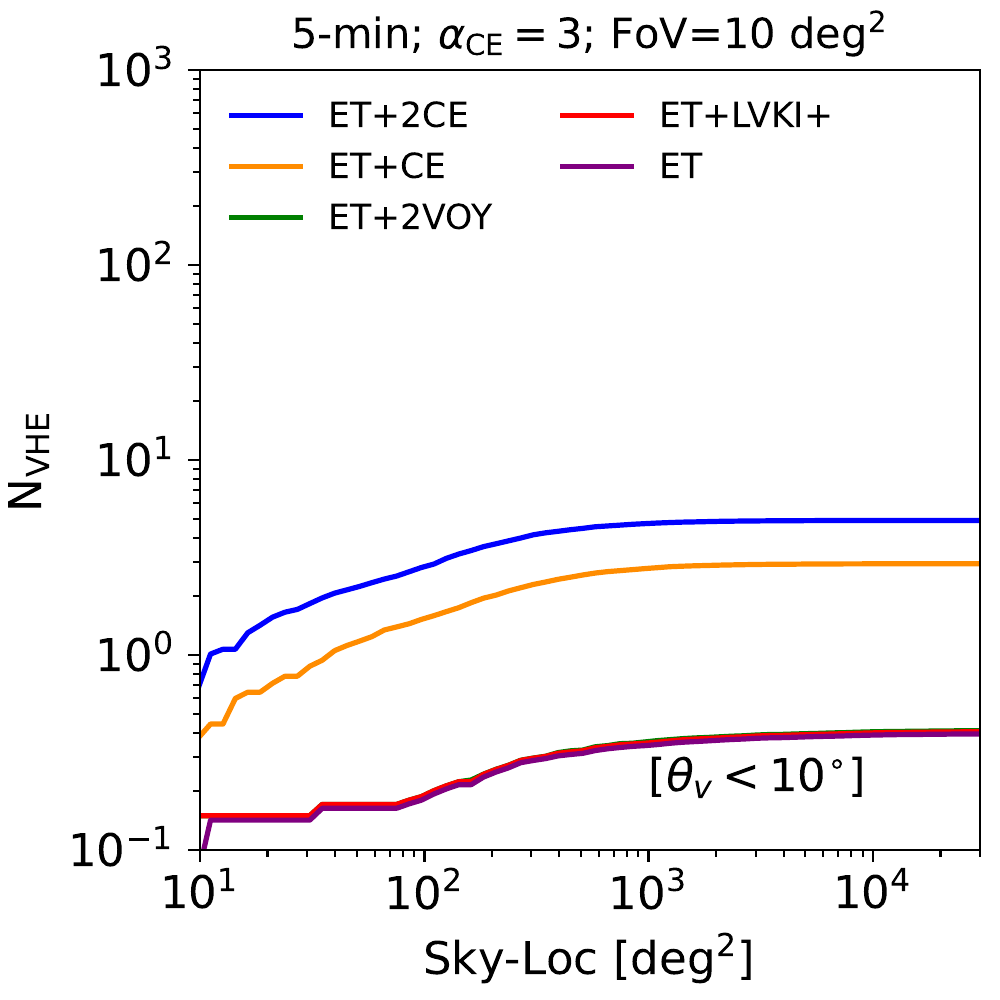}
            \includegraphics[width=0.33 \linewidth, height=7cm]{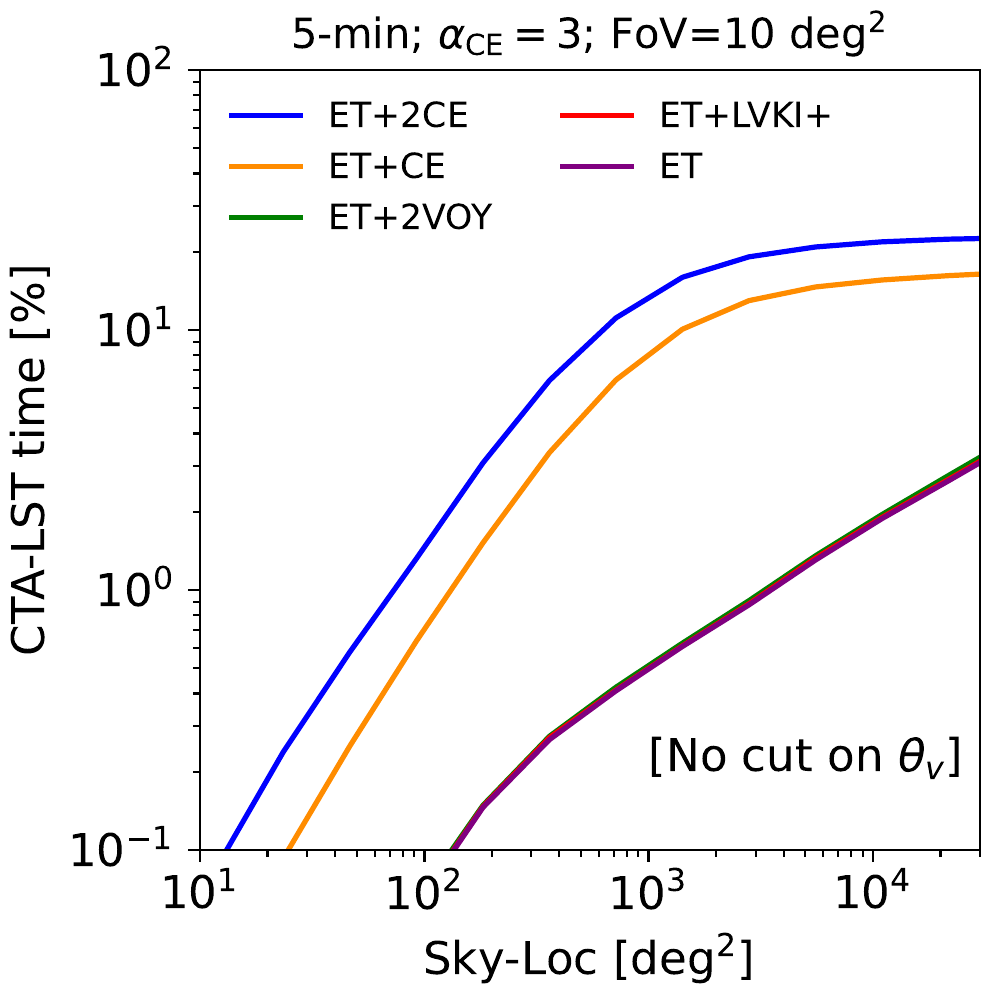}
            \includegraphics[width=0.33 \linewidth, height=7cm]{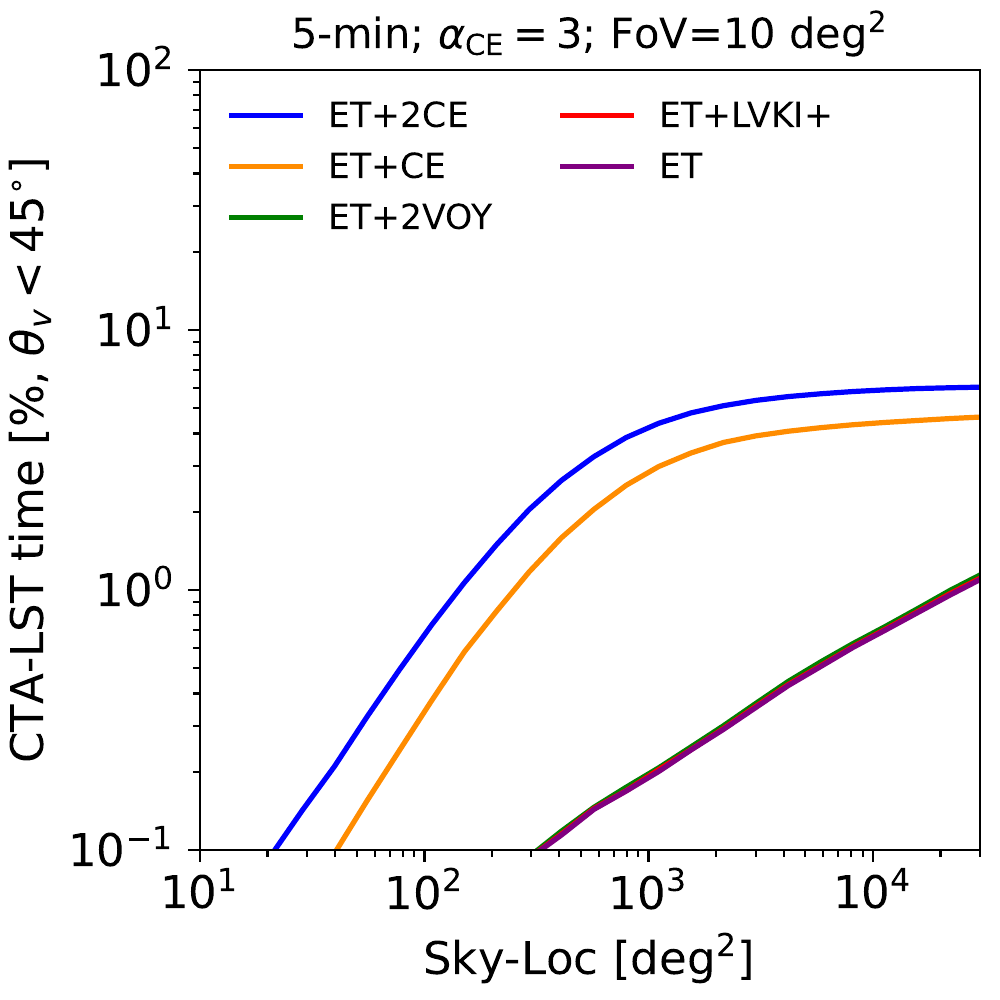}

            \includegraphics[width=0.33 \linewidth, height=7cm]{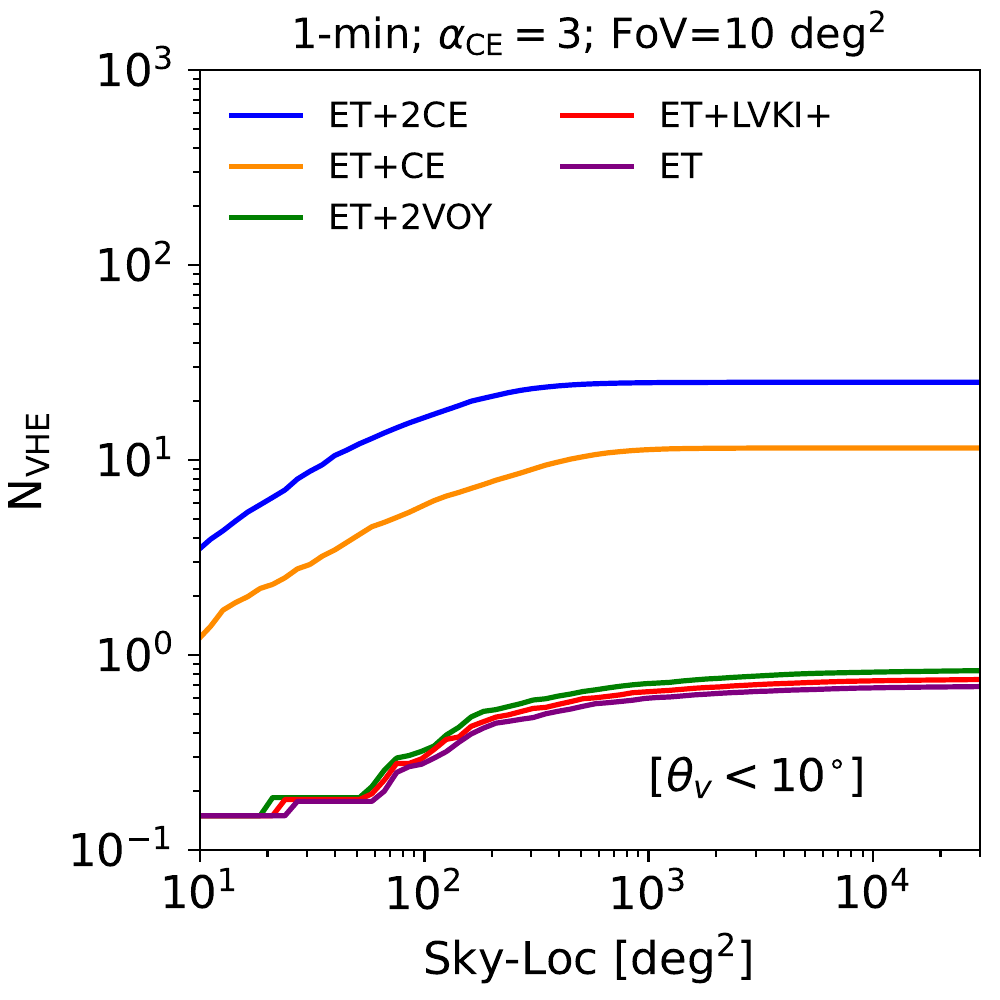}
            \includegraphics[width=0.33 \linewidth, height=7cm]{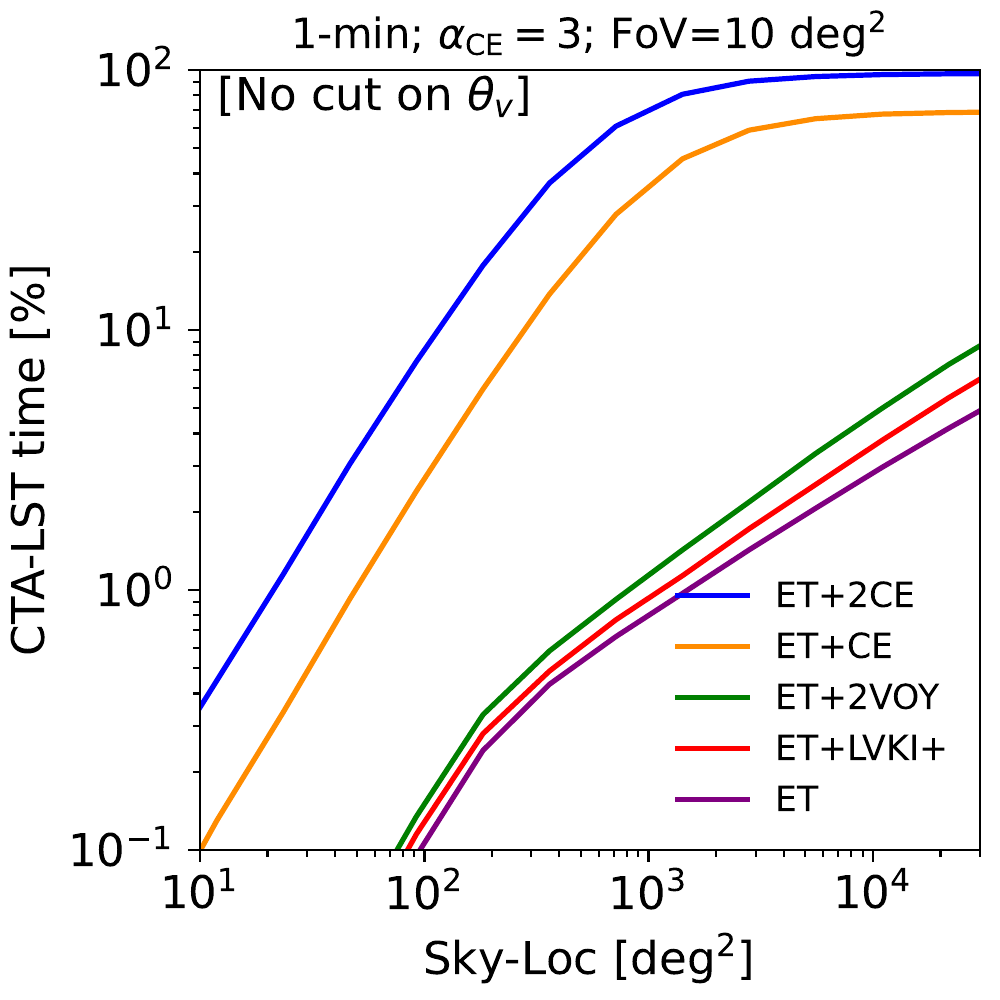}
            \includegraphics[width=0.33 \linewidth, height=7cm]{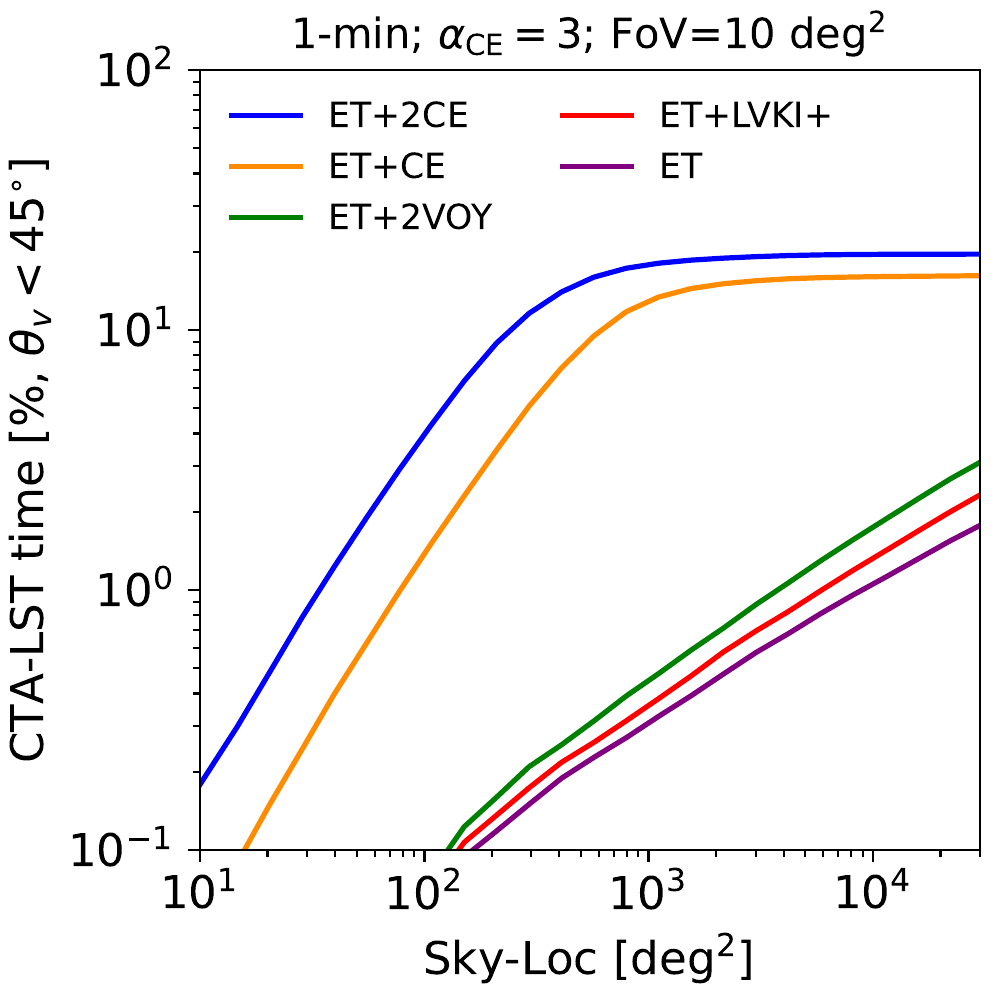}
            
            \caption{Same as Fig. \textcolor{red}{7} but considering LSTs. We use a FoV of of $\sim$10 deg$^2$ and observational time for each event  (t$_{\rm obs}$) of 50 \,s given by the slewing time (t$_{\rm slew}$) of 20\,s, an additional repositioning time of 10\,s, and the exposure time (t$_{\rm exp}$) of 20\,s. Thanks to the rapid slew time, for LST we show also the scenario of directly following the 1-minute pre-merger alerts.}
            \label{fig:FigLSTPrediction}
        \end{figure*} 

\begin{figure*}
\centering
            \includegraphics[width=0.32\linewidth, height=5cm]{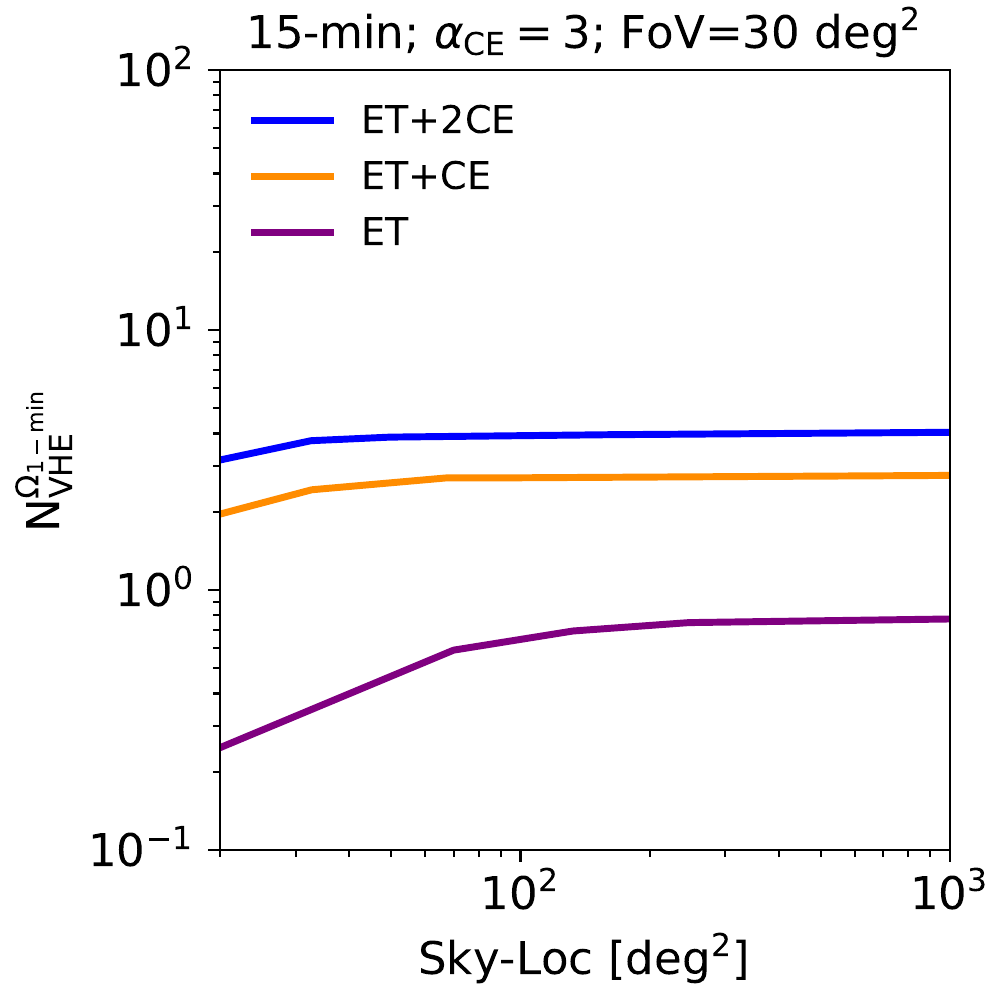}
            \includegraphics[width=0.32 \linewidth, height=5cm]{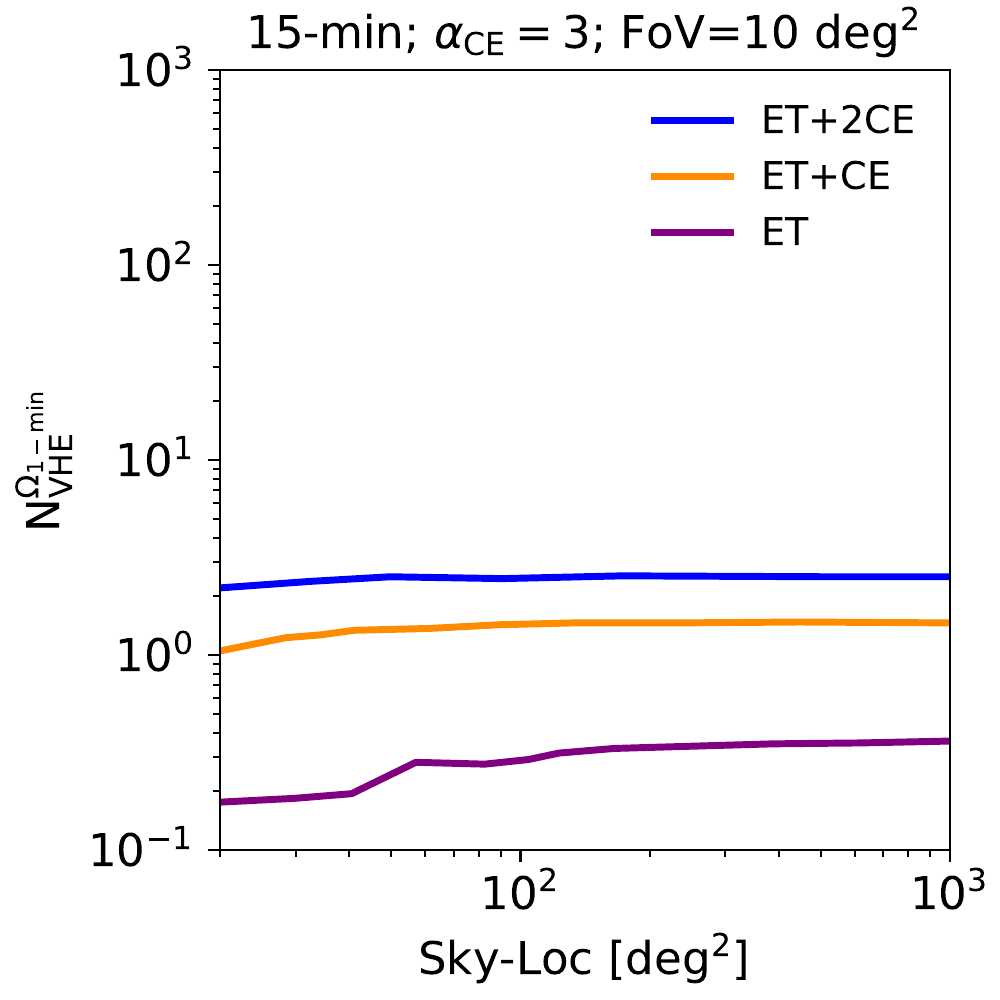}
            
            \includegraphics[width=0.32\linewidth, height=5cm]{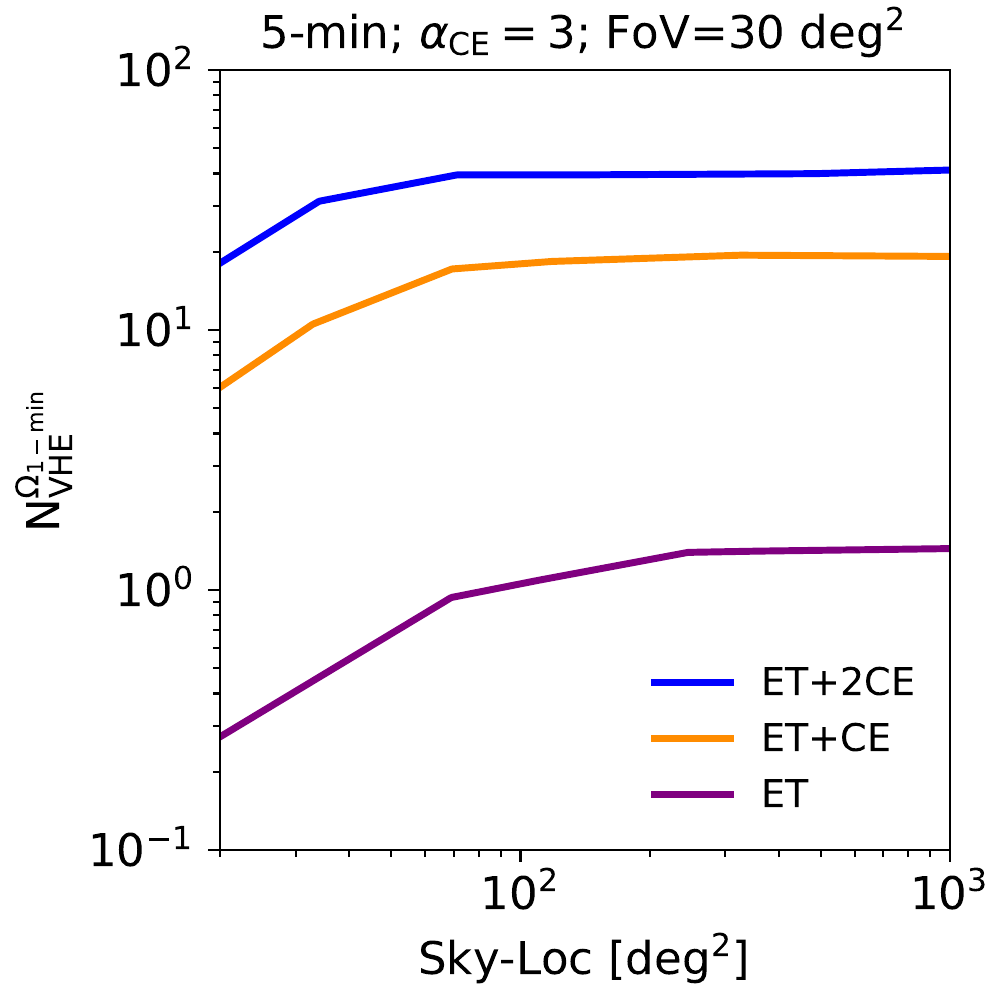}
            \includegraphics[width=0.32\linewidth, height=5cm]{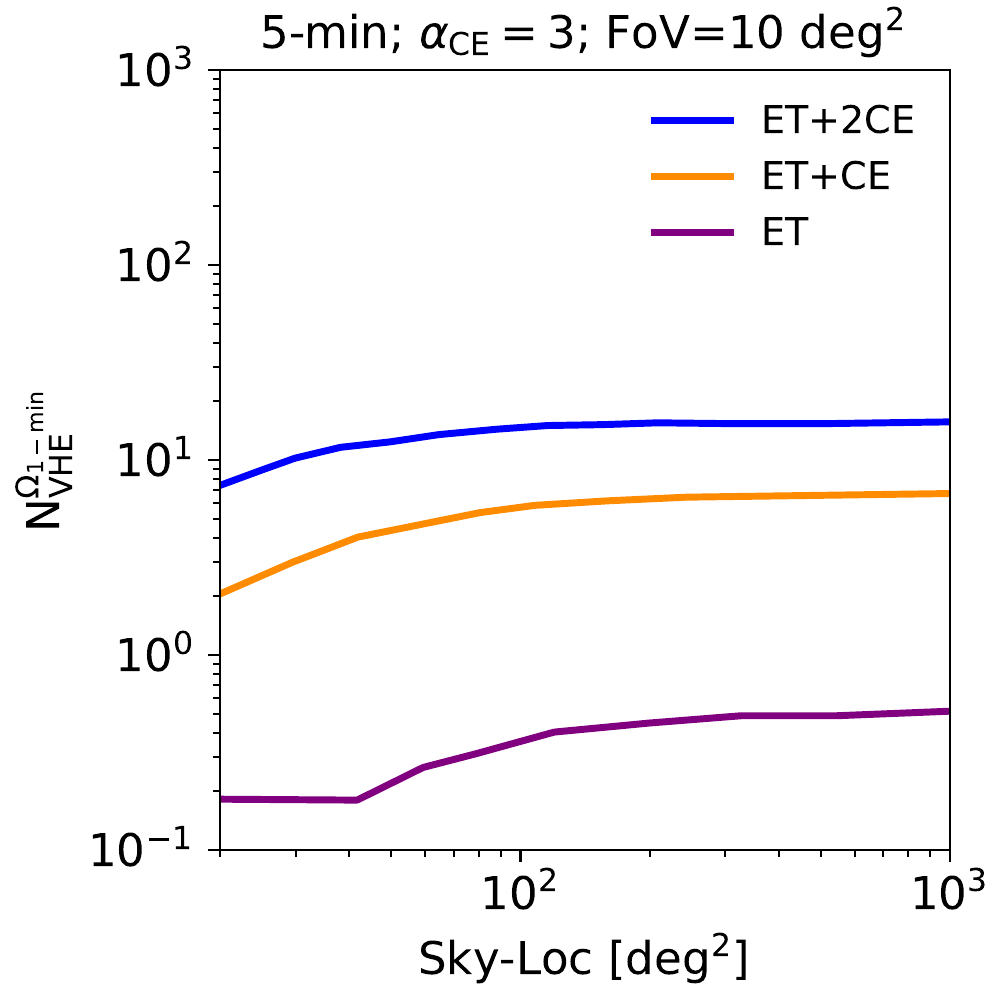}
            \caption{Same as Fig.s \ref{fig:Fig7MSTPrediction} and \ref{fig:FigLSTPrediction} by following-up 15 and 5 minutes pre-merger alerts but using the updated sky-localization obtained one minute before the merger. For this scenario, we consider only events detected 15 minutes or 5 minutes before the merger with sky-localization less than 1000 $\rm deg^2$. 
            The expected number of VHE possible detections includes the CTA visibility, the fact that LST or MST antennas 
            {are not able to observe below the}
            elevation angle of 30$^\circ$ 
            {(zenith larger than 60$^{\circ}$)}. The plots in the right column show the LST array and the plots in the left column the MST array.}
            \label{fig:NdetReal}
        \end{figure*} 
        
\begin{figure*}
\centering
            \includegraphics[width=0.32 \linewidth, height=5cm]{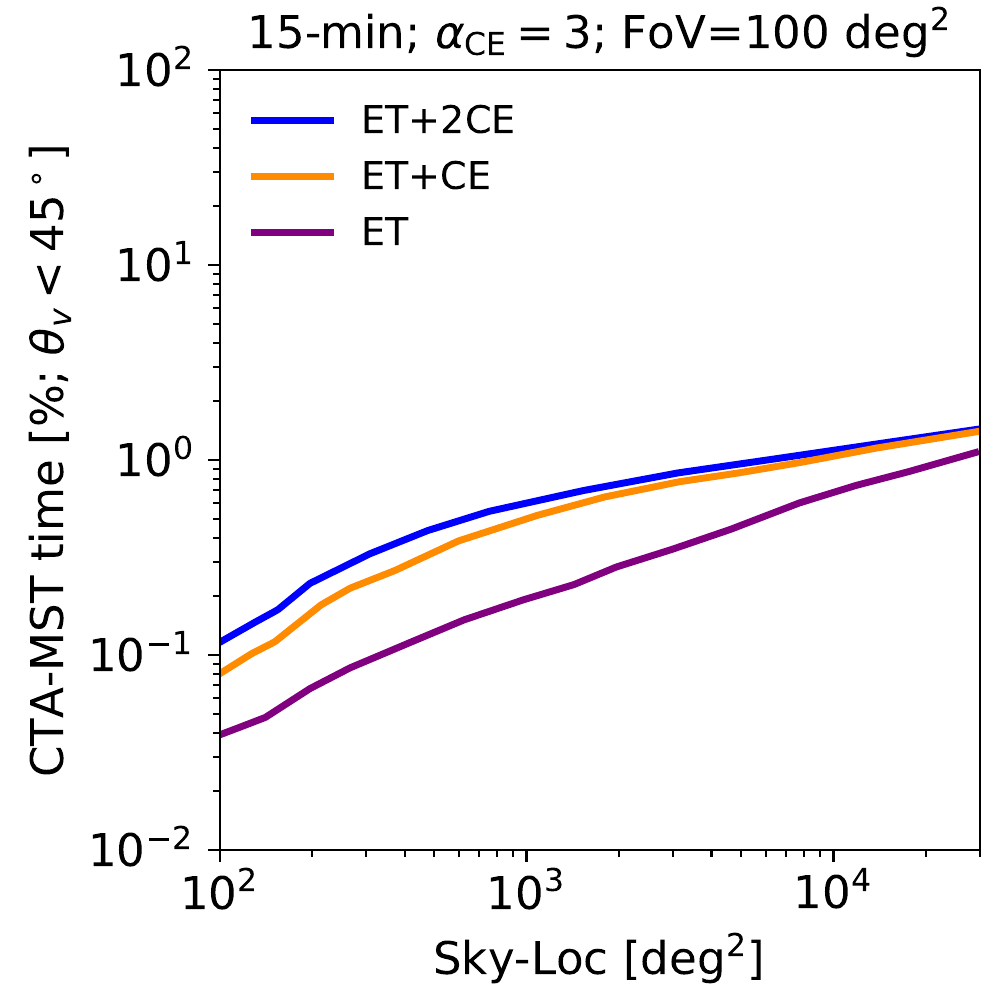}
            \includegraphics[width=0.32 \linewidth, height=5cm]{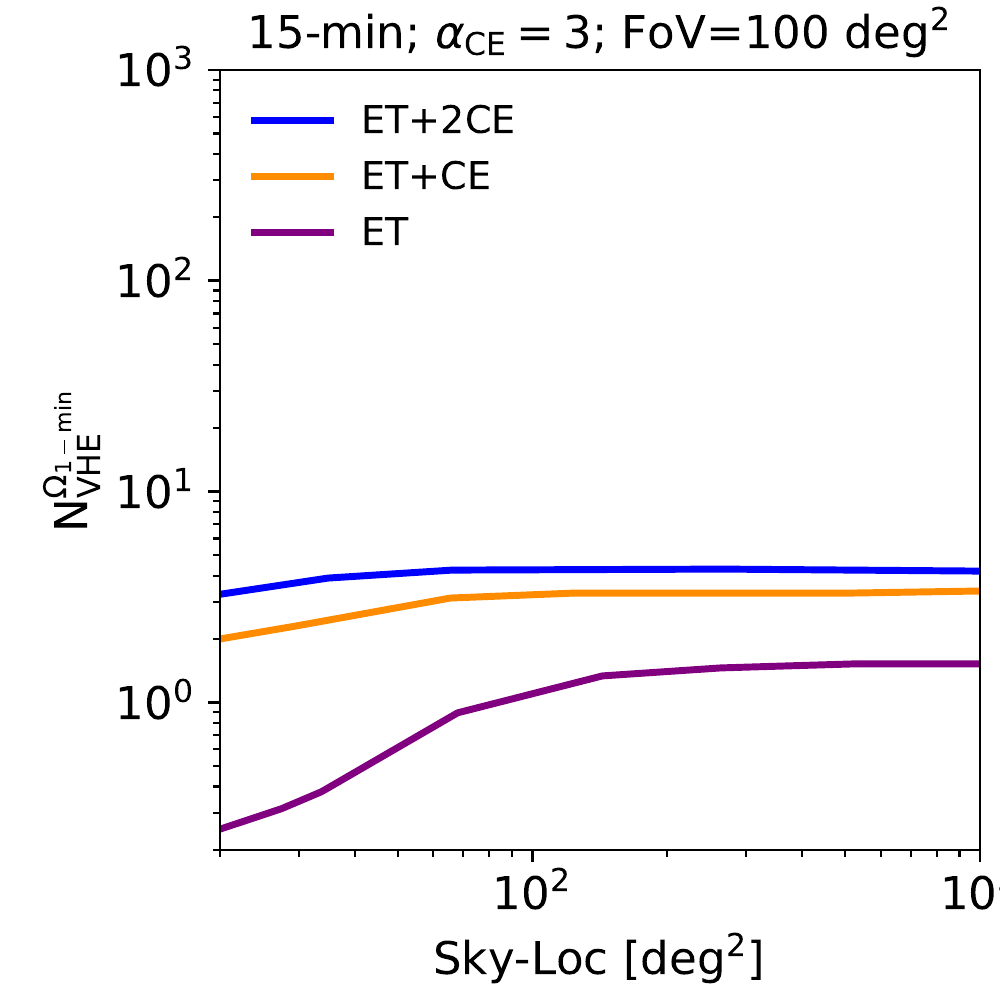}
            
            \includegraphics[width=0.32 \linewidth, height=5cm]{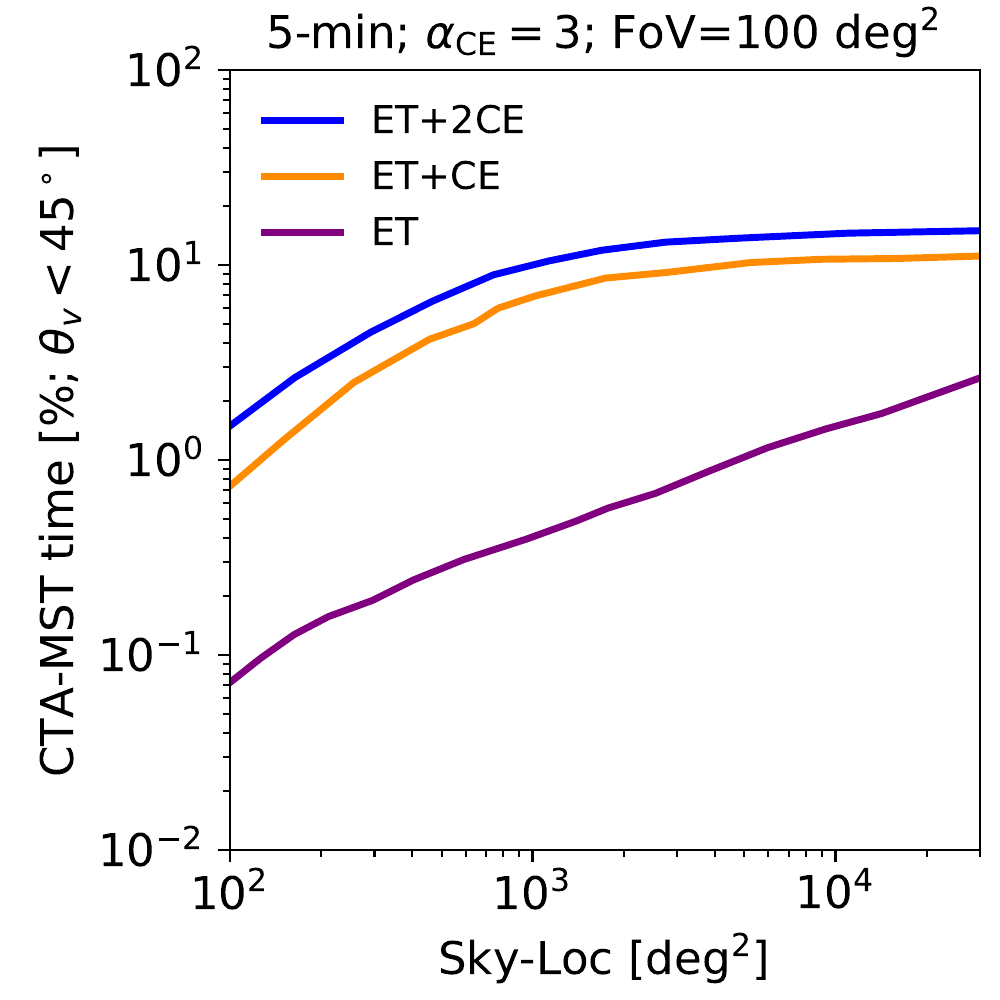}
            \includegraphics[width=0.32 \linewidth, height=5cm]{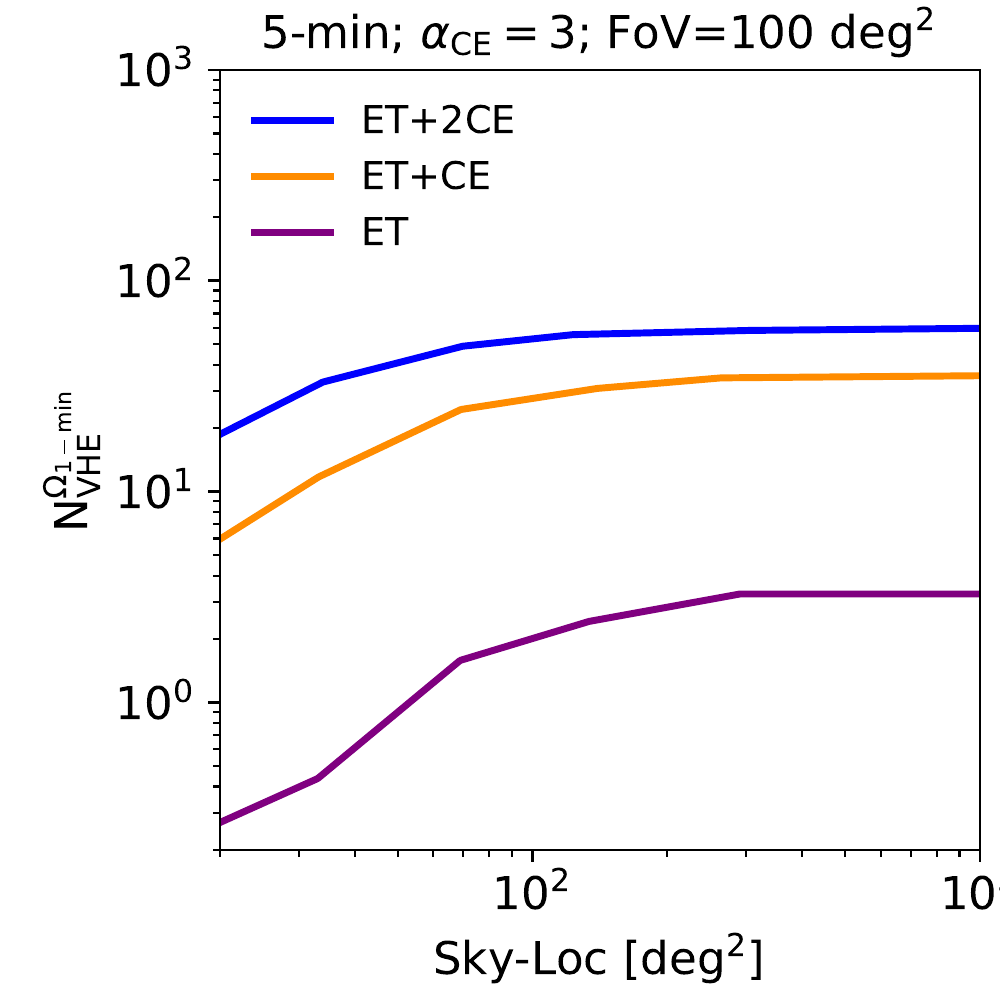}
            \caption{Same as Fig. \ref{fig:Fig7MSTPrediction} with a FoV of 100\,deg$^2$ using CTA-MST divergent pointing operation, where the individual telescopes in a subarray 
            {are} 
            pointed with an offset (i.e., 3$^\circ$) to obtain a larger FoV.}
            \label{fig:DivergentPointing}
        \end{figure*}         

\section{Models producing early TeV emission}\label{sec:theory}

\subsection{GRB prompt emission}

The prompt emission dissipation models are very uncertain (see \cite{Piran2004},\cite{Kumar2015}, and \cite{ZhangBook} for a review). This is partially due to the fact the dominant radiative processes responsible for the 
observed GRB spectra are not identified. Both time-resolved and time-integrated spectra in the 10\,keV--10\,MeV energy band are typically well accounted for by two power laws smoothly connected at their peak energy $E_{\gamma}$ of $\nu F_{\nu}$ \citep{Band1993}. The photon index below $E_{\gamma}$ has a typical value of -1 for long GRBs and -0.7 for short GRBs (e.g. see \citet{Nava2012}). The spectra with these photon indices physically are harder than simple fast-cooling synchrotron emission spectra, and they are much softer than thermal spectra \citep{Preece1998,Ghisellini2000}. One can generally divide the prompt emission models into those that invoke standard synchrotron-based models with dissipation occurring above the GRB jet photosphere and those models that invoke sub-photospheric dissipation. The most discussed model is the internal shocks model, which suggests an internal dissipation of a jet above the photosphere with a Lorentz factor gradient. This model is based on the assumption that the jet is dominated by kinetic energy. Alternatively, the GRB jet could be highly magnetized and the dissipation may occur via magnetic reconnection. Due to the huge uncertainty in these models and the absence of a clear preference for one over another, we are forced to rely on simplified models that can account for the basic spectral and temporal features. 

The most standard model for prompt emission assumes that synchrotron radiation from non-thermal electrons makes the GRB emission. It has become more clear that in order to explain the GRB spectra by the synchrotron model, one is forced to assume a marginally or slow cooling regime of radiation \citep{2009A&A...498..677B, 2008MNRAS.384...33K, 2013ApJ...769...69B, 2017ApJ...846..137O, 2019A&A...625A..60R}. The most recent studies on the broad-band prompt emission spectra have found a low-energy hardening of the GRB spectra at 2-20 keV but even at higher energies. The low-energy breaks are found only for long GRBs, while it was shown that short GRBs are best described by a simple power-law below the spectral peak, with an index of -0.7, which corresponds to a slow cooling synchrotron regime of radiation \citep{2019A&A...625A..60R}. This is not true for a recent GRB 211211A with a kilonova emission, i.e. associated with a compact binary merger, where the spectra show a clear presence of low-energy breaks \citep{Gompertz2022}. 
Nevertheless, in a slow or marginally fast cooling regime of radiation, in the electron synchrotron scenario, the parameter space for the production of the prompt emission is quite at odds with our naive expectation from the GRB dissipation site. It requires that (1) the magnetic field in the GRB emitting region is very weak $\sim 10$ G, (2) only a fraction of electrons are accelerated (total number typically required $\sim 10^{49}$ electrons), (3) extremely high energy of the accelerated electron (typical Lorentz factor of electrons of $\gamma_m \sim 10^{5}$), (4) very large size of the dissipation region of $R_{\gamma} \sim 10^{16}$ cm and extreme Lorentz factors of the jet of $\Gamma > 400$. Given the extreme energies of electrons, the SSC emission will be deeply in the Klein-Nishina regime, since the characteristic Lorentz factor of electrons that reach the Klein-Nishina threshold $\gamma_{KN}=m_e c^{2} \Gamma / E_{\gamma} (1+z) \approx 260 \Gamma_{2.7} E_{\gamma,2.7}^{-1} << \gamma_m$, where we assumed z=1 and $E_{\gamma}=500$ keV, typical for short GRBs. Therefore, the peak energy of SSC will be approximately at $\approx \gamma_m \Gamma m_e c^2 \sim 30 \, \gamma_{m,5} \Gamma_{2.7}$ TeV. The relative, TeV to MeV flux can be roughly estimated by the Compton parameter $Y \approx \frac{4}{3} \tau \gamma_m^{2} \xi_{KN}$, where $\tau \sim N_e \sigma_T /4 \pi R_{\gamma}^{2}$ is the optical depth and $\xi_{KN} \approx \left(\frac{\gamma_{KN}}{\gamma_m}\right)^{\frac{1}{2}}$ is the suppression factor due to the Klein-Nishina cross-section \citep{Ando2008}. By implying the above-mentioned parameters, we have an estimate of $Y \sim 0.5$, i.e. SSC VHE component with comparable luminosity as the keV - MeV prompt emission. This does mean that we expect always to observe TeV emission comparable with MeV prompt emission. The very presence of $TeV/\Gamma$ photons in the jet initiates the pair production, which further suppresses the SSC component. \citealt{2004ApJ...613.1072R} derives analytically two characteristic energy thresholds for the internal attenuation of VHE photons within the prompt emission region. The first threshold $E_1 = m_e^{2} c^{4} \Gamma^{2}/ 2 E_{\gamma} \approx 260 \, \Gamma_{3}^{2} E_{\gamma,2.7}^{-1}$ GeV. Photons above $E_1$ will be suppressed by the pair production with $E_{\gamma}$, i.e. with most of the photons produced by GRB. The second threshold is $E_2 = 3 \Delta L_{\rm iso} \sigma_T m_e^{2} c^2 / 64\pi \Gamma^{2} \delta t E_{\gamma 0}^{2} \approx 2 \times 10^{4} \, \Delta_{3} L_{iso,52} \Gamma_{3}^{-2} \delta t_{0}^{-1} E_{\gamma 0,\rm 10 \, keV}^{-2}$ GeV comes from the pair production of VHE photons with lowest energy photons (with $E_{\gamma 0}$) in the GRB spectrum, where $\delta t$ is the minimum variability time-scale measured in the rest frame of the GRB host and $\Delta = ln(2(2E_{\gamma 0}E_{\gamma})^{1/2}/m_e \Gamma)-1$. Photons above $E_2$ survive due to the decrease of the pair production cross section for extremely high-energy photons. Clearly, the suppression of the TeV component strongly depends on the bulk Lorentz factor of a GRB and the low-energy characteristics of GRB spectra, i.e. below 10 keV. Therefore, we would expect very different TeV signals (or no TeV emission at all) from a GRB to a GRB \citep{2013ApJ...769...69B}. 

There are other channels to produce VHE photons from the prompt emission. Shock and reconnection acceleration would result in efficient acceleration of protons in GRB jets. Shock accelerated electrons that radiate at maximum $h m_e c^{3}/2 \pi^{2} e^{2} = 22$ MeV photons (in the comoving frame of the jet) via the synchrotron radiation, while protons can produce photons up to 41 GeV. Therefore, TeV photons can be produced by the proton synchrotron mechanism \citep{Aharonian2000}. Proton-dominated jets, i.e. when the ratio between the fraction of protons to electrons exceeds 100, then the TeV component from the protons can be as luminous as the MeV prompt emission (about $10^{52} \rm {erg/s}$) component \citep{Asano:2008tc}. The proton synchrotron component is very sensitive to $\Gamma$, the magnetic field, and the proton-to-the-electron ratio. Recently, it was suggested that the usual MeV component can be produced by the proton synchrotron radiation \citep{Ghisellini2020} if the magnetic field is strong ($B\sim 10^{6}$ G) (see also \citealt{Florou:2022okf}). The TeV photons are also expected from the products of the photo-meson interaction or the proton-synchrotron radiation by itself. There are two channels for the photo-meson process: 
\begin{align}
    \label{eq:pgamma_interactions}
    p+\gamma\rightarrow \Delta^+\rightarrow\begin{cases}
        p+\pi^0 \ \ \ \ \ \ \ \\
        n+\pi^+ \ \ \ \ \ \ 
        \end{cases}
\end{align}
The ratio between the first and the second channels is 2:1 at the resonant energy and equal out of the resonance. To obtain $p\gamma$ interaction, the photon in the comoving frame of the proton should reach the energy threshold of $\sim 300$ MeV, which corresponds to protons with the Lorentz factor of $\gamma_p \sim 3 \times 10^{4} \Gamma_{2} E_{\gamma,2.7}^{-1}$ 
assuming the peak of the GRB spectrum as the main source of target photons. The typical energy of a neutral pion (in the comoving frame of the jet) is correspondingly 
$E_{\pi_{0}}^{\prime} \sim  0.2 E_p^\prime$, 
i.e. $\gamma_{\pi_{0}} \sim 1.5 \times 10^{5} \Gamma_{2} E_{\gamma,2.7}^{-1}$. 
A neutral pion decays into two photons of energy of $\gamma_{\pi_{0}} E_{\pi_{0}}^{\prime} \Gamma / 2(1+z) \sim 500 \Gamma_{2}^{2} E_{\gamma,2.7}^{-1}$ TeV. The luminosity of the VHE component from the decay of neutral pions can be roughly estimated by the photo-meson cooling time. 
This returns a quantity $f_{\rm p\gamma} \approx 0.4 \chi(\alpha,\beta) L_{iso,52} E_{\gamma,2.7}^{-1} \Gamma_{2}^{-4} \delta t^{-1}$ which is the fraction of protons making to photo-meson process  
\citep[e.g., see][]{2022arXiv220206480K} and $\chi(\alpha,\beta)$ is a function which depends on GRB spectral indices $\alpha$ and $\beta$. Assuming typical 
$\alpha=-1$ and $\beta=-2.3$, we obtain $f_{\rm p\gamma} \approx 0.06 L_{\rm iso,52} E_{\gamma,2.7}^{-1} \Gamma_{2}^{-4} \delta t^{-1}$. In the formulae, for $f_{\rm p\gamma}$ we used the relation between the size of the dissipation region $R_{\gamma}$ and its bulk Lorentz factor $\Gamma$, $R_{\gamma}\approx 2 c \Gamma^{2} \delta t$ which assumes that the GRB variability $\delta t$ is driven by the radial or angular spread of the emission. The luminosity of the VHE component due to the neutron pions decay will be approximately $\sim 0.5 L_{\rm iso} f_{\rm p\gamma} f_{\rm p} \xi_{\rm p}$, where $f_{\rm p}$ is the fraction of protons with energies suitable for the photo-meson interactions and $\xi_{\rm p}$ is the baryon loading fraction of the GRB jet, i.e. the ratio between the energy in the non-thermal protons and the emitted energy in the MeV prompt emission. If we take into account the non-detection of TeV neutrinos from GRBs \citep{Lucarelli2022}, then $f_{\rm p} \xi_{\rm p} < 1$ and the TeV component would have a luminosity of $< 0.03 L_{\rm iso,52}$ for the abovementioned parameters. One needs to carefully take into account the internal suppression of the VHE component, as discussed above (see recent developments by \citealt{2022arXiv221200765R} for the internal shock model with hadrons).  

Alternatively, to the internal shocks model, another possible scenario is that of a magnetically dominated jet, where most of the energy of the jet is in the magnetic field and some dissipation process occurs to transfer the magnetic field energy to the accelerated particles, e.g. via magnetic reconnection \citep{Drenkhahn2002, Lyutikov2003, Zhang2011}. Recent first-principle simulations of magnetically dominated plasma turbulence show that electrons are impulsively accelerated to Lorentz factors $\gamma\approx \sigma_e$ by magnetic reconnection in large-scale current sheets, where $\sigma_e=U_B/(n_e m_e c^2)$ is the plasma magnetization with respect to the electron rest mass and $U_B$ is the magnetic energy density. Since the accelerating electric field is nearly aligned with the local magnetic field, the distribution of the particles' pitch angle $\theta$ is strongly anisotropic, and synchrotron emission is suppressed. Then {inverse Compton (IC)} scattering may be the dominant cooling process, even in magnetically dominated plasma. It was already known that the typical spectral slope of the GRB prompt emission can be produced by synchrotron if the emitting electrons radiate most of their energy via IC in the Klein-Nishina regime. However, if the particle pitch angle distribution is isotropic (as it was usually assumed), this would require the radiation energy density to be much larger than the magnetic energy density, which is not possible in the magnetically dominated jet. Instead, if the pitch angle $\theta$ is small, the condition for the IC cooling to be dominant become $u_s \gg \theta^2 U_B$, where $u_s$ is the energy density of synchrotron photons, which may be easily satisfied even in magnetically dominated plasma. The luminosity of the IC component is a fraction $\eta\approx 0.3 L_{52}^{1/8}\Gamma_{300}^{1/2}R_{15}^{-1/4}E_{pk,1\,MeV}^{-3/4}\theta_{-1}^{-3/4}$ of the synchrotron luminosity, where $E_{pk}$ is the peak energy of the synchrotron spectrum. The spectrum of the IC peaks at $E_{pk,IC}\approx 4\, L_{52}^{-1/4}\Gamma_{300}R_{15}^{1/2}E_{pk,1\,MeV}^{1/2}\theta_{-1}^{-1/2}$ TeV, with two breaks at $E_{IC,b}\approx 1\, L_{52}^{-1/8}\Gamma_{300}^{3/2}R_{15}^{1/4}E_{pk,1\,MeV}^{-1/4}\theta_{-1}^{-5/4}$ TeV and $E_{IC,KN}\approx 4\, \Gamma_{300}^2E_{pk,1\,MeV}^{-1}$ GeV. Photons with energy $E_{IC}>E_{IC,KN}$ can annihilate before escaping, reducing the luminosity of the VHE component. The derivation of the emission of the secondary component from the created pairs  requires a complex analysis considering also the effect of the created pairs on the jet magnetization \citep{SobacchiSironiBeloborodov2021}. 
{A detailed description of the} production of the VHE emission from GRB can be found in \citealt{Gill2022} and references therein. 

\subsection{Afterglow emission}
A few TeV sources during the afterglow emission have been detected by MAGIC and H.E.S.S. \citep[][and references therein]{2022Galax..10...67B}. The afterglow TeV emission is interpreted as the SSC component from the electrons accelerated in the forward shock caused by the propagation of the GRB jet in the circumburst medium. This emission component depends on the microphysical parameters of the external shock, as well as on the density of the circumburst medium. So far, TeV emission has been identified at relatively late times. The fastest slewing time so far is around 25 seconds for GRB 160821B \citep{2021ApJ...908...90A} Even though the SSC is the most obvious interpretation for the late TeV component, there are alternatives due to unclear observational distinction between synchrotron and SSC components. Surprisingly, the SSC component has a comparable amplitude to the synchrotron one. Apart from being a coincidence, due to microphysical parameters (low magnetic fields and small fraction of accelerated electrons), this can be also an indication of non-trivial acceleration processes \citep{2021Sci...372.1081H} or self-regulation of the external shock by the pairs \citep{Derishev:2021ivd}. Therefore, it is extremely important to detect the TeV component from the forward shock from the early times. This will allow 
{us} 
to (1) constrain the total energetics of the jet, (2) initial bulk Lorentz factor, and (3) trace the evolution of the micro-physical parameters \citep{Derishev:2021ivd}. For more details on the modeling of the VHE afterglow component from the forward shock accelerated electrons see a recent review by \citet{2022Galax..10...66M}. 

Yet a less explored mechanism for the production of early VHE component is the reverse shock. The reverse shock forms at the earliest stages of the deceleration of the jet in the circumburst medium. Since the GRB jet is denser than the circumburst medium, the reverse shock is expected to accelerate electrons to lower energies. The synchrotron emission from the reverse shock is expected to produce a bright optical flash in the first tens of seconds from GRB detection \citep{Meszaros1997}. Several bright optical flashes have been interpreted to arise from the reverse shock (see the list of GRBs with optical flashes in \citet{Oganesyan2021}). However, some GRBs lack these optical flashes, even if well monitored at very early times. One interesting possibility is that the reverse shock develops in these jets, but the ongoing MeV prompt emission produced behind the reverse shock cools down the hot electrons, extracting their energy by the external inverse Compton (EIC) mechanism, rather than the synchrotron emission \citep{Beloborodov:2005nd}. This is possible only if the reverse shock occurs in the relativistic regime, i.e. the duration of GRB is $> 1 R_{\rm dec,17} \Gamma_{3}^{-2} (1+z)/2$\,s. We never witnessed an early optical flash from SGRBs simply due to instrumental difficulty to follow up a GRB of the second duration. However, given that a large fraction of SGRBs has soft extended emission, one can still have a source of prompt emission photons to cool down reverse shock accelerated electrons via EIC. There are other promising sources of EIC emission in the presence of long-lasting central engines \citep{Murase2018,Zhang2021a} or the presence of the cocoon \citep{Kimura2019}.

\subsection{Delayed pair echoes emission}
VHE photons emitted either in the prompt or afterglow emission can annihilate with photons from extragalactic background light (EBL) with energy $E_{\rm EBL}= 2(m_ec^2)^2/(1+z)^2 E_{\rm VHE}\sim 0.5 (1+z)^{-2} E_{1\, \rm TeV}^{-1}$ eV over the mean free path length $\lambda_{\gamma\gamma}=1/\sigma_{\gamma\gamma}n_{\rm EBL}(E_{EBL})\sim 19 n_{EBL,-1}^{-1}$ Mpc. The electron and positron produced share the energy of the VHE photon, and they can upscatter  the photons from the cosmic microwave background (CMB) via IC up to energies $E_{\rm echo}\approx \gamma_e^2 E_{\rm CMB}(z)(1+z)^{-1}\approx 0.6 (1+z)^2 E_{1\, TeV}^2$ GeV. Here we used that $\gamma_e=E_{VHE}(1+z)/2m_ec^2\approx10^6 (1+z)E_{1\, TeV}$ and $E_{\rm CMB}(z)\approx 6.35\times 10^{-4} (1+z) $ eV.

This secondary HE emission is called \emph{pair echoes} \citep{1994ApJ...423L...5A, 1995Natur.374..430P, 2008ApJ...687L...5T,Murase2009}, and it arrives with a characteristic time delay with respect to the primary VHE component due to the deflection of the pairs by the intergalactic magnetic field (IGMF). The pairs IC cool over a characteristic distance $\lambda_{\rm IC}\approx 0.731(1+z)^{-5} E_{1\,\rm TeV}^{-1}$ Mpc. Assuming that the pair front expands spherically over a distance $\lambda_{\rm IC}$ with particles with Lorentz factors $\gamma_e$, the radial delay of the pair echoes with respect to the primary VHE emission is of order $t_{delay}\approx(1+z)\lambda_{\rm IC}/2\gamma_e^2c\approx 10$ s for $z=1$. Assuming maximum energy of the intrinsic GRB spectrum  of $E_\gamma^{max}=10$ TeV and $z=1$, the maximum energy of the produced pair is $\sim 5 $ TeV and the up-scattered CMB photons can reach $\approx 100$ GeV. The detection of the pair echoes would allow 
{us} 
to reconstruct the characteristics of the primary VHE component and 
{probe} 
the structure of the IGMF. Non-observation of GeV photons from persistent sources such as TeV blazars allowed 
{putting a} lower limit of $B_{\rm IGMF}> 10^{-16}$ G for a coherence length of 10 kpc \citep{Neronovetal2010, Ackermannetal2018}. However, for persistent sources, it is difficult to discriminate if the GeV photon observed  are produced via pair echoes or via the intrinsic emission mechanism of the blazar, while the impulsive nature of the GRB can allow a clear temporal separation between these two components and serve as a better probe of the IGMF.

\section{Conclusions}
In this work, we explored the possibility of detecting the VHE prompt/early emission associated with a binary system of neutron stars. So far, no prompt VHE emission has been detected from short GRBs, and the use of GW signal from BNS can represent a unique way to search for it effectively. While the ability to detect such emissions is largely limited for current GW and VHE observatories, our analysis shows that the future generation of GW detectors, such as ET and CE, operating in synergy with the next generation of VHE instruments, such as CTA, will provide the instrumental capabilities that make prompt/early VHE detections a reality. The search for the VHE counterpart will benefit from the GW pre-merger alerts made possible by accessing lower GW frequency observations, the much better sensitivity of next-generation GW and VHE observatories, and the large FoV, fast response, and slewing time of the VHE instruments.\\
\\

We summarize the key points of our study as follows:
\begin{itemize}
\item \textit{Pre-merger alerts and sky-localization capabilities of ET and network of GW detectors.} A Fisher Matrix \citep[{\it GWFish};][]{Dupletsa2023} approach has been used to estimate the capabilities of the next generation of gravitational-wave detectors to detect and localize binary-neutron star mergers.  We explored the scenario to detect ad localize BNSs pre-merger (during the inspiral phase) considering ET observing as a single detector, ET observing in a network with the current generation of detectors (LIGO, Virgo, KAGRA, and LIGO-India), ET with Voyager, ET with one or two CE. ET as a single observatory is able to detect several tens of BNS per year with sky-localization smaller than 100 deg$^2$ 15 minutes before the merger, the
{number} of detections increases to hundreds 5 and 1 minute(s) before the merger. While the presence of the current GW detectors (LVKI) or two Voyager operating with ET do not change the ET capability 15 and 5 minutes before the merger, one minute before the merger ET+LVKI (ET+2 VOY) increase by 17\% (40\%) the events localized better than  100 deg$^2$ with respect to ET alone. The number of well-localized ($<$ 100 deg$^2$) events pre-merger is significantly higher when ET observes in a network of two or three third-generation detectors; hundreds of relatively well-localized detections 15 minutes before the merger, and several thousand detections 5 and 1 minute(s) before the merger. The absolute numbers are given in Table~\ref{table:skylocall}, where the values corresponding to a pessimistic and optimistic BNS population reproducing the range of the local rate of BNSs measured by LIGO and Virgo  are also given in brackets. In terms of redshifts, 5 minutes before the merger the well-localized  ($<$ 100 deg$^2$) events reach a redshift of 0.4 for ET alone, 0.5 ET+CE, and 0.6 ET+2CE. The reached redshifts increase to 0.5 (ET alone), 1.0 (ET+CE), and 1.3 (ET+2CE) for events detected 1 minute before the merger. {\it ET alone is already able to detect a large number of well-localized pre-merger events, however,  
operating in a network of next-generation detectors will significantly increase this number and the redshifts up to which well-localized detections are possible.}  

\vspace{0.2cm}
\item  \textit{CTA capability to detect VHE prompt emission from GRBs.} The next generation IACT, CTA 
{(consisting} 
of three sub-arrays; LST, MST, SST) is expected to reach an order of magnitude better sensitivity than current VHE facilities. 
We evaluated the minimum energy requirement (E$^{\rm TeV}_{\rm ISO}$) of a short-lived burst for 10\,s to be observed with CTA for a range of redshift from 0.01 to 1.5 in the energy band of 0.2-1\,TeV as the one observed by MST and LST. 
As described in Sect~\ref{sect:VHEmethod}, we found that the minimum isotropic energy required 
for an event to get detected is $\sim 10^{43}$\,erg, $\sim 10^{45}$\,erg, $\sim 10^{47}$\,erg, and $\sim 10^{51}$\,erg at the redshift of 0.001, 0.01, 0.1, and 1.0, respectively. This energy range is consistent with the VHE emission scenarios described in Sect~\ref{sec:theory}.

\vspace{0.2cm}
\item \textit{CTA observational strategies to detect prompt/early VHE emission from BNS mergers.} Taking into account the FoV and slewing time of MST and LST, we propose three observational strategies to follow up on pre-merger alerts given by the GW detectors;
They are: a)\, the {\it direct pointing} strategy following up all the events with sky-localization smaller than the FoV, that is 30 deg$^2$ for MST and 10 deg$^2$ for LST, b)\, the {\it one-shot\,observational} strategy consisting of following-up a large number of triggers using a single observation randomly located within the sky-localization uncertainty of the GW signals; and c)\, {\it mosaic strategy} tiling the sky-localization more effective to detect the afterglow emission. Due to the longer slewing time of MST (t$_{\rm slew}$=90\,s), the one minute of pre-merger alerts are followed up only with LST (t$_{\rm slew}$=20\,s).  For the {\it one-shot\,observational} strategy, Table~\ref{table:obsstrategyMST} and Table~\ref{table:obsstrategyLST} summarize the different paths to follow up pre-merger alert events with CTA-MST and –LST, 
{respectively.}
We consider: following up 1) all the events detected 15 minutes before the merger, 2) all the events detected 5 minutes before the merger, 3) all the events detected 1 minute before the merger (only LST), and 4) using the improved sky-localization updated 1 minute before the merger for events detected 15 and 5 minutes  pre-merger. Since the VHE emission is expected along the relativistic jet, we evaluate a strategy to prioritize alerts and increase the probability of detecting on-axis events. Taking into account the uncertainty on the viewing angle estimated pre-merger, we select to be followed up all the events with an observed viewing angle smaller than $45^\circ$.

\vspace{0.2cm}
\item \textit{Expected number of detection by MST and LST using the direct pointing strategy.}
Following up the events detected with sky-localization smaller than 10 deg$^2$ one minute before the merger  (a few hundred for ET+CE and several hundred for ET+2CE, see Table~\ref{table:skylocall}) using LST, a few (ET+CE) to several (ET+CE) tens are expected to be on-axis (see Table~\ref{table:skylocall}), namely events with a viewing angle smaller than $10^{\circ}$ from which we expect to observe the VHE emission. The number of on-axis events is negligible for ET alone, ET in a network with current generation detectors or upgraded instruments such as Voyager. It is also negligible following up 15 and 5 minutes pre-merger alerts by ET in a network of two/three third-generation detectors. MST is able to observe a few tens of on-axis events following up a few hundred (several hundred) events detected 5 minutes before the merger with sky-localization smaller than 30 deg$^2$. Taking into account the CTA duty cycle of 15\% and the CTA visibility which 
{reduces} the observable sky by a factor 2, also assuming that all the BNS produce a jet, a few joint detections are expected using the direct pointing strategy only when ET is operating within a network of three third-generation detectors. {\it The direct pointing strategy results to be not so effective also for ET in a network of third generation detectors.} 

\vspace{0.2cm}
\item \textit{Expected number of detection by MST and LST using the one-shot observation strategy}. The expected numbers of possible VHE detection by MST and LST per year and the corresponding CTA time required using the one-shot observation strategy are shown in Figures \ref{fig:Fig7MSTPrediction} and \ref{fig:FigLSTPrediction}, respectively.  The plotted numbers take into account the CTA duty cycle of 15\% and the CTA visibility. The CTA time includes slewing time, repositioning time, and exposure time. Using pre-merger 5 minutes alerts, we expect to detect around ten VHE possible 
{counterparts}
per year by CTA-MST operating with ET+CE and ET+2CE by following events with sky-localization smaller than $\rm 10^3 deg^2$ at the cost of 25\%  (ET+CE) and 40\% (ET+2CE) CTA time. Another ten (twenty) possible VHE counterparts are expected using one-minute pre-merger alerts by following the events with LST with sky-localization smaller than about $\rm 200 deg^2$ detected by ET+CE (ET+2CE) at the expense of 20\% (10\%) of the CTA time. Only following pre-merger alerts of 1 minute can give a few detections for ET as a single observatory or operating in the network ET+LVKI+ and LVK+2VOY. 

\vspace{0.2cm}
{\it A significant reduction of the required CTA time can be obtained by prioritizing the alerts to be followed} on the basis of the viewing angle estimate (expected to be given in the GW alert) and its uncertainty; following up the events with a viewing angle smaller then 45$^{\circ}$ reduces the required CTA time of about a factor 3.

\vspace{0.2cm}
{\it The effectiveness of this search increases by using updated information on the source parameters}; that is using updated sky localization available 1 minute before the merger following up 15 or 5 minutes pre-merger alerts. Using MST, the expected number of possible VHE counterparts using updated information becomes 20 per year for ET+CE and 40 per year for ET+2CE following 5 minutes pre-merger alerts with sky-localization smaller 
than
$10^3 \rm deg^2$. For LST, following 5 minutes pre-merger alerts and updated information enable us to detect 10 (20) possible VHE counterparts are expected with CTA operating with ET+CE (ET+2CE).

\vspace{0.2cm}
In addition, the usage of {\it the divergent pointing (single telescopes pointing slightly offset and thus significantly enlarging the FoV) will be highly beneficial for observation, particularly in the case of large localizations} 100-1000 deg$^2$. With the expense of the same amount of MST-time of around 10\% ($<1$\%) for a network of ET and CE (ET alone), divergent pointing is expected to 
follow up a factor of 6 (4) more on-axis events than MST (see Fig. \ref{fig:DivergentPointing}) bringing the possible VHE detection to 60 (4) per year. 
However, the number of real detection will be influenced by the sensitivity, and  the divergent pointing sensitivity is compromised with respect to the MST array as individual telescopes will be operated separately (the expected sensitivity reduction is of 20-25\% for an offset of 3$^\circ$). 

\vspace{0.2cm}
\item \textit{Expected number of detection by MST and LST using mosaic observational strategy}.
The observational mosaic strategy which tries to rapidly cover the entire sky-localization can be more effective than the one-shot observation strategy for signals longer than 20 seconds, such as an afterglow emission. In the case of long 
signals,
this strategy can significantly benefit from the presence of 
keV-MeV detection by high-energy satellites able to better localize the source.

\vspace{0.2cm}
\item \textit{Origin of VHE emission.} Several emission scenarios envisage the possibility of producing a VHE component of emission in the prompt phase of GRBs. In the standard fireball model, while the MeV component is produced by electron synchrotron radiation in a marginally fast cooling regime, the same electrons are expected to emit a VHE component via SSC of intensity comparable to that of the MeV component. Another possibility is given in the scenario in which the prompt emission is produced by proton synchrotron, which can reach GeV energies in the jet comoving frame. The TeV component is the expected product of photo-meson interactions, emitting VHE photons via pion decay. In both these scenarios the VHE component is strongly dependent on the bulk Lorentz factor and magnetic field, as well as the electron-proton ratio and the lower-energy characteristics of the spectrum, so different VHE emission is expected for different GRBs. Furthermore, pair production within the jet and, at higher redshift, with the extragalactic background light and cosmic microwave background is expected to attenuate its intensity. In the afterglow phase, the observed VHE photons can be interpreted as SSC from the forward shock, but a synchrotron origin associated with non-trivial acceleration processes is not excluded. The reverse shock may also be able to produce VHE emission via external inverse-Compton, offering an interpretation of the GRBs observed only in the VHE with no counterpart in the keV-MeV band (orphan GRBs). Furthermore, VHE photons coming either from prompt or afterglow emission can annihilate with lower energy photons, but the pair produced can up-scatter photons from the cosmic microwave background up to $\sim 100$ GeV. This delayed component can be used to reconstruct the primary VHE emission as well as to probe the intergalactic magnetic field.
\end{itemize}

In summary, our work demonstrates that the next generation of GW observatories operating in synergy with VHE arrays, such as CTA, provides a unique opportunity to detect the prompt/ early VHE energy counterpart of binary neutron star merger gravitational-wave signal and short GRBs.  The results show that pre-merger alerts and rapid communication, response, and slewing time are essential. The presence of 
a network of third-generation
detectors can significantly increase the effectiveness of this search by largely increasing the number of possible detections per year with respect to 
a single third-generation
detector operating alone or in a network of 
second-generation
GW detectors. Prioritizing the events to be 
followed up 
on the basis of the source parameters estimated by the GW signals and giving 
updates on these parameters
before the merger can greatly enhance the chance of detection reducing the time request on EM observatories. Detecting the VHE prompt emission is crucial to understand the physics governing the GRB engine.

\begin{acknowledgements}
We acknowledge Stefano Bagnasco, Federica Legger, Sara Vallero, and
the INFN Computing Center of Turin for providing support
and computational resources. We thank Yann Bouffanais for the work and help on the population of BNSs. BB and MB acknowledge financial support from the Italian Ministry of University and Research (MUR) for the PRIN grant METE under contract no. 2020KB33TP. BB acknowledges financial support from the Italian Ministry of University and Research (MUR) (PRIN 2017 grant 20179ZF5KS). MB and GO acknowledge financial support from the AHEAD2020 project (grant agreement n. 871158). MM and FS acknowledge financial support from the European Research Council for the ERC Consolidator grant DEMOBLACK, under contract no. 770017.
\end{acknowledgements}

\bibliographystyle{aa} 
\bibliography{references} 

\begin{appendix}
\onecolumn

\section{Pessimistic and optimistic BNS population scenarios}

\begin{figure*}
\centering

            \includegraphics[width=0.24 \linewidth, height=5cm]{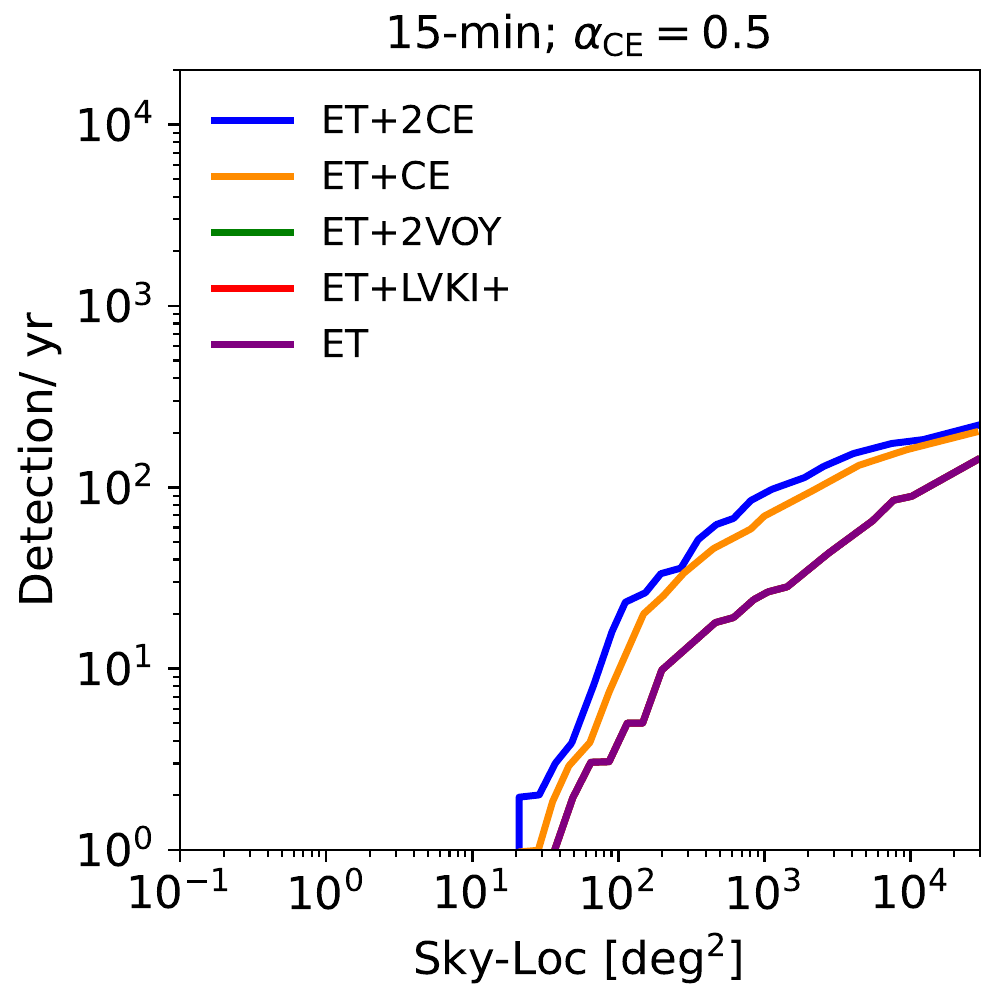}
            \includegraphics[width=0.24 \linewidth, height=5cm]{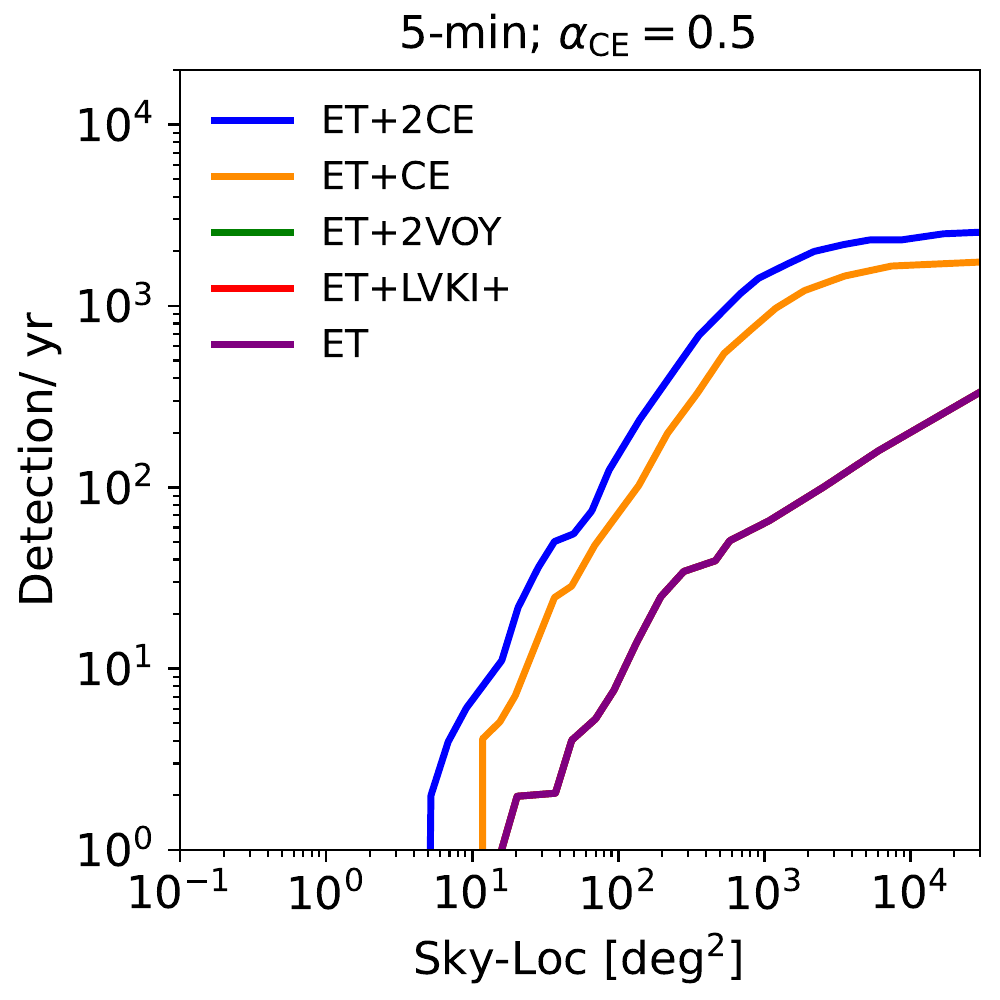}
            \includegraphics[width=0.24 \linewidth, height=5cm]{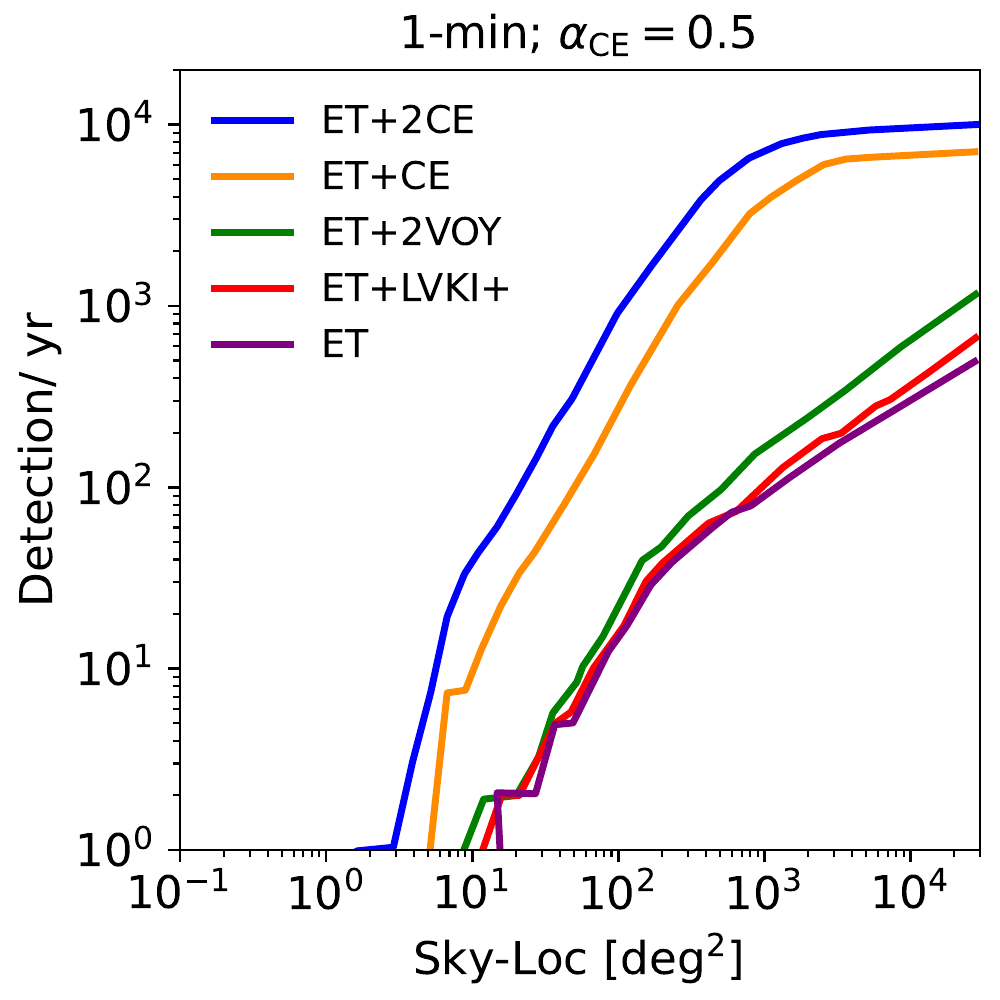}
            \includegraphics[width=0.24 \linewidth, height=5cm]{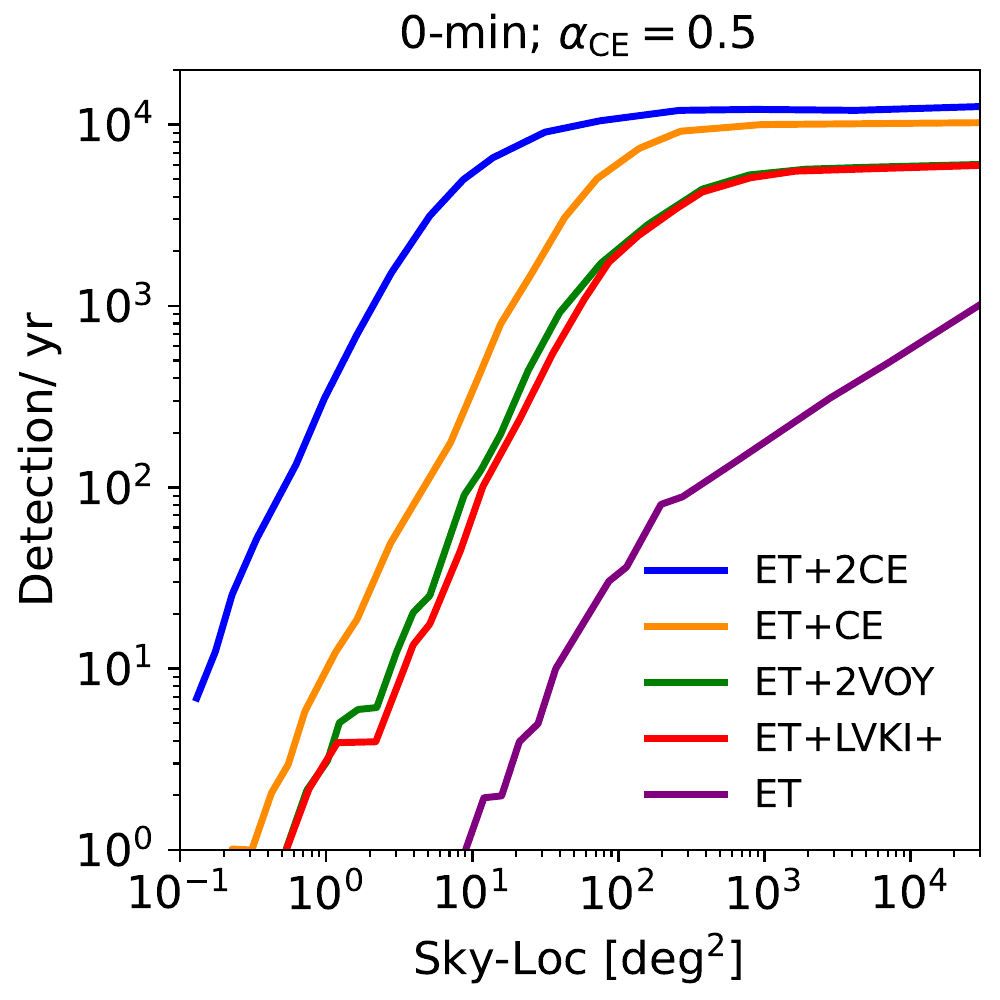}
            
            \includegraphics[width=0.24 \linewidth, height=5cm]{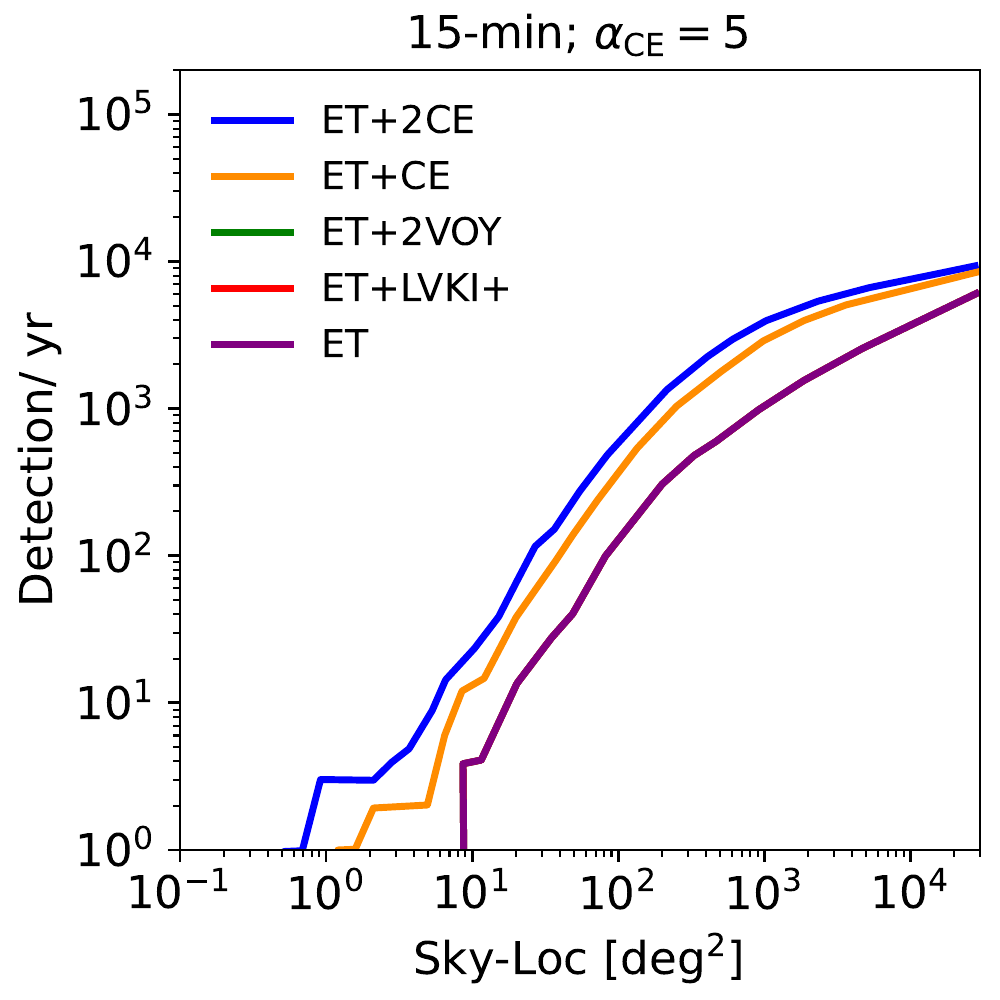}
            \includegraphics[width=0.24 \linewidth, height=5cm]{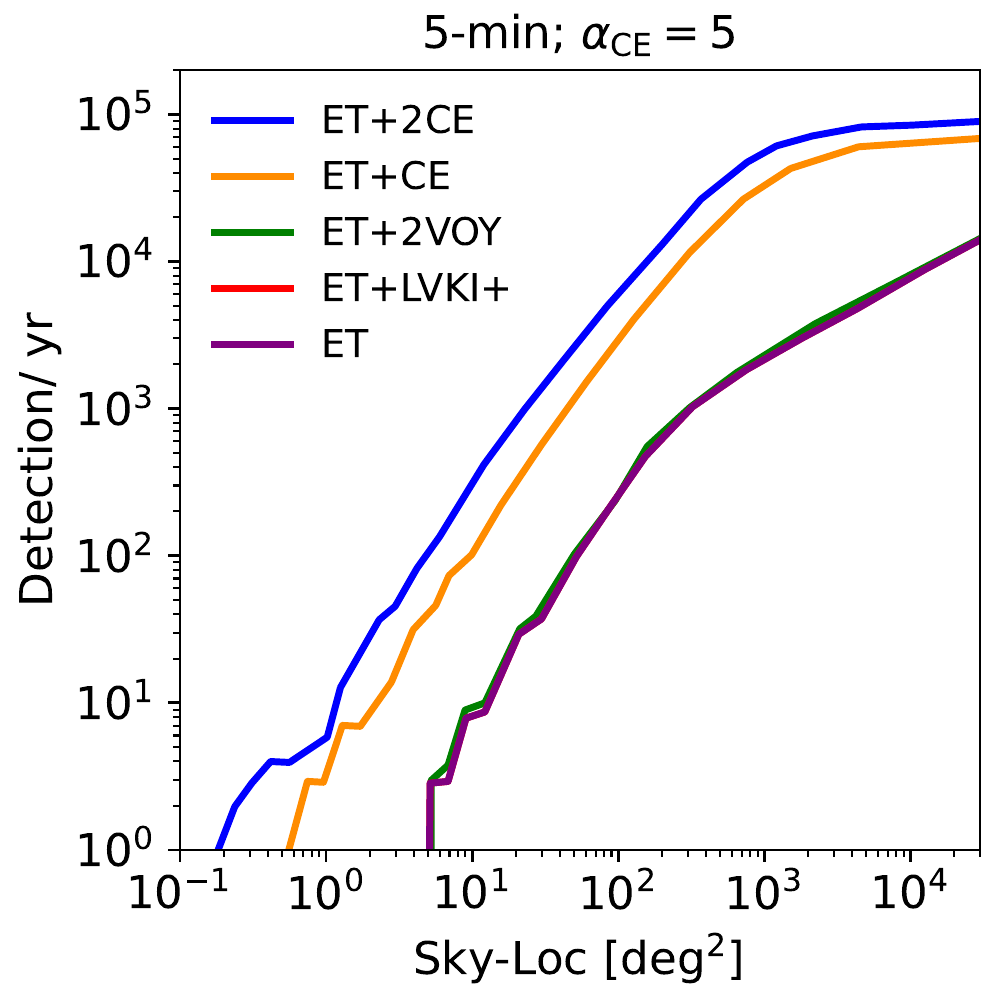}
            \includegraphics[width=0.24 \linewidth, height=5cm]{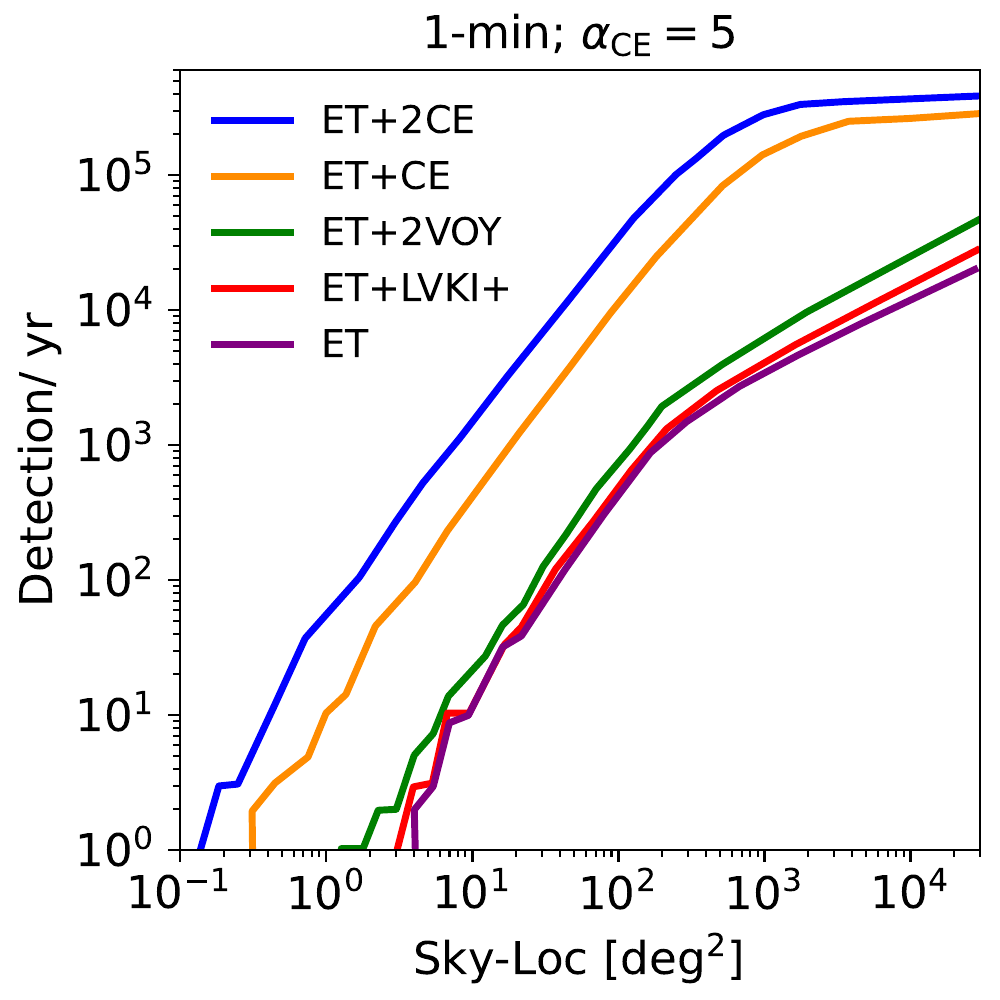}
            \includegraphics[width=0.24 \linewidth, height=5cm]{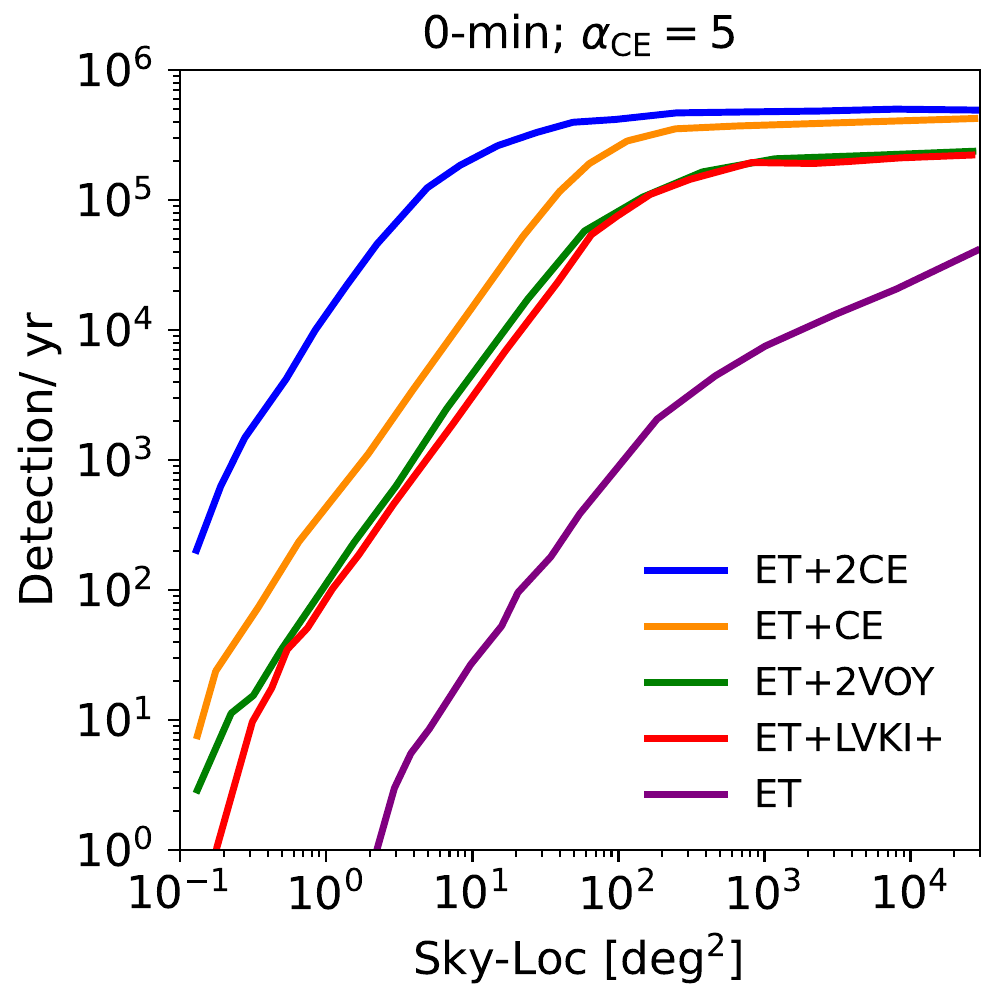}
            
            \caption{The cumulative number of detections (SNR$>$8) per year for different networks of GW detectors considering 15, 5, and 1 minutes before the merger and at the merger time considering the pessimistic (top row) and optimistic (bottom row) BNS merger population scenarios. The number of injected BNSs has been set to $2.0\times10^{4}$ and $4.0\times10^{5}$ within redshift z=1.5 for the pessimistic and optimistic scenarios, respectively. The plots show the detections considering BNS systems with all orientations.}
            \label{fig:BNSrateAll_all}
        \end{figure*}

\begin{figure*}
\centering
            \includegraphics[width=0.24 \linewidth, height=5cm]{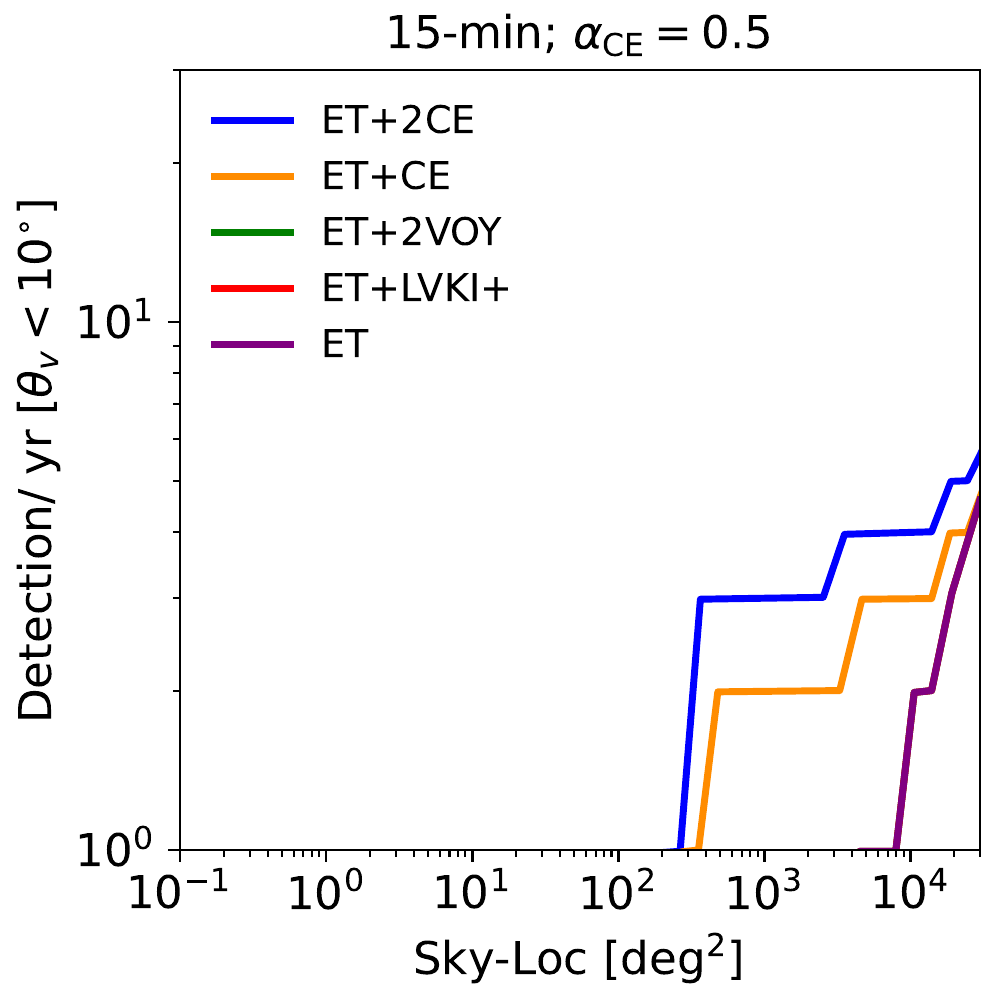}
            \includegraphics[width=0.24 \linewidth, height=5cm]{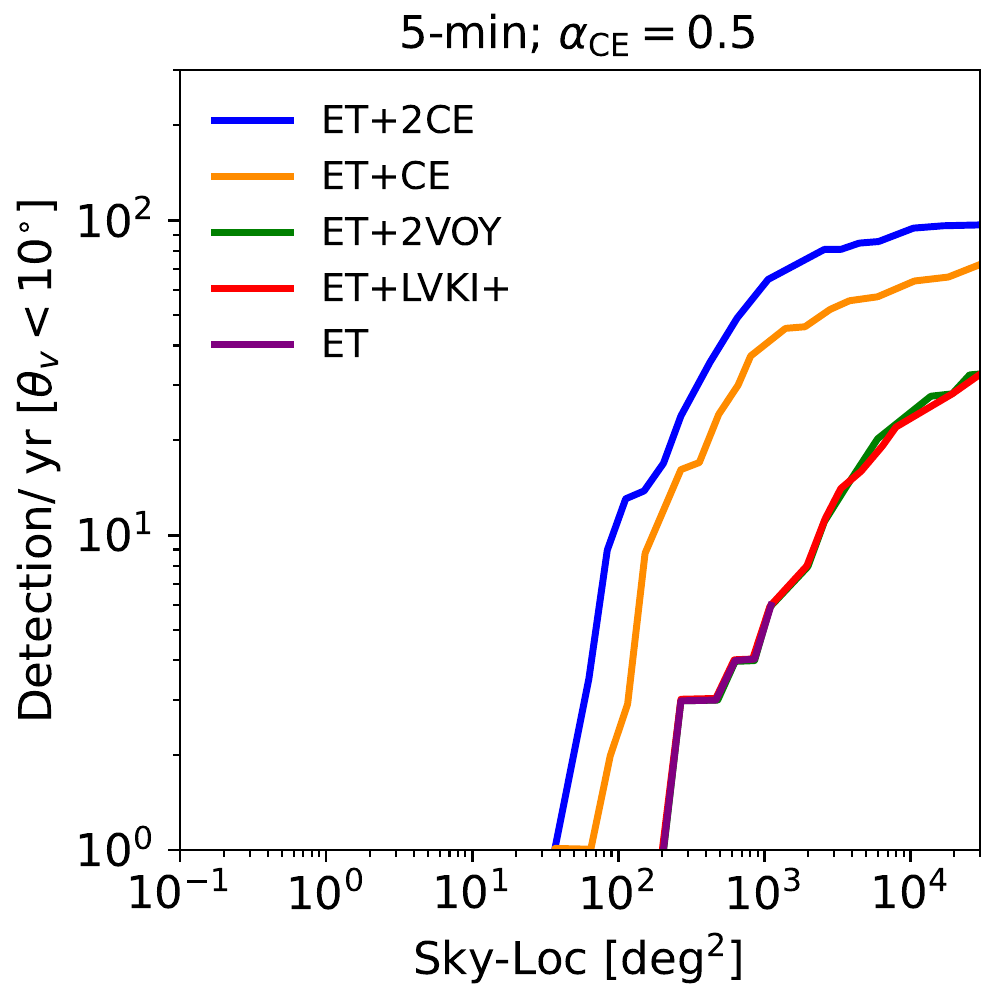}
            \includegraphics[width=0.24 \linewidth, height=5cm]{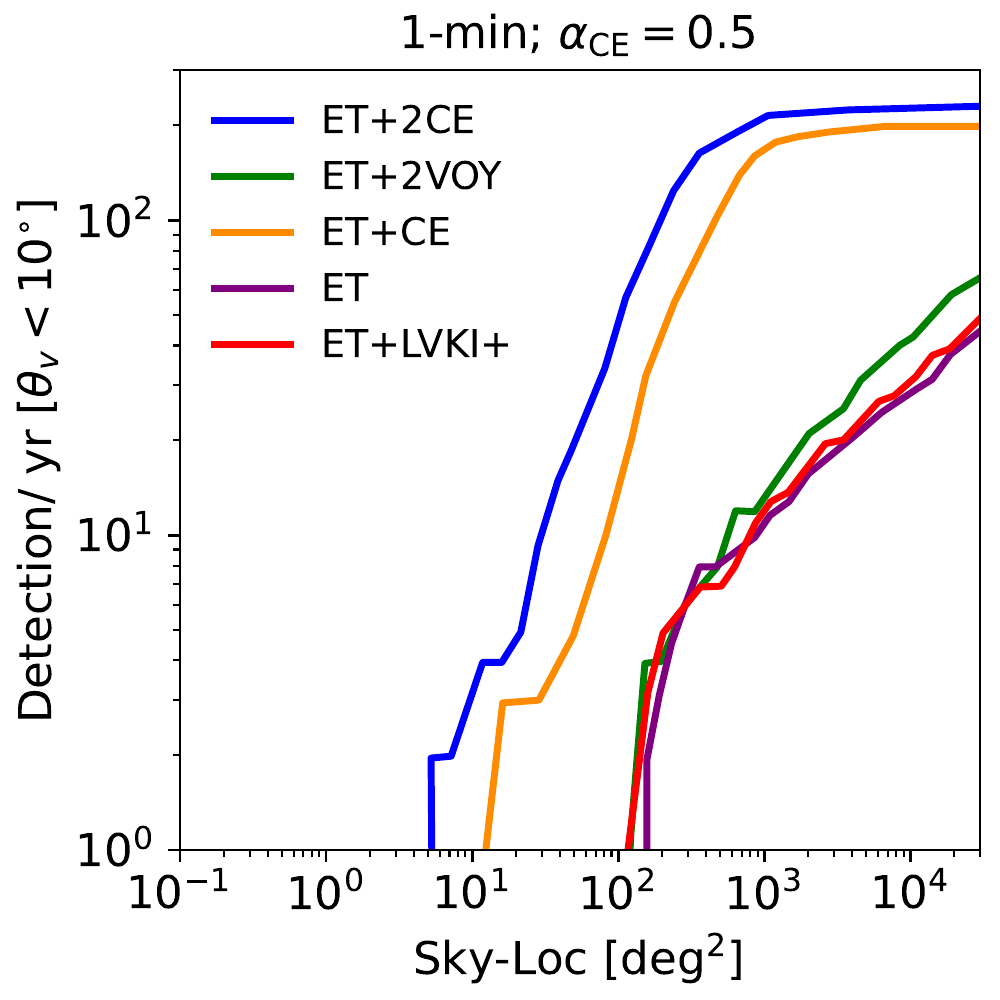}
            \includegraphics[width=0.24 \linewidth, height=5cm]{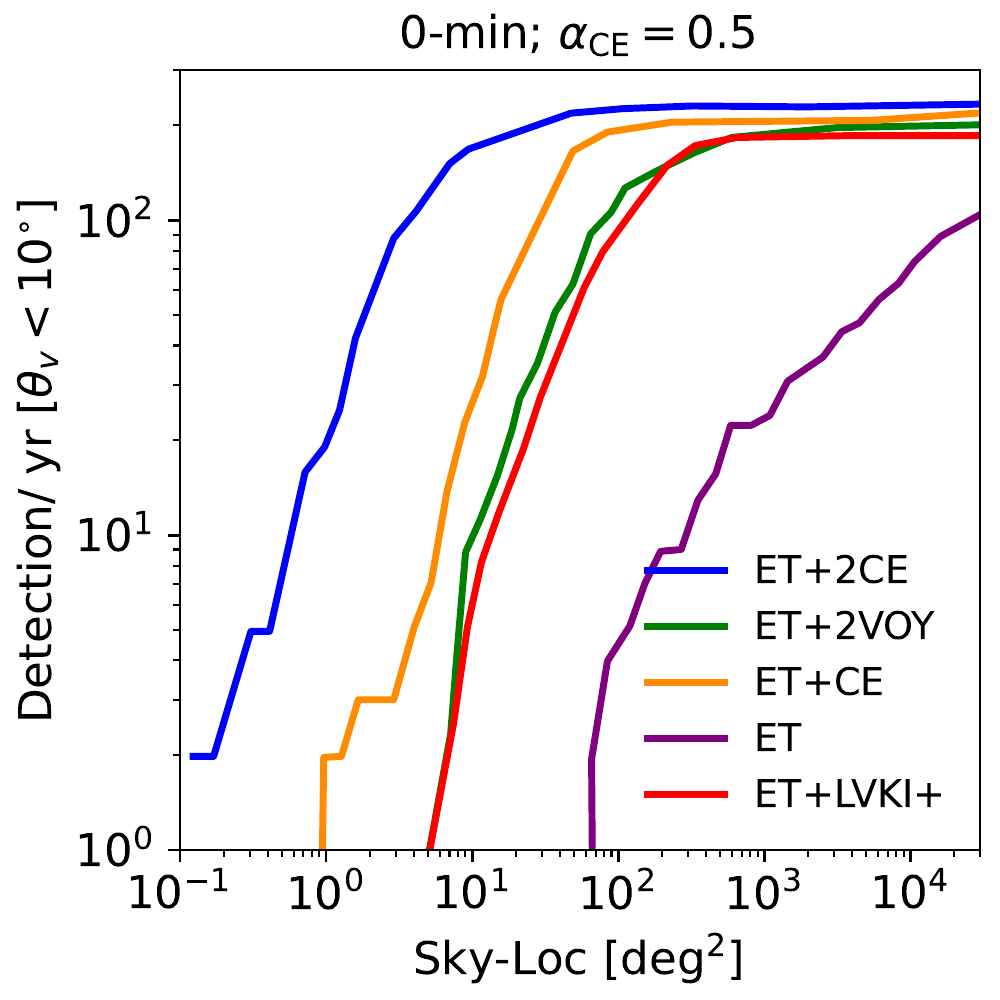}
            
            \includegraphics[width=0.24 \linewidth, height=5cm]{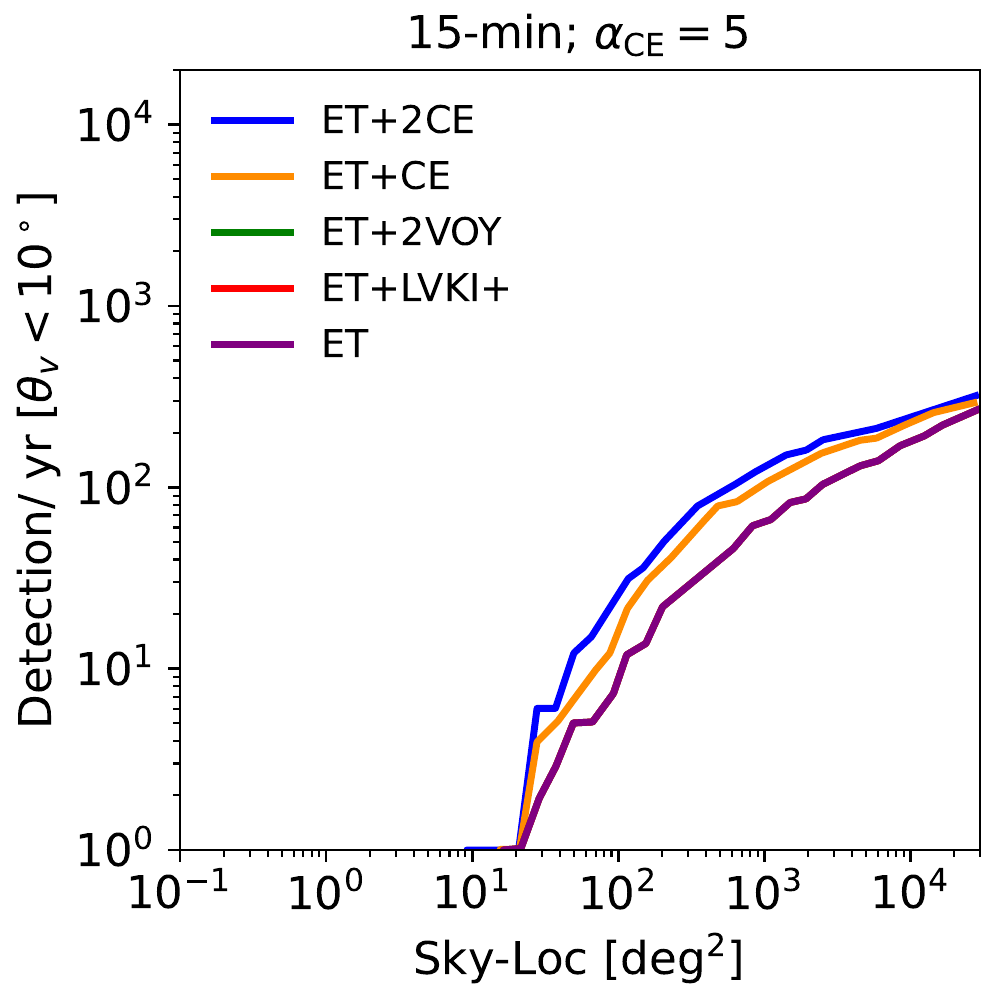}
            \includegraphics[width=0.24 \linewidth, height=5cm]{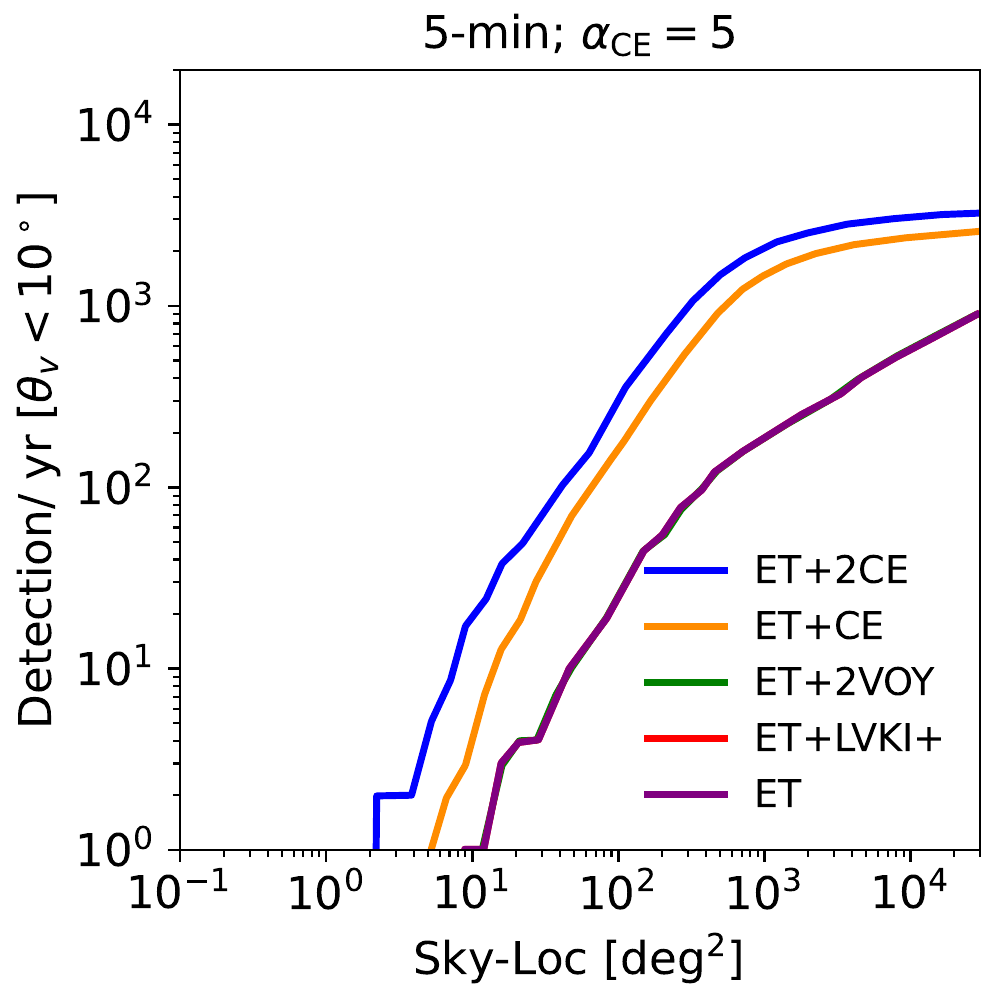}
            \includegraphics[width=0.24 \linewidth, height=5cm]{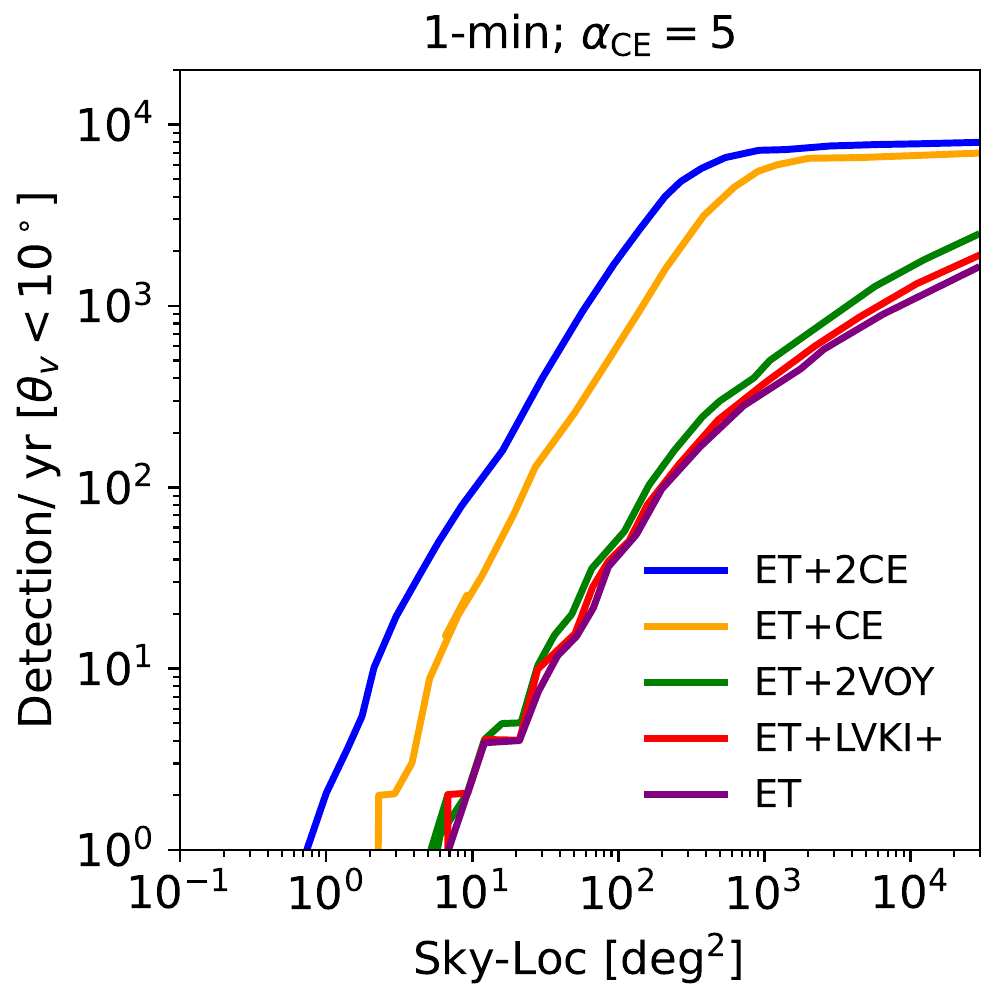}
            \includegraphics[width=0.24 \linewidth, height=5cm]{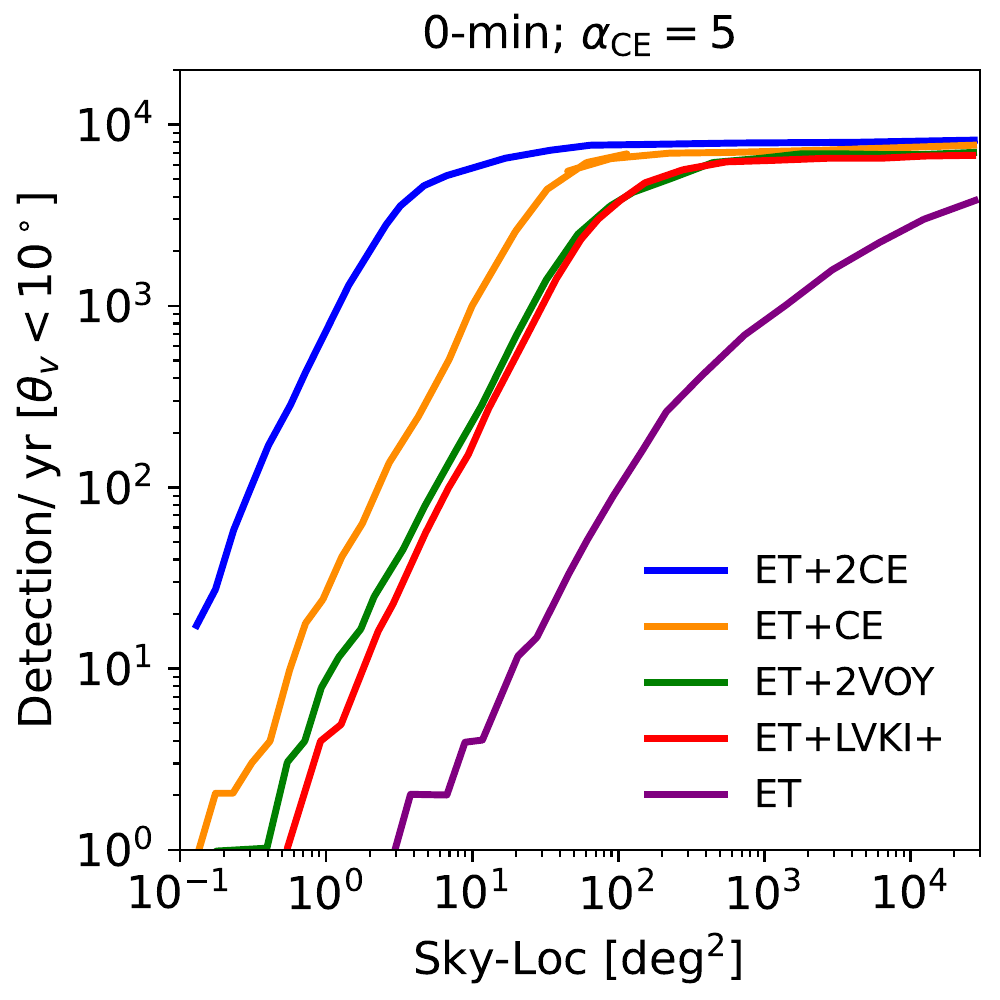}
            
            \caption{Same as Fig. \ref{fig:BNSrateAll_all} but showing the detections of BNS systems with a  viewing angle smaller than 10$^{\circ}$ (on-axis events), a fraction of which are expected to produce detectable VHE emissions.}
            \label{fig:BNSrate_onaxis}
        \end{figure*}

{Our results through the paper are given for a fiducial scenario that adopts a common envelope ejection efficiency parameter, $\alpha_{CE}$, equal to 3  (see section \ref{pop}). Here we show the detection capabilities of ET as a single observatory and in network with other facilities such as LVKI+, CE, and 2CE using two populations, the pessimistic and optimistic BNS merger populations, obtained with a common envelope ejection efficiency parameter of $\alpha=$0.5 and $\alpha=$5. Fig. \ref{fig:BNSrateAll_all} shows the number of detections per year at different times before the merger, namely 15, 5, and 1 minute(s) and at the merger time for ET as a single observatory and in the network of detectors. Figure \ref{fig:BNSrate_onaxis} shows the number of detections of on-axis BNS systems per year.}

\section{Gravitational-Wave alert prioritization based on the viewing angle}
\begin{figure*}
\centering
            
            \includegraphics[width=0.33 \linewidth, height=6cm]{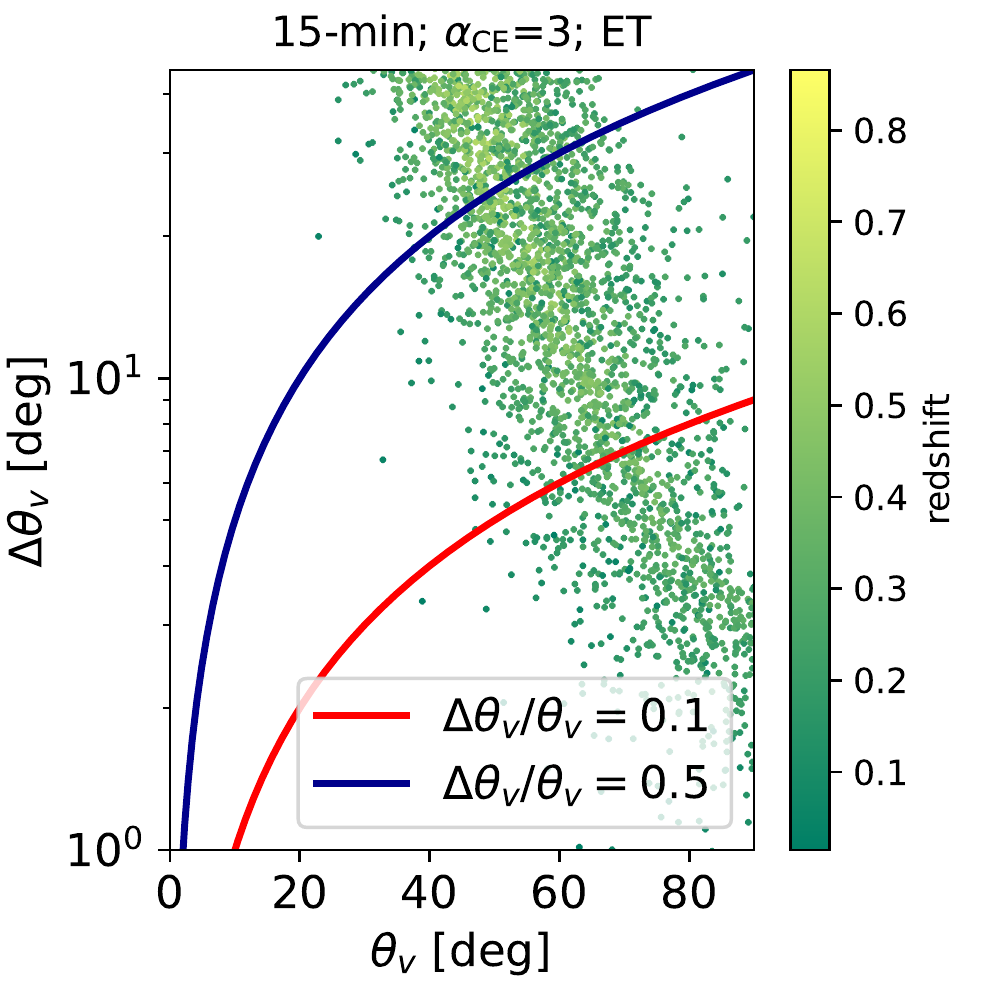}
            \includegraphics[width=0.33 \linewidth, height=6cm]{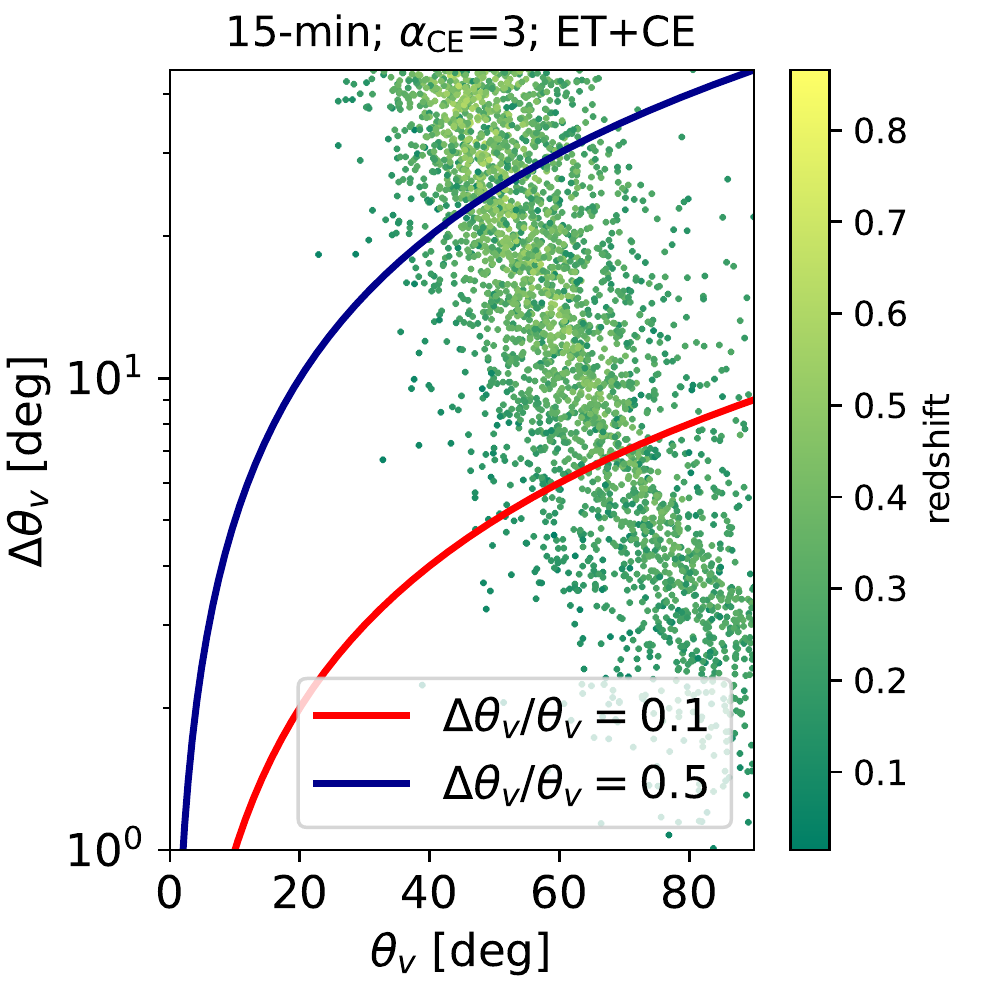}
            \includegraphics[width=0.33 \linewidth, height=6cm]{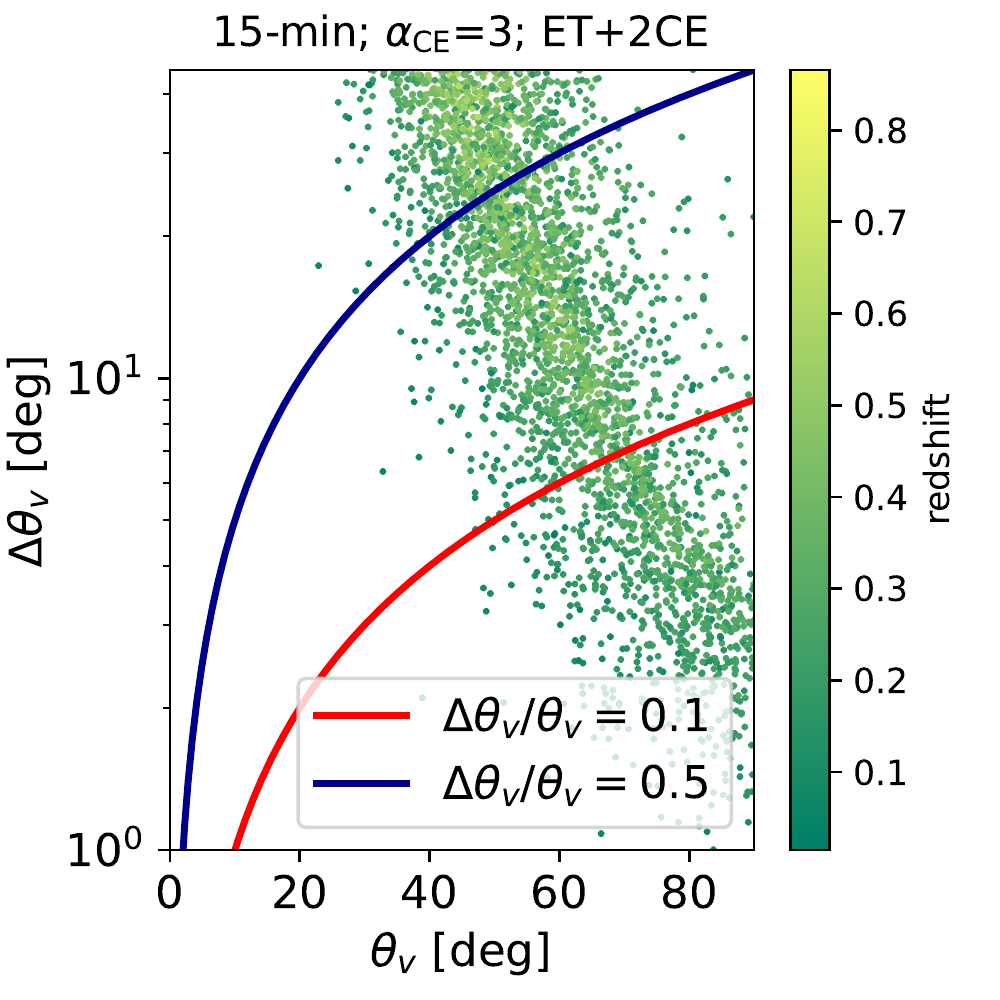}
            
            \includegraphics[width=0.33 \linewidth, height=6cm]{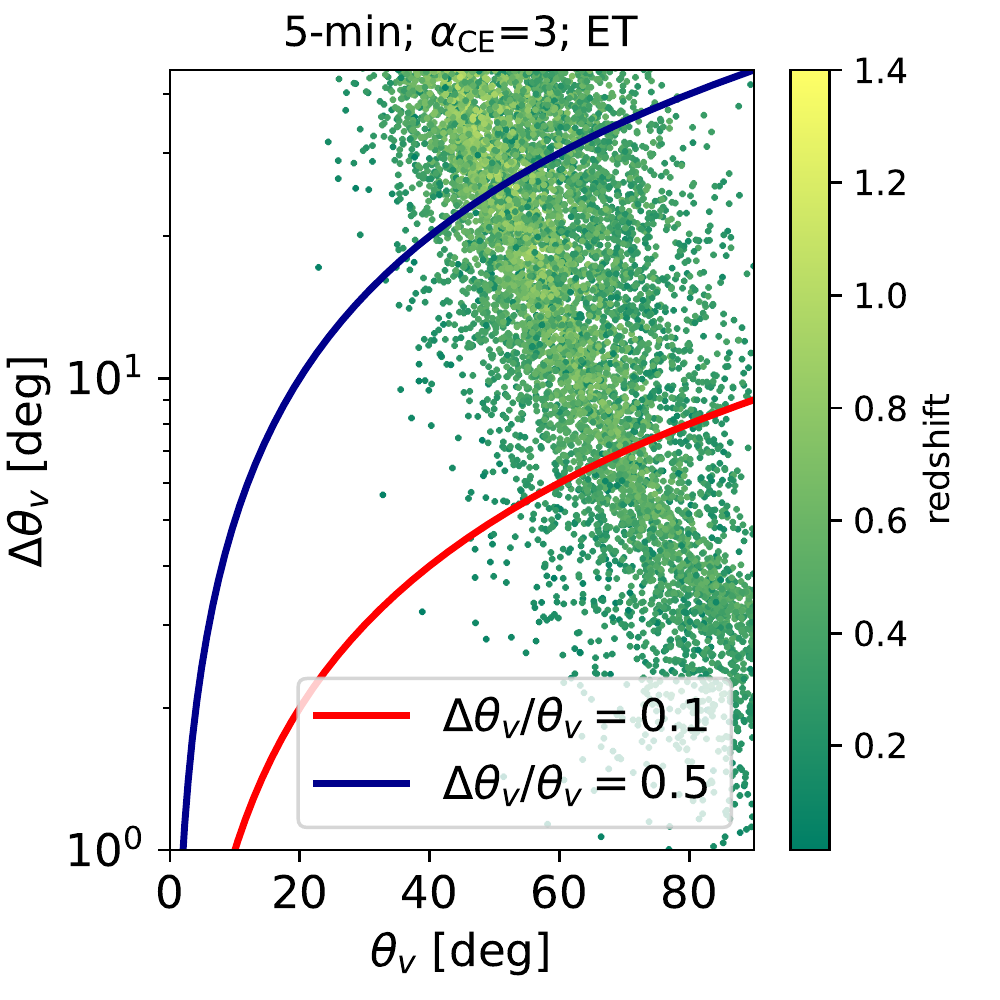}
            \includegraphics[width=0.33 \linewidth, height=6cm]{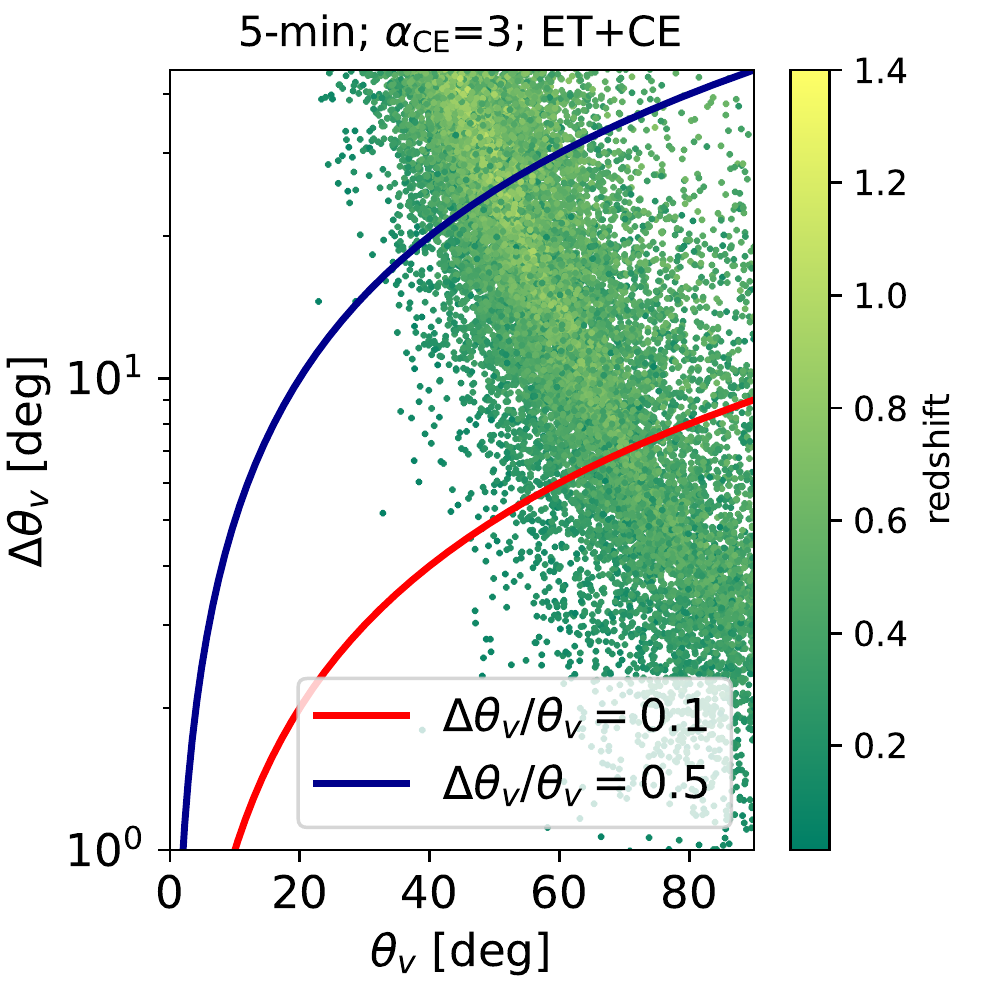}
            \includegraphics[width=0.33 \linewidth, height=6cm]{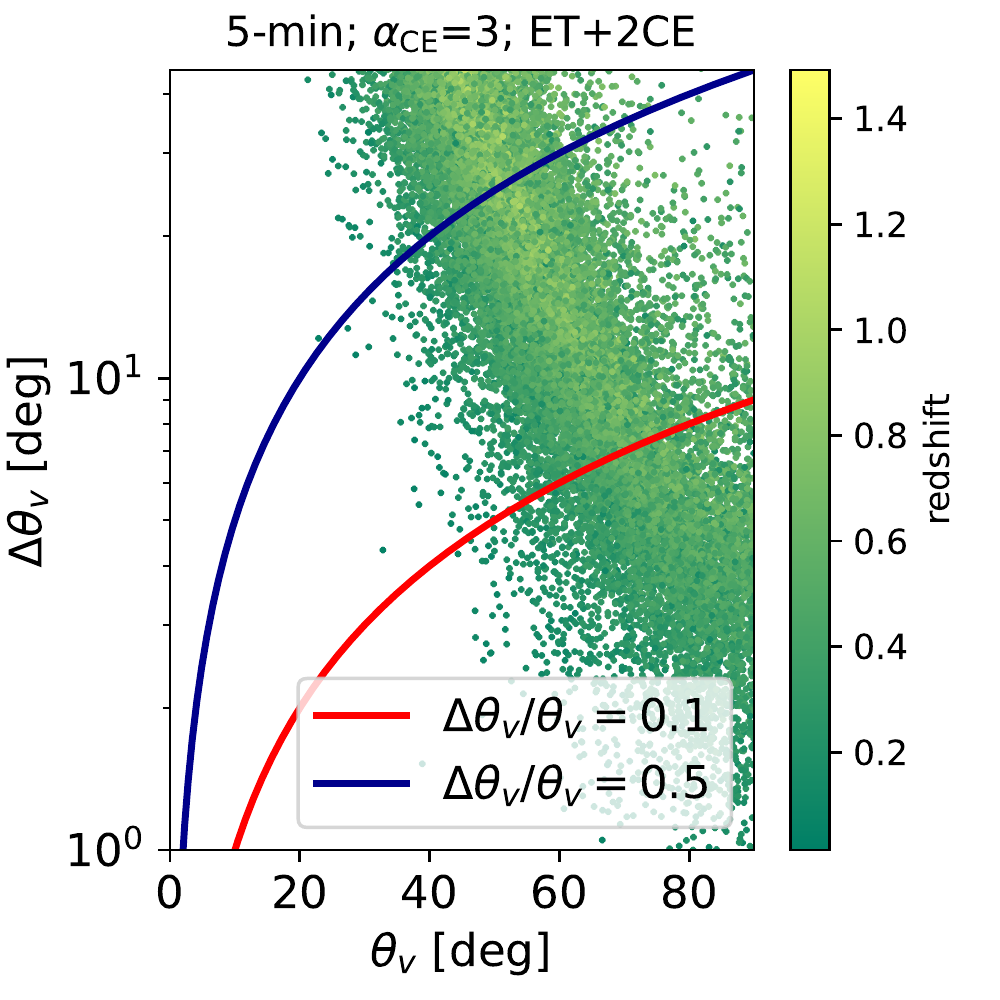} 
            
            \includegraphics[width=0.33 \linewidth, height=6cm]{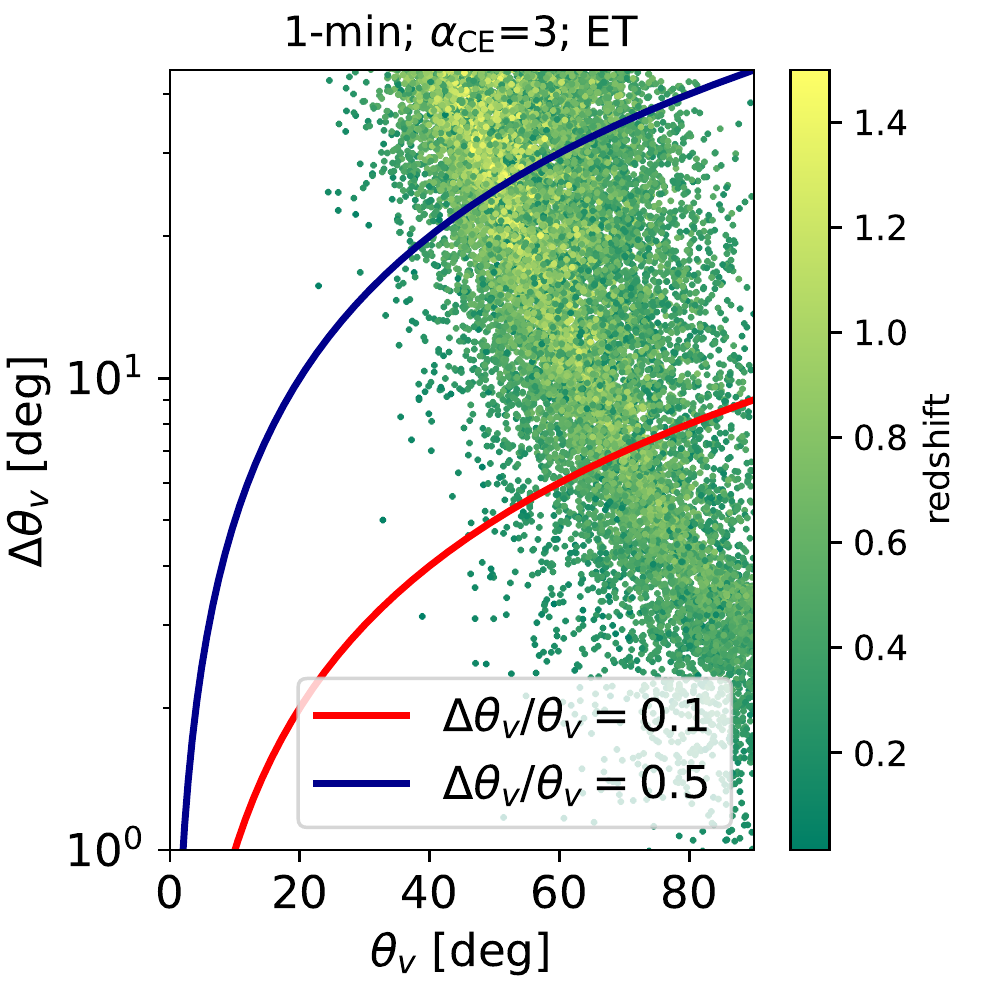}
            \includegraphics[width=0.33 \linewidth, height=6cm]{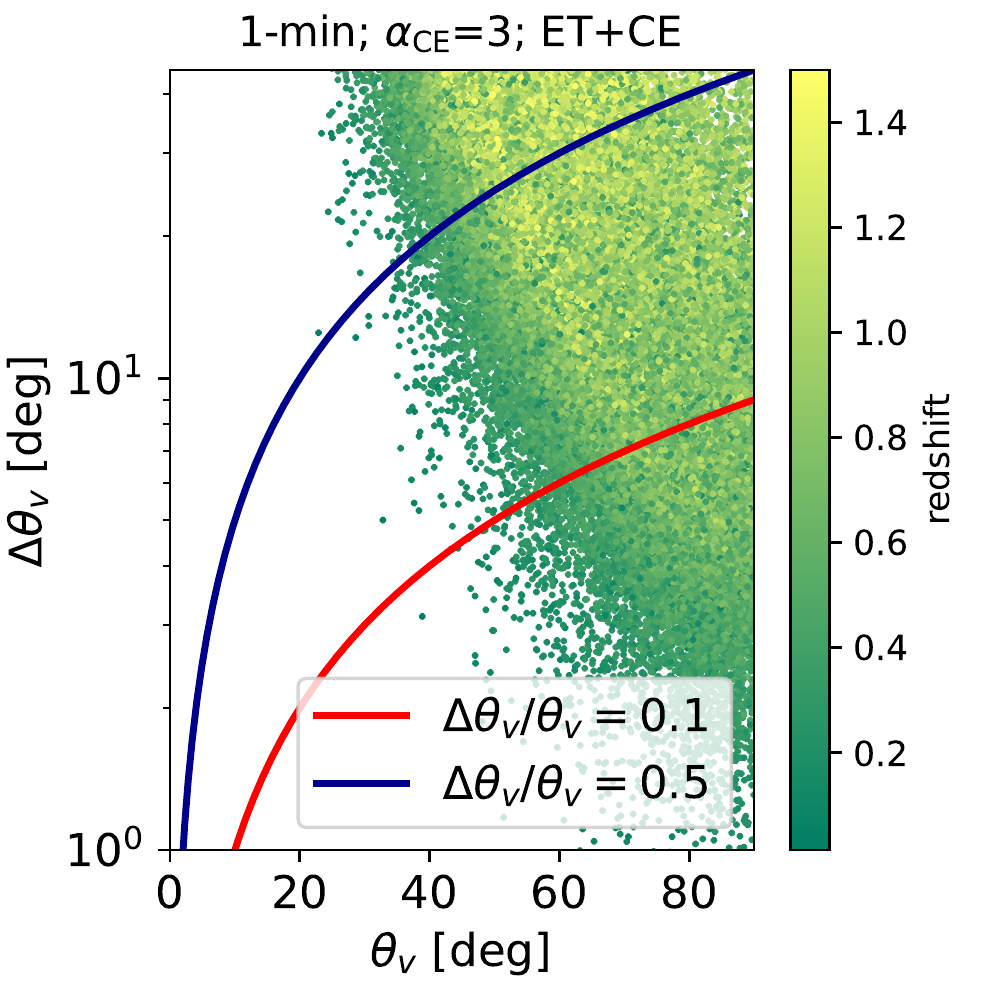}
            \includegraphics[width=0.33 \linewidth, height=6cm]{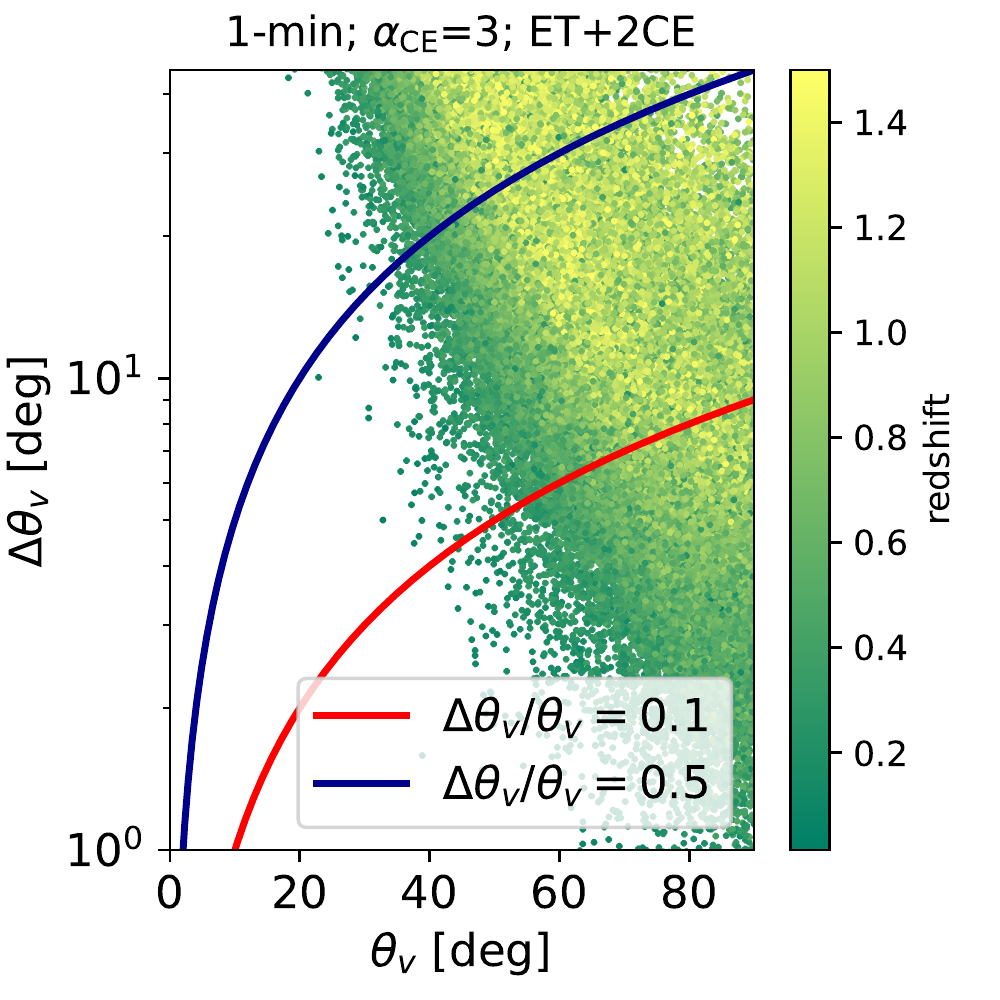}

            \caption{The distribution of uncertainties on the viewing angle ($\Delta \theta_{\rm v}$) as estimated by {\it GWFish} vs the injected $\theta_{\rm v}$  of our fiducial population of BNS mergers for different detector configurations (ET, ET+CE, and ET+2CE) and for different pre-merger alert times (15 min, 5 min, 1 min) and at the time of the merger. The solid lines correspond to $\Delta \theta_{\rm v}/\theta_{\rm v}$=0.1 and 0.5. The vertical color bar indicates the redshift of each event. The plots show only the BNS mergers below redshift 1.5. The uncertainty on the viewing angle, $\Delta \theta_{\rm v}$, is larger in the case of on-axis events.
}
            \label{fig:FigdTvsT}
        \end{figure*}

{The next generation of GW observatories will detect a large number of events. To optimize the required observational time required by the EM observatories to follow them will be necessary to prioritize the events to be followed. In the case of VHE prompt emission expected from on-axis events a parameter that can be used is the viewing angle. In this appendix, we evaluate the precision of determining the viewing angle from GW observations.  Figure \ref{fig:FigdTvsT} shows the uncertainty on the viewing angle $\theta_{v}$ ($\Delta\theta_{\rm v}$) estimated by {\it GWFish} versus the injected viewing angle ($\theta_{v}$) for different configuration of the network of GW detectors and at a different time before the merger. As can be seen from the figure the uncertainties are larger for on-axis events. }

\section{Improvement on sky-localization of the pre-merger events approaching the merger time}
\begin{figure*}
\centering
            \includegraphics[width=0.33 \linewidth, height=6cm]{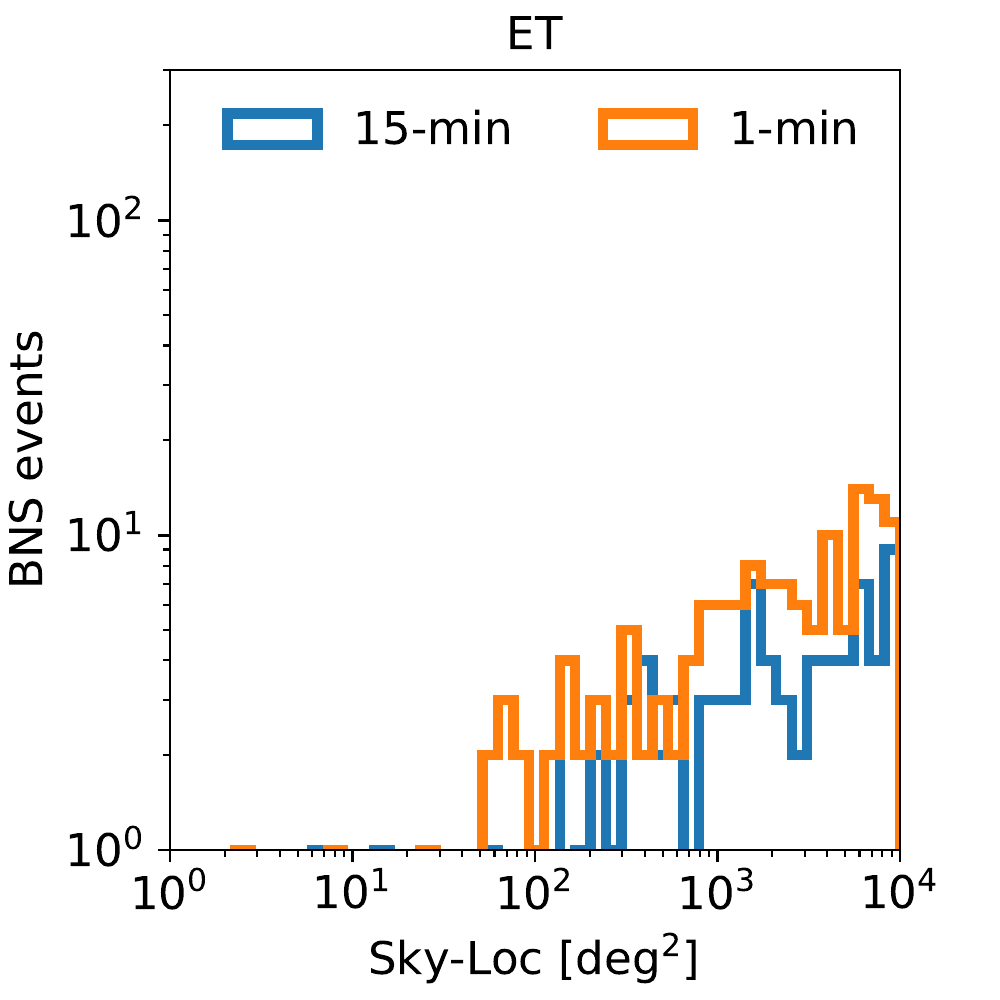}
            \includegraphics[width=0.33 \linewidth, height=6cm]{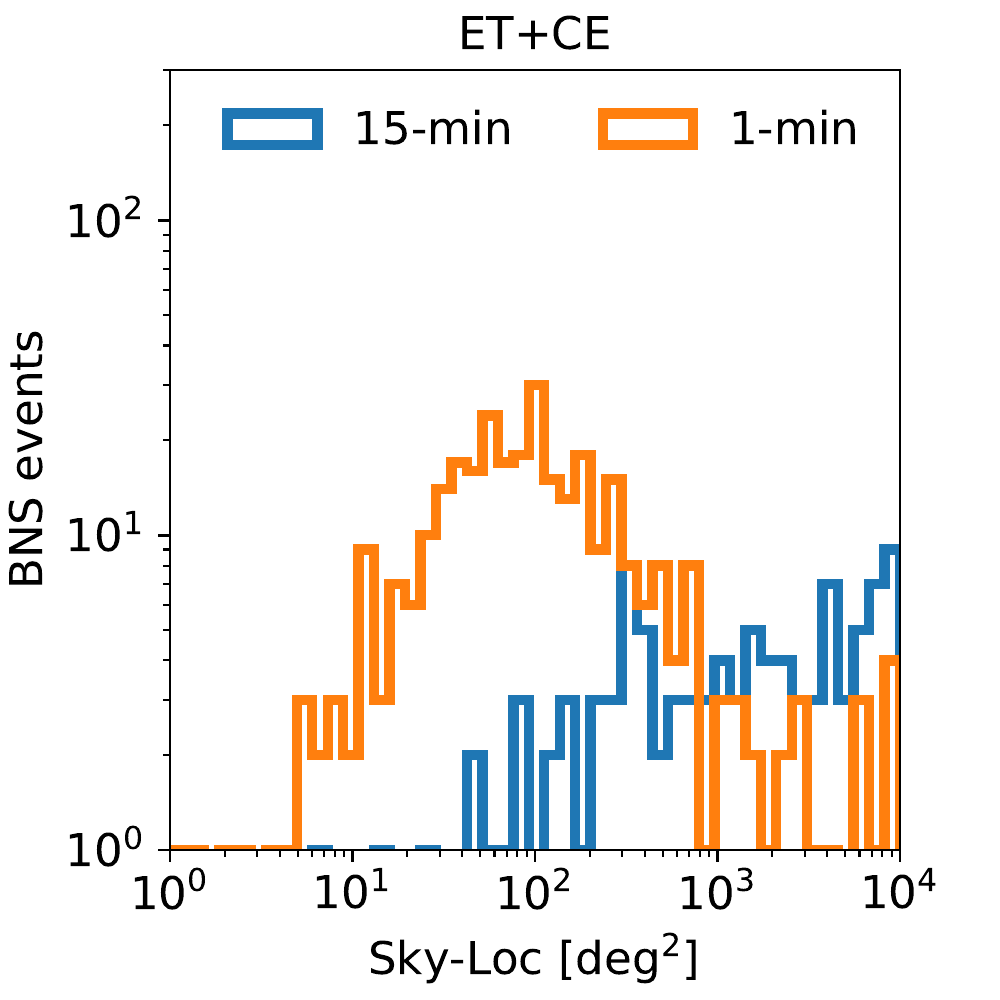}
            \includegraphics[width=0.33 \linewidth, height=6cm]{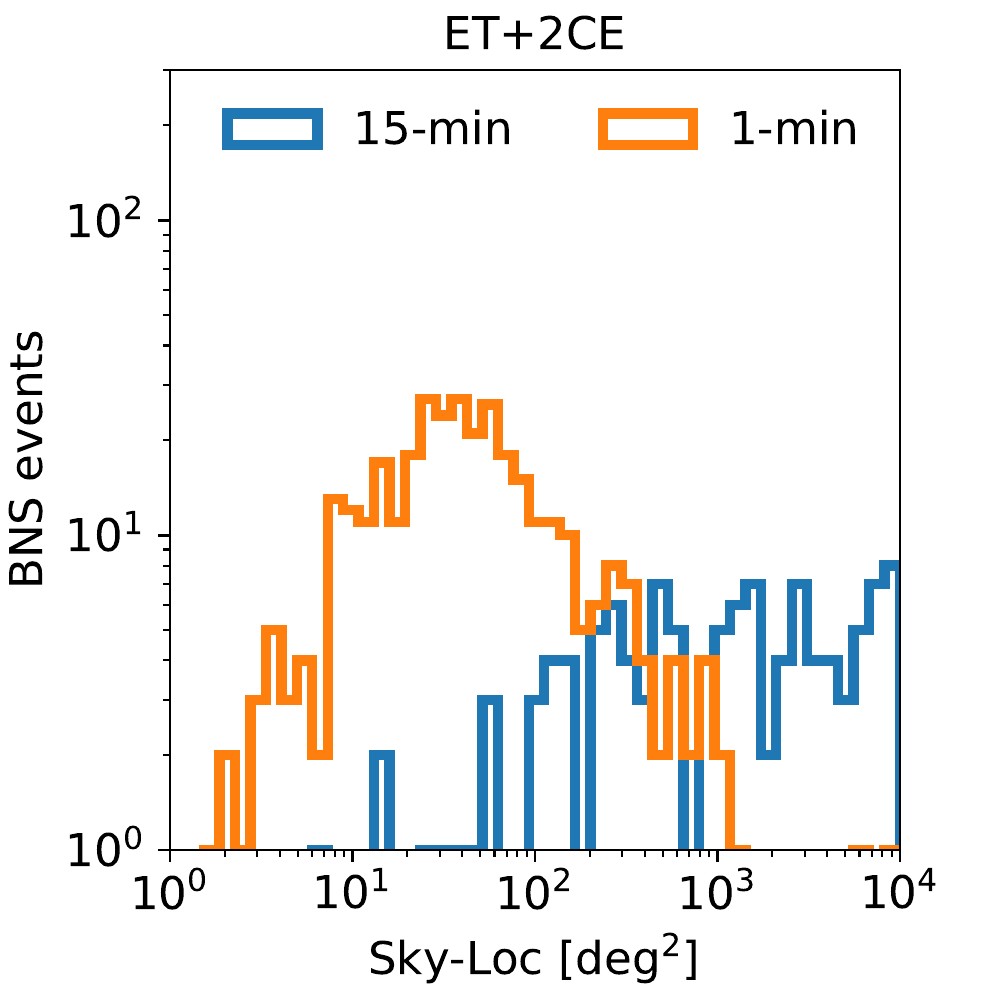}
            \includegraphics[width=0.33 \linewidth, height=6cm]{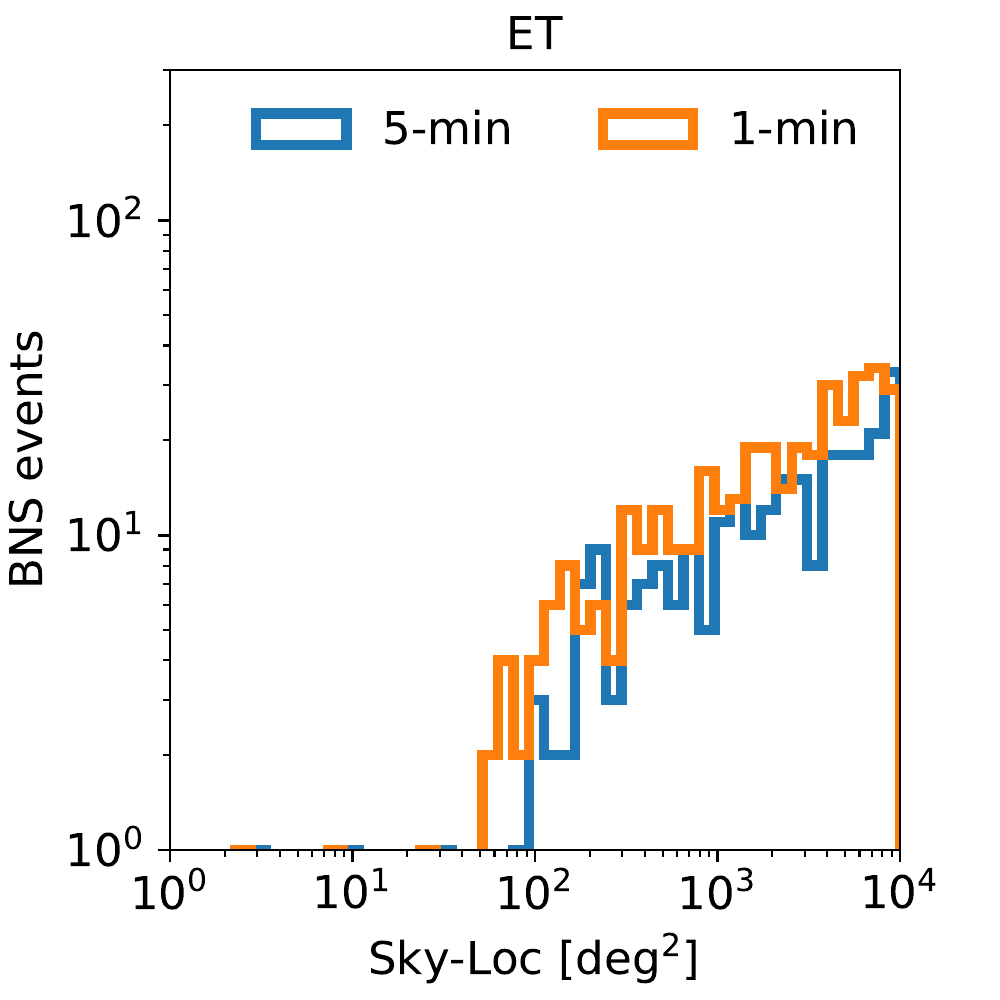}
            \includegraphics[width=0.33 \linewidth, height=6cm]{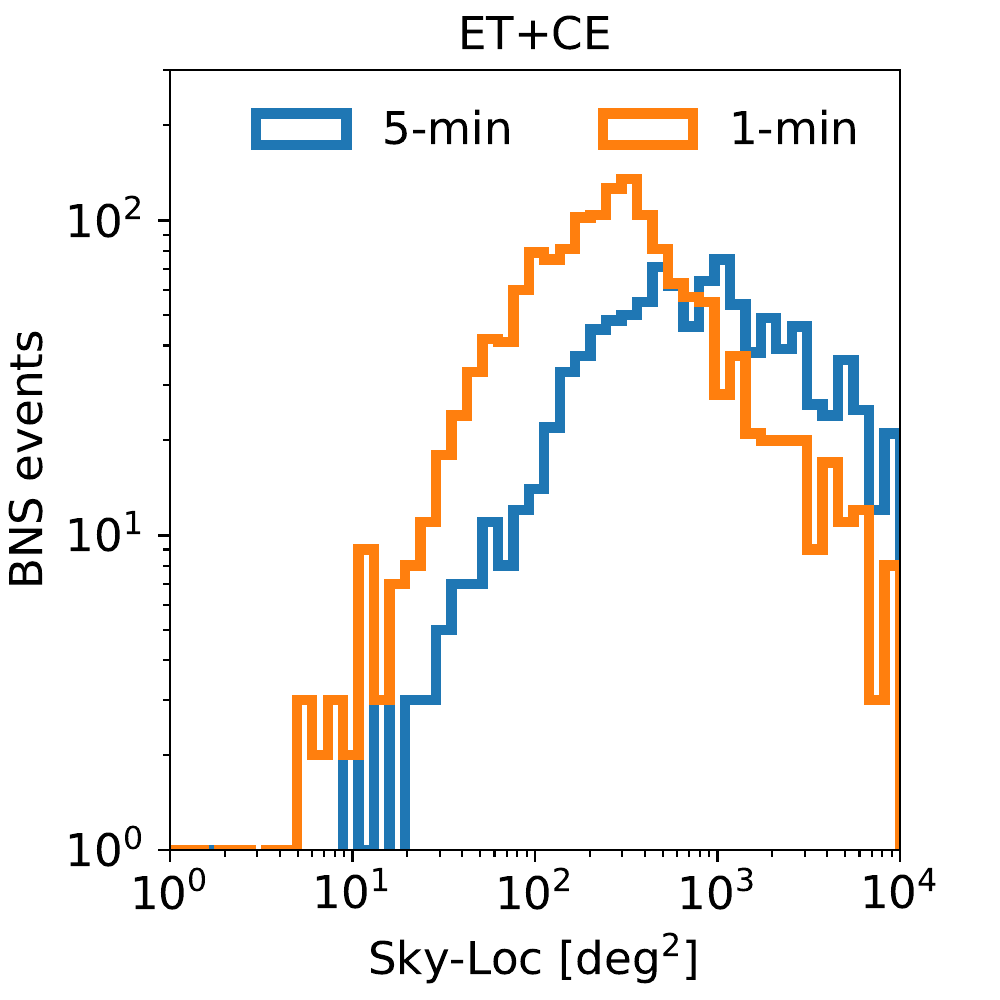}
            \includegraphics[width=0.33 \linewidth, height=6cm]{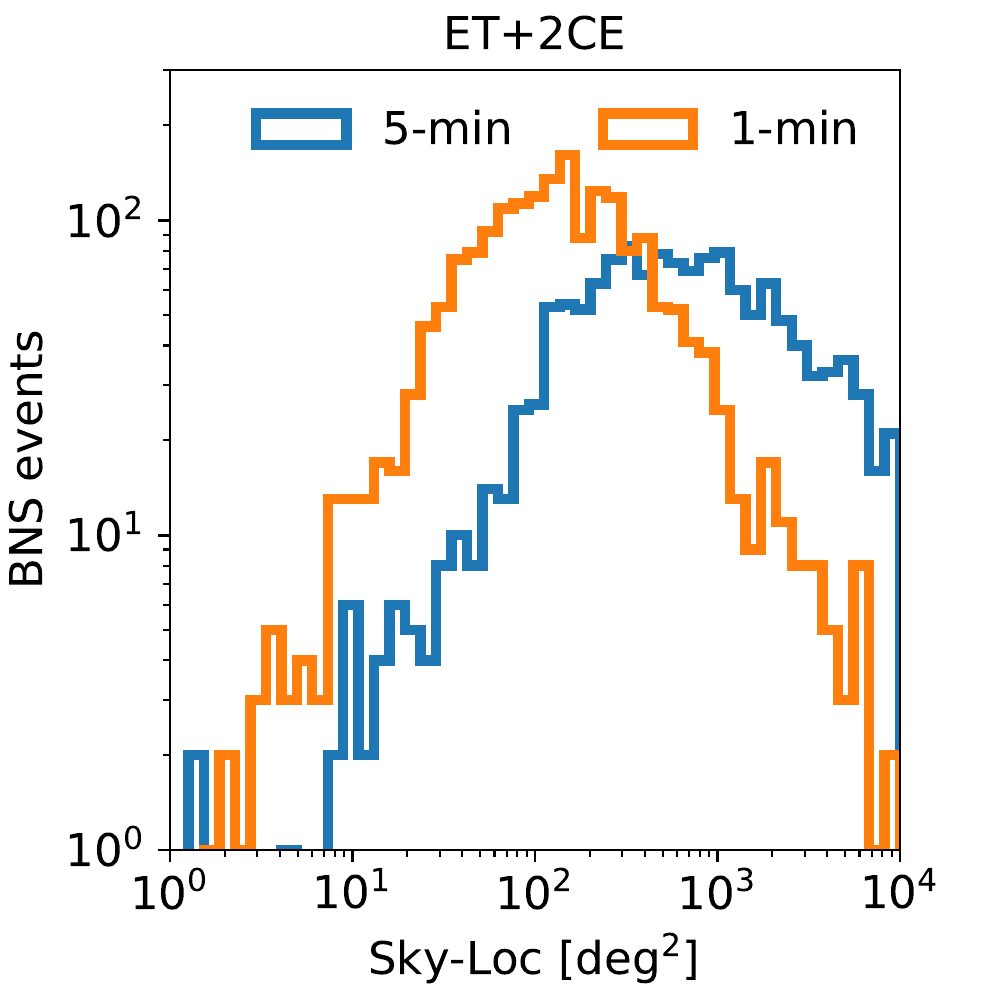}
       \caption{The improvement of the sky-localization estimates getting closer to the merger time. The top (bottom) plots show the sky localizations obtained 15 minutes (5 minutes) before the merger (blue histograms). The 1-minute (orange) histograms show the sky-localization estimates for the same events detected 15 minutes (5 minutes) before the merger (blue histograms). Re-positioning CTA within the updated (smaller) sky-localization released one minute before the merger increases the chance of VHE detection.}
            \label{fig:Update1min}
        \end{figure*}

{For an event detected pre-merger, the estimate of the sky-localization improves as the event gets closer to merger time. Figure \ref{fig:Update1min} shows the systematic improvement of the sky-localization one minute before the merger with respect to the sky-localizations obtained 15-min and 5-min for the same events.}

\end{appendix}
\end{document}